\documentclass[showpacs,preprintnumbers,amsmath,amssymb]{revtex4-2}

\usepackage{graphicx}
\usepackage{dcolumn}
\usepackage{bm}
\usepackage{amsmath}
\DeclareMathOperator\arctanh{arctanh}

\usepackage{amsmath,amssymb,amsbsy,amsfonts,latexsym,graphicx,hyperref,subfigure}
\newmuskip\pFqmuskip

\newcommand\fH[1]{\sbox0{#1}\dimen0=\ht0 \advance\dimen0 -1ex
	\sbox2{\'{}}\sbox2{\raise\dimen0\box2}%
	{\ooalign{\hidewidth\kern.1em\copy2\kern-.5\wd2\box2\hidewidth\cr\box0\crcr}}}

\newcommand*\pFq[6][8]{%
	\begingroup 
	\pFqmuskip=#1mu\relax
	\mathchardef\normalcomma=\mathcode`,
	\mathcode`\,=\string"8000
	\begingroup\lccode`\~=`\,
	\lowercase{\endgroup\let~}\pFqcomma
	{}_{#2}F_{#3}{\left[\genfrac..{0pt}{}{#4}{#5};#6\right]}%
	\endgroup
}
\newcommand{\pFqcomma}{{\normalcomma}\mskip\pFqmuskip} 

\newcommand{\bq}{\begin{equation}}
\newcommand{\eq}{\end{equation}}
\newcommand{\bqa}{\begin{eqnarray}}
\newcommand{\eqa}{\end{eqnarray}}
\newcommand{\nn}{\nonumber \\}

\def\be     {\begin{equation}}
\def\ee     {\end{equation}}
\def\bea        {\begin{eqnarray}}
\def\eea        {\end{eqnarray}}
\def\bnn    {\begin{eqnarray*}}
\def\enn    {\end{eqnarray*}}

\begin{document}

\title{Holographic dual effective field theory in the Luttinger-Ward functional approach: Application to an SYK model}
\author{Yoon-Seok Choun$^{a}$, Hyeon Jung Kim$^{a}$, and Ki-Seok Kim$^{a,b}$ }
\affiliation{$^{a}$Department of Physics, POSTECH, Pohang, Gyeongbuk 37673, Korea \\ $^{b}$Asia Pacific Center for Theoretical Physics (APCTP), Pohang, Gyeongbuk 37673, Korea}

\email[Ki-Seok Kim: ]{tkfkd@postech.ac.kr} 
\email[Yoon-Seok Choun: ]{ychoun@gmail.com}
\email[Hyeon Jung Kim: ]{hkim7218@postech.ac.kr}
%

%
%

%
%

%
%

%
%
%
%
%

\date{\today}

\begin{abstract}
We construct an emergent holographic dual description in the Luttinger-Ward functional approach, where the renormalization group (RG) flows of collective bi-local fields appear manifestly in the bulk effective action with an emergent extra dimension. This holographic dual effective field theory reproduces $1/N$ quantum corrections in a self-consistent manner when we take the UV limit in the bulk effective action. Going into the IR regime in the extra dimension, we observe that a partial class of the field theoretic $1/N$, $1/N^{2}$, ... quantum corrections are resummed in the all-loop order and reorganized to form a holographic dual effective field theory in a large $N$ fashion living on the one-higher dimensional spacetime. In this study, we apply this theoretical framework into an Sachdev-Ye-Kitaev (SYK) model. 

Taking the large $N$ limit in the holographic dual effective field theory, we obtain nonlinearly coupled second-order bulk differential equations of motion for the three bi-local order-parameter fields of fermion self-energy, Green's function, and polarization function. Here, both UV and IR boundary conditions are derived self-consistently from the boundary effective action. We solve these highly intertwined nonlinear differential equations based on the so called matching method. Our ansatz for the bi-local order-parameter fields coincide with the conformally invariant solution of the field theoretic large $N$ limit in the UV limit, but their overall coefficients $RG-flow$ along the extra dimensional space, respectively, reflecting effects of higher-order quantum corrections. As a result, we find an insulating behavior, where the self-energy diverges at IR. 

To confirm this insulating physics, we investigate thermodynamics. We obtain an effective free energy functional in terms of such bi-local dual order-parameter fields, which satisfy the Hamilton-Jacobi equation of the holographic dual effective field theory. Based on the insulating solution, we find that the density of states vanishes at IR. This indicates that the RG flows of the collective dual order-parameter fields give rise to deviation from the Bekenstein-Hawking entropy behavior of the field theoretic large $N$ limit.
\end{abstract}


\maketitle 

\section{Introduction} 

Black hole entropy problem \cite{BH_entropy_0,BH_entropy_I,BH_entropy_II,BH_entropy_III,BH_entropy_IV,BH_entropy_V,BH_entropy_VI} is at the heart of holographic duality conjecture between strongly coupled (weakly correlated) quntum field theories and semiclassical (quantum) gravity descriptions \cite{Holographic_Duality_I,Holographic_Duality_II,Holographic_Duality_III,Holographic_Duality_IV,Holographic_Duality_V,Holographic_Duality_VI,Holographic_Duality_VII}. The Bekenstein-Hawking entropy formula, i.e., the area law of the black hole entropy \cite{BH_entropy_0,BH_entropy_I,BH_entropy_II,BH_entropy_III,BH_entropy_IV} is modified by quantum gravity corrections \cite{BH_entropy_string_theory_Review_I,BH_entropy_string_theory_Review_II}. Not only `bulk' gravitational path integrals in anti-de Sitter space ($AdS$) but also `boundary' conformal field theory calculations are shown to give the same entropy formula, that is, the leading area law plus universal subleading logarithmic correction, higher-order perturbative corrections in the $1/N_{c}$ expansion, and even non-perturbative quantum corrections \cite{BH_entropy_string_theory_I,BH_entropy_string_theory_II,BH_entropy_string_theory_III,BH_entropy_string_theory_IV,BH_entropy_string_theory_V,BH_entropy_string_theory_VI,BH_entropy_string_theory_VIII}.

Various types of $AdS$ black holes \cite{GR_Textbook} with more than mass, for example, charge or angular momentum turn out to show that an extremal $AdS_{2}$ black hole emerges from a higher dimensional $AdS_{D+1}$ with $D \geq 2$ \cite{BH_entropy_string_theory_Review_I,BH_entropy_string_theory_Review_II}. As a result, the entropy of the higher dimensional AdS black hole is essentially given by that of the extremal $AdS_{2}$ black hole. Recent gravitational path integral calculations confirmed this result but with corrections to the original entropy formula of the $AdS_{2}$ black hole, where the Kaluza-Klein reduction for the higher dimensional gravity theory gives rise to Jackiw–Teitelboim (JT) effective gravity theory modified by Kaluza-Klein modes \cite{BH_entropy_AdS2_I,BH_entropy_AdS2_II}. On the other hand, it is not clear how most entropy contributions of higher dimensional conformal field theories can be realized or `localized' into the one-dimensional subsector of them, i.e., the `nearly' conformal quantum mechanics.

Effective `conformal', more precisely, scale-invariant quantum mechanics theory has been proposed in the vicinity of quantum critical points of strongly coupled quantum field theories, referred to as dynamical mean-field theory (DMFT) in the context of many body theory \cite{DMFT_Review}. For example, suppose metal-insulator Mott transitions. Here, not only Fermi-surface electrons but also whole electrons deep inside the Fermi surface are involved with the Mott localization, regarded to be genuine nature of strong correlations. In the vicinity of the metal-insulator transition, `almost' localized electrons start to show their emergent magnetic moments, not showing Pauli paramagnetism but exhibiting the Curie-like behavior. These emergent localized magnetic moments are such low lying critical degrees of freedom, responsible for effective critical quantum mechanics. Until now, it is not clear how strongly coupled quantum field theories above one spacetime dimension RG-flow into such `conformal' quantum mechanics near quantum criticality although the DMFT framework is justified in the infinite dimension limit. 

Recently, we could obtain an effective $AdS_{2}$ extremal black hole type geometry in the IR from the $AdS_{3}$ geometry in the UV \cite{Emergent_AdS2_BH_RG}. Here, we derived an effective holographic dual description for the Gross–Neveu model in two spacetime dimensions, considering Wilsonian RG transformations. The RG flows of all the coupling functions manifest in the level of an effective bulk action, where the RG scale plays the role of an emergent extra dimension \cite{RG_Flow_Holography_Monotonicity,Nonperturbative_Wilson_RG_Disorder,Nonperturbative_Wilson_RG,Einstein_Klein_Gordon_RG_Kim,Einstein_Dirac_RG_Kim,RG_GR_Geometry_I_Kim,RG_GR_Geometry_II_Kim,Kondo_Holography_Kim, Kitaev_Entanglement_Entropy_Kim,RG_Holography_First_Kim}. We found an RG flow from a weakly-coupled chiral-symmetric UV fixed point to a strongly-correlated chiral-symmetry broken IR fixed point, where the renormalized velocity of Dirac fermions vanishes most rapidly and effective critical quantum mechanics appears in the IR. We translated these RG flows of the coupling functions into those of emergent metric tensors and extracted out geometrical properties of the emergent holographic spacetime constructed from the UV- and IR-regional solutions. Interestingly, we obtained the `volume' law of entanglement entropy in the Ryu-Takayanagi formula, which implies appearance of an $AdS_{2}$ black hole type solution even at zero temperature. We critically discussed our field theoretic interpretation for this solution in terms of potentially gapless multi-particle excitation spectra.

In this study, we propose an effective holographic dual description directly from a quantum mechanics theory. Here, we consider an SYK (Sachdev–Ye–Kitaev) model \cite{SYK_Model_I,SYK_Model_II,SYK_Model_III,SYK_Model_IV,SYK_Model_V}. Historically, it was proposed as an effective lattice model for an insulating phase of Si:P, i.e., P doped Si, where randomly doped Ps give rise to emergent localized magnetic moments in Si and their random positions cause random Heisenberg spin-exchange interactions between such localized magnetic moments \cite{MIT_SiP_I,MIT_SiP_II}. This corresponds to a complex SYK model or an SY model. To avoid a spin glass solution given by replica symmetry breaking, the Majorana fermion representation of spin has been considered, referred to as the SYK model, resulting in a conformal invariant saddle-point solution in the large $N$ limit \cite{SYK_Model_I,SYK_Model_II,SYK_Model_III,SYK_Model_IV,SYK_Model_V}. Here, $N$ is the flavor number of Majorana fermion fields.

Recently, the stability of this conformal invariant saddle-point solution was investigated, where Nambu-Goldstone bosons are described by an effective Schwarzian action \cite{SYK_Model_II,SYK_Schwarzian_I,SYK_Schwarzian_II,SYK_Schwarzian_III,SYK_Schwarzian_IV,SYK_Schwarzian_V,SYK_Schwarzian_VI}. These low lying excitations live in the $Diff(1) / SL(2,R)$ coset space, where $Diff(1)$ represents one-dimensional reparametrization symmetry for time (Diffeomorphism invariance before symmetry breaking) and $SL(2,R)$ denotes the invariant subgroup of the conformal invariant saddle-point (symmetry broken) solution. It turns out that $1/N$ quantum corrections given by such collective modes destroy the conformal invariance. As a result, the finite residual entroy given by the saddle-point solution turns out to vanish, and accordingly, the finite density of states also disappears by such quantum corrections.

Although the Schwarzian effective action can be derived from the JT gravity \cite{SYK_Model_II,SYK_Schwarzian_I,SYK_Schwarzian_II,SYK_Schwarzian_III,SYK_Schwarzian_IV,SYK_Schwarzian_V}, it is not completely clear how the JT gravity description \cite{JT_GR_I,JT_GR_II,JT_GR_III,JT_GR_IV} is dual to the SYK quantum mechanics. Indeed, it turns out that there exist additional corrections beyond the Schwarzian effective field theory in the SYK quantum mechanics \cite{SYK_beyond_Schwarzian}. In this respect it is desirable to derive an effective holographic dual description directly from the SYK quantum mechanics theory. Here, we do not ask the mother theory of the SYK effective model Hamiltonian. It is quite a difficult problem to reveal the general path way from higher-dimensional strongly coupled conformal field theories to emergent conformal quantum mechanics theory.

We construct an emergent holographic dual description for an SYK model, where the renormalization group (RG) flows of collective bi-local fields appear manifestly in the bulk effective action with an emergent extra dimension. This holographic dual effective field theory reproduces $1/N$ quantum corrections in a self-consistent manner when we take the UV limit in the bulk effective action. Going into the IR regime in the extra dimension, we observe that a partial class of the field theoretic $1/N$, $1/N^{2}$, ... quantum corrections are resummed in the all-loop order and reorganized to form a holographic dual effective field theory in a large $N$ fashion living on the one-higher dimensional spacetime. 

Taking the large $N$ limit in the holographic dual effective field theory, we obtain nonlinearly coupled second-order bulk differential equations of motion for the three bi-local order-parameter fields of fermion self-energy, Green's function, and polarization function. Here, both UV and IR boundary conditions are derived self-consistently from the boundary effective action. We solve these highly intertwined nonlinear differential equations based on the so called matching method. Our ansatz for the bi-local order-parameter fields coincide with the conformally invariant solution of the field theoretic large $N$ limit in the UV limit, but their overall coefficients $RG-flow$ along the extra dimensional space, respectively, reflecting effects of higher-order quantum corrections. As a result, we find an insulating behavior, where the self-energy diverges at IR. 

To confirm this insulating physics, we investigate thermodynamics. We obtain an effective free energy functional in terms of such bi-local dual order-parameter fields, which satisfy the Hamilton-Jacobi equation of the holographic dual effective field theory. Based on the insulating solution, we find that the density of states vanishes at IR. This indicates that the RG flows of the collective dual order-parameter fields give rise to deviation from the Bekenstein-Hawking entropy behavior of the field theoretic large $N$ limit.

We emphasize that our holographic dual effective field theory for the SYK model has to be regarded as an example of a more general theoretical framework, that is, emergent dual holography in the Luttinger-Ward functional approach, where effective low energy degrees of freedom are given by self-energy and Green's functions. In the Luttinger-Ward functional approach, both self-energy and vertex corrections can be taken into account self-consistently, satisfying the Ward identity, in other words, the Luttinger-Ward functional approach being referred to as a conserving approximation \cite{LW_Functional_1,LW_Functional_2}. In a more general context, the present study is to generalize the Luttinger-Ward functional approach and to take higher order quantum corrections in the sense of the functional renormalization group (FRG) approach, as discussed above. Recently, it has been discussed how the FRG framework can be reformulated as the RG flows of self-energy and Green's functions in the Luttinger-Ward functional approach \cite{LW_Functional_FRG}. To investigate the non-perturbative structure of this general framework would be an interesting future direction beyond the scope of the present study.

\section{Emergent dual holographic description for an SYK model}

\subsection{A review on the SYK model in the replica symmetric ansatz}

We consider an SYK model in the replica symmetric ansatz as follows
\bqa  \mathcal{Z}_{R} = && \int D \Sigma_{\tau}^{\tau'} D G_{\tau}^{\tau'} D \eta_{\alpha}(\tau)  \exp\Big[ - R \int_{0}^{\beta} d \tau \int_{0}^{\beta} d \tau' \Big\{ \eta_{\alpha}(\tau) \delta(\tau-\tau') \partial_{\tau'} \eta_{\alpha} (\tau') \nn && + q_{d} \Gamma_{J} \Sigma_{\tau}^{\tau'} \Big( \eta_{\alpha}(\tau) \eta_{\alpha}(\tau') -  N G_{\tau'}^{\tau} \Big)  -  N q_{d} \Gamma_{J} \Big( G_{\tau}^{\tau'} G_{\tau'}^{\tau} \Big)^{2} \Big\} \Big] . \eqa
Here, $R$ is the replica index to keep the identity, $\ln Z = \lim_{R \rightarrow 0} \frac{Z^{R} - 1}{R}$, where $Z^{R}$ is replaced with $\mathcal{Z}_{R}$. $\eta_{\alpha}(\tau)$ is a Majorana fermion field with its flavor index $\alpha$, running from $1$ to $N$. $\Sigma_{\tau}^{\tau'}$ is a Lagrange multiplier field to keep the constraint of $G_{\tau'}^{\tau} = \frac{1}{N} \sum_{\alpha = 1}^{N} \eta_{\alpha}(\tau) \eta_{\alpha}(\tau')$,
where the Einstein convention has been used in the effective action. Introducing this expression into the last term, we obtain the SYK model. It is clear that the bi-local (in-time) collective field $\Sigma_{\tau}^{\tau'}$ ($G_{\tau}^{\tau'}$) plays the role of the self-energy (Green's function) of the Majorana fermions in this effective partition function.

Performing the Hubbard-Stratonovich transformation for the last interaction term, we obtain
\bqa \mathcal{Z}_{R} = && \int D \Sigma_{\tau}^{\tau'} D \Pi_{\tau}^{\tau'} D G_{\tau}^{\tau'} D \eta_{\alpha}(\tau) \exp\Big[ - R \int_{0}^{\beta} d \tau \int_{0}^{\beta} d \tau' \Big\{ \eta_{\alpha}(\tau) \delta(\tau-\tau') \partial_{\tau'} \eta_{\alpha} (\tau') \nn &&   + q_{d} \Gamma_{J} \Sigma_{\tau}^{\tau'} \Big( \eta_{\alpha}(\tau) \eta_{\alpha}(\tau') -  N G_{\tau'}^{\tau} \Big) +  N q_{d} \Gamma_{J} \Pi_{\tau'}^{\tau}  G_{\tau}^{\tau'} G_{\tau'}^{\tau} + \frac{ N q_{d} \Gamma_{J} }{4} \Pi_{\tau'}^{\tau}\Pi_{\tau'}^{\tau} \Big\} \Big] . \eqa
Here, the collective bi-local field $\Pi_{\tau'}^{\tau} $ corresponds to a polarization function, taking most singular corrections at the free fermion fixed point in the large $N$ limit. This will be clarified below.

%
%

Performing the path integral for Majorana fermion fields, we obtain
\bqa  \mathcal{Z}_{SYK} = && \int D \Sigma_{\tau}^{\tau'} D \Pi_{\tau}^{\tau'} D G_{\tau}^{\tau'} \exp\Big[ - N \int_{0}^{\beta} d \tau \int_{0}^{\beta} d \tau' \Big\{ - \frac{1}{2} \ln \Big( \delta(\tau-\tau') \partial_{\tau'} + q_{d} \Gamma_{J} \Sigma_{\tau}^{\tau'} \Big) \nn &&  - q_{d} \Gamma_{J} \Sigma_{\tau}^{\tau'} G_{\tau'}^{\tau} + q_{d} \Gamma_{J}\Pi_{\tau'}^{\tau}  G_{\tau}^{\tau'} G_{\tau'}^{\tau} + \frac{ q_{d} \Gamma_{J} }{4} \Pi_{\tau'}^{\tau}\Pi_{\tau'}^{\tau} \Big\} \Big] . \label{SYK_EFT} \eqa 
Here, we set $R = 1$ for the replica symmetric solution. Taking the large $N$ limit, we consider the saddle-point approximation. As a result, we find that the Green's function, the self-energy, and the polarization function should satisfy
\bqa && \Big( \delta(\tau-\tau'') \partial_{\tau''} + q_{d} \Gamma_{J} \Sigma_{\tau}^{\tau''}  \Big) G_{\tau''}^{\tau'} = - \frac{1}{2} \delta(\tau - \tau') , \nn && \Sigma_{\tau}^{\tau'} = 2\Pi_{\tau'}^{\tau}  G_{\tau}^{\tau'} , ~~~~~\Pi_{\tau'}^{\tau}  = - 2 G_{\tau'}^{\tau} G_{\tau}^{\tau'} . \nonumber \eqa
It turns out that the first-order time-derivative term is irrelevant in the long-time limit. As a result, the saddle-point equations are given by 
\bqa && q_{d} \Gamma_{J} \Sigma_{\tau}^{\tau''}  G_{\tau''}^{\tau'} = - \frac{1}{2} \delta(\tau - \tau') , ~~~ \Sigma_{\tau}^{\tau'} = 2\Pi_{\tau'}^{\tau}  G_{\tau}^{\tau'} , ~~~\Pi_{\tau'}^{\tau}  = - 2 G_{\tau'}^{\tau} G_{\tau}^{\tau'} . \nonumber \eqa
It is not so difficult to check out that these coupled equations are invariant under the SL(2,R) transformation. The SL(2,R) invariant saddle-point solution is given by 
\bqa && \Sigma_{\tau}^{\tau'} = \mbox{sgn}(\tau-\tau') \frac{\mathcal{C}_{\Sigma}}{|\tau-\tau'|^{\Delta_{\Sigma}}} , ~~~ G_{\tau}^{\tau'} = \mbox{sgn}(\tau-\tau') \frac{\mathcal{C}_{G}}{|\tau-\tau'|^{\Delta_{G}}} , ~~~ \Pi_{\tau}^{\tau'} = \frac{\mathcal{C}_{\Pi}}{|\tau-\tau'|^{\Delta_{\Pi}}} . \nonumber \eqa
Here, both the critical exponents of $\Delta_{\Sigma}$, $\Delta_{G}$, and $\Delta_{\Pi}$ and the coefficients of $\mathcal{C}_{\Sigma}$, $\mathcal{C}_{G}$, and $\mathcal{C}_{\Pi}$ can be determined by these self-consistent equations. In this study, we investigate possible non-perturbative quantum corrections to this solution based on our holographic dual effective field theory.

%
%

To invetigate the stability of this conformally invariant fixed point, we consider quantum corrections as follows
\bqa && \Sigma_{\tau}^{\tau'} \longrightarrow \Sigma_{\tau}^{\tau'} + \delta \Sigma_{\tau}^{\tau'} , \\ && \Pi_{\tau}^{\tau'} \longrightarrow \Pi_{\tau}^{\tau'} + \delta \Pi_{\tau}^{\tau'} , \\ && G_{\tau}^{\tau'} \longrightarrow G_{\tau}^{\tau'} + \delta G_{\tau}^{\tau'} . \eqa
Here, $\Sigma_{\tau}^{\tau'} $, $\Pi_{\tau}^{\tau'}$, and $G_{\tau}^{\tau'}$ are collective dual `order-parameter' fields to be determined self-consistently through their `Landau-Ginzburg' self-consistent equations, which take into account effects of quantum fluctuations from $\delta \Sigma_{\tau}^{\tau'}$, $\delta \Pi_{\tau}^{\tau'}$, and $\delta G_{\tau}^{\tau'}$ in the Gaussian order.

Introducing these fluctuations into Eq. (\ref{SYK_EFT}) and expanding the log-term up to the second order in $\delta \Sigma_{\tau}^{\tau'}$, we find
\bqa && \mathcal{Z}_{SYK}[\Sigma_{\tau}^{\tau'} , \Pi_{\tau}^{\tau'}, G_{\tau}^{\tau'}] = \bar{\mathcal{Z}}_{SYK}[\Sigma_{\tau}^{\tau'} , \Pi_{\tau}^{\tau'}, G_{\tau}^{\tau'}] \int D \delta \Sigma_{\tau}^{\tau'} D \delta \Pi_{\tau}^{\tau'} D \delta G_{\tau}^{\tau'} \nn && \exp\Big[ - N \int_{0}^{\beta} d \tau \int_{0}^{\beta} d \tau' \Big\{ q_{d} \Gamma_{J} \delta \Sigma_{\tau}^{\tau'} \Big( \mathcal{G}_{\tau'}^{\tau}  - G_{\tau'}^{\tau} \Big) - q_{d} \Gamma_{J} \Big( \Sigma_{\tau}^{\tau'} - 2\Pi_{\tau'}^{\tau}  G_{\tau}^{\tau'} \Big) \delta G_{\tau'}^{\tau} \nn && \ \ \ \ \ \ \ + q_{d} \Gamma_{J} \delta\Pi_{\tau'}^{\tau}  \Big( G_{\tau}^{\tau'} G_{\tau'}^{\tau} + \frac{1}{2}\Pi_{\tau'}^{\tau}  \Big) + (q_{d} \Gamma_{J})^{2} \int_{0}^{\beta} d \tau'' \delta \Sigma_{\tau}^{\tau'} \Big( \int_{0}^{\beta} d \tau''' \mathcal{G}_{\tau'}^{\tau'''}  \mathcal{G}_{\tau'''}^{\tau''}  \Big) \delta \Sigma_{\tau''}^{\tau}  \nn && \ \ \ \ \ \ \ - q_{d} \Gamma_{J} \delta \Sigma_{\tau}^{\tau'} \delta G_{\tau'}^{\tau} + q_{d} \Gamma_{J}\Pi_{\tau'}^{\tau}  \delta G_{\tau}^{\tau'} \delta G_{\tau'}^{\tau} + 2 q_{d} \Gamma_{J} \delta\Pi_{\tau'}^{\tau}  \delta G_{\tau}^{\tau'} G_{\tau'}^{\tau} + \frac{ q_{d} \Gamma_{J} }{4} \delta \Pi_{\tau'}^{\tau}\delta \Pi_{\tau'}^{\tau} \Big\} \Big] . \label{SYK_Correction} \eqa
Here, $\bar{\mathcal{Z}}_{SYK}[\Sigma_{\tau}^{\tau'} , \Pi_{\tau}^{\tau'}, G_{\tau}^{\tau'}]$ is the partition function, given by
\bqa \bar{\mathcal{Z}}_{SYK}[\Sigma_{\tau}^{\tau'} , \Pi_{\tau}^{\tau'}, G_{\tau}^{\tau'}] =  \exp\Big[- N \int_{0}^{\beta} d \tau \int_{0}^{\beta} d \tau' \Big\{ - \frac{1}{2} \ln \Big( - \frac{1}{2} \frac{1}{\mathcal{G}_{\tau}^{\tau'}}  \Big) - q_{d} \Gamma_{J} \Sigma_{\tau}^{\tau'} G_{\tau'}^{\tau} + q_{d} \Gamma_{J}\Pi_{\tau'}^{\tau}  G_{\tau}^{\tau'} G_{\tau'}^{\tau}  +  \frac{ q_{d} \Gamma_{J} }{4} \Pi_{\tau'}^{\tau}\Pi_{\tau'}^{\tau} \Big\} \Big] , \label{Partition_Ft_Saddle} \nn \eqa
%
%
and $\mathcal{G}_{\tau}^{\tau'}$ is the fermion Green's function, given by
\bqa && \Big( \delta(\tau-\tau'') \partial_{\tau''} + q_{d} \Gamma_{J} \Sigma_{\tau}^{\tau''}  \Big) \mathcal{G}_{\tau''}^{\tau'} = - \frac{1}{2} \delta(\tau - \tau') . \eqa
It is straightforward to see that the partition function of Eq. (\ref{SYK_Correction}) is reduced into the saddle-point partition function, neglecting the gaussian quantum fluctuations.

Performing the path integrals for quantum fluctuations ($\int D \delta \Sigma_{\tau}^{\tau'} D \delta \Pi_{\tau}^{\tau'} D \delta G_{\tau}^{\tau'}$), we obtain an effective potential with $1/N$ quantum corrections for $\Sigma_{\tau}^{\tau'} $, $\Pi_{\tau}^{\tau'}$, and $G_{\tau}^{\tau'}$. In appendix section \ref{SYK_1_N_QCs}, we perform the path integrals for such Gaussian quantum fluctuations and obtain an effective potential with $1/N$ quantum corrections. We point out that this effective potential determines collective dual order-parameter fields in a self-consistent way beyond the introduction of $1/N$ quantum corrections into the large $N$ saddle-point solution. In this study, we will not try to solve these complex self-consistent equations. Instead, we generalize this framework to take into account even more higher-order quantum corrections, manifesting RG flows of the collective dual order-parameter fields in the bulk effective action.

\subsection{Emergent dual holographic description for an SYK model} 

Based on the self-consistent construction of Eq. (\ref{SYK_Correction}) in the previous subsection, we propose an effective holographic dual field theory as follows 
\bqa && \mathcal{Z}_{SYK}(z_{f}) = \int D \Sigma_{\tau}^{\tau'}(z)  D \Pi_{\tau}^{\tau'}(z)  DG_{\tau}^{\tau'}(z)  \int D \eta_{\alpha}(\tau) \exp\Bigg[ - \int_{0}^{\beta} d \tau \int_{0}^{\beta} d \tau' \Bigg\{ \eta_{\alpha}(\tau) \delta(\tau-\tau') \partial_{\tau'} \eta_{\alpha} (\tau') \nn && + q_{d} \Gamma_{J} \Sigma_{\tau}^{\tau'}(z_{f})     \Big( \eta_{\alpha}(\tau) \eta_{\alpha}(\tau') - N G_{\tau'}^{\tau}(z_{f})  \Big) + N q_{d} \Gamma_{J} \Pi_{\tau'}^{\tau}(0)  G_{\tau}^{\tau'}(0)  G_{\tau'}^{\tau}(0)  + \frac{ N q_{d} \Gamma_{J} }{4} \Pi_{\tau'}^{\tau}(0)  \Pi_{\tau'}^{\tau}(0)  \Bigg\} \Bigg] \nn && \exp\Bigg[ - N \int_{0}^{z_{f}} d z \int_{0}^{\beta} d \tau \int_{0}^{\beta} d \tau' \Bigg\{ 2 q_{d} \Gamma_{J}  \Pi_{\tau'}^{\tau}(z)   \Big(\partial_{z}G_{\tau}^{\tau'}(z) \Big) G_{\tau'}^{\tau}(z)   + q_{d} \Gamma_{J} \Big(\partial_{z}  \Pi_{\tau'}^{\tau}(z)  \Big) G_{\tau}^{\tau'}(z) G_{\tau'}^{\tau}(z) \nn && \ \ \ \ \ \ \ + \frac{ q_{d} \Gamma_{J} }{2} \Big(\partial_{z}  \Pi_{\tau'}^{\tau}(z)  \Big)  \Pi_{\tau'}^{\tau}(z) + q_{d} \Gamma_{J} \mathcal{G}_{\tau'}^{\tau}(z) \Big(\partial_{z} \Sigma_{\tau}^{\tau'}(z)  \Big)  - q_{d} \Gamma_{J} \Big(\partial_{z} \Sigma_{\tau}^{\tau'}(z) \Big) \Big(\partial_{z} G_{\tau'}^{\tau}(z)  \Big)\nn && \ \ \ \ \ \ \ + (q_{d} \Gamma_{J})^{2} \int_{0}^{\beta} d \tau'' \int_{0}^{\beta} d \tau''' \Big(\partial_{z} \Sigma_{\tau}^{\tau'}(z) \Big) \mathcal{G}_{\tau'}^{\tau'''}(z) \mathcal{G}_{\tau'''}^{\tau''}(z) \Big(\partial_{z} \Sigma_{\tau''}^{\tau}(z)  \Big) + q_{d} \Gamma_{J}  \Pi_{\tau'}^{\tau}(z)   \Big(\partial_{z}G_{\tau}^{\tau'}(z) \Big) \Big(\partial_{z} G_{\tau'}^{\tau}(z)  \Big) \nn && \ \ \ \ \ \ \ + 2 q_{d} \Gamma_{J} \Big(\partial_{z}  \Pi_{\tau'}^{\tau}(z)  \Big) \Big(\partial_{z}G_{\tau}^{\tau'}(z) \Big) G_{\tau'}^{\tau}(z)  + \frac{ q_{d} \Gamma_{J} }{4} \Big(\partial_{z}  \Pi_{\tau'}^{\tau}(z)  \Big)^{2}  - \frac{1}{2} \ln \Big( - \frac{1}{2} [\mathcal{G}_{\tau}^{\tau'}(z)]^{-1} \Big) \Bigg\} \Bigg] .  \label{HDEFT} \eqa 
Here, $\mathcal{G}_{\tau}^{\tau'}(z)$ is the Majorana fermion Green's function, given by
\bqa && \Big( \delta(\tau-\tau'') \partial_{\tau''} + q_{d} \Gamma_{J} \Sigma_{\tau}^{\tau''}(z)   \Big) \mathcal{G}_{\tau''}^{\tau'}(z) = - \frac{1}{2} \delta(\tau-\tau') . \eqa
$z$ denotes the coordinate of an extra dimension, identified with an RG scale. Accordingly, $z_{f}$ indicates an IR boundary. Considering the $z_{f} = 0$ case, we obtain 
\bqa && \bar{\mathcal{Z}}_{SYK}[\Sigma_{\tau}^{\tau'}(0)  ,\Pi_{\tau'}^{\tau}(0) ,G_{\tau'}^{\tau}(0) ] = \int D \eta_{\alpha}(\tau) \exp\Big[ - \int_{0}^{\beta} d \tau \int_{0}^{\beta} d \tau' \Big\{ \eta_{\alpha}(\tau) \delta(\tau-\tau') \partial_{\tau'} \eta_{\alpha} (\tau') \nn && + q_{d} \Gamma_{J} \Sigma_{\tau}^{\tau'}(0)   \Big( \eta_{\alpha}(\tau) \eta_{\alpha}(\tau') - N G_{\tau'}^{\tau}(0)  \Big) + N q_{d} \Gamma_{J} \Pi_{\tau'}^{\tau}(0)  G_{\tau}^{\tau'}(0)  G_{\tau'}^{\tau}(0)  + \frac{ N q_{d} \Gamma_{J} }{4} \Pi_{\tau'}^{\tau}(0)  \Pi_{\tau'}^{\tau}(0)  \Big\} \Big] , \nonumber \eqa 
which reproduces the saddle-point partition function of the SYK model in the large $N$ limit.

Although this holographic dual effective field theory looks quite complicated at first glance, we claim that one can read its mathematical structure from the one-loop effective partition function (\ref{SYK_Correction}) of the previous subsection. We point out that there exists a concrete renormalization structure in this expression. In this and next subsections, we figure out the RG flow structure and discuss its non-perturbativeness.

To understand the role of the emergent extra dimension in the bulk effective action, we take the $z_{f} \rightarrow 0$ limit instead of $z_{f} = 0$. We introduce an expansion for each collective dual order-parameter field in the first order with resect to $z$ as follows
\bqa && \Sigma_{\tau}^{\tau'}(z)  = \Sigma_{0,\tau}^{\tau'}   + z \Sigma_{1,\tau}^{\tau'}    , \\ && \Pi_{\tau}^{\tau'}(z)  = \Pi_{0,\tau}^{\tau'}    + z \Pi_{1,\tau}^{\tau'}    , \\ &&G_{\tau}^{\tau'}(z)  = G_{0,\tau}^{\tau'}    + z G_{1,\tau}^{\tau'}   . \eqa
The Majorana fermion Green's function is also expanded as $\mathcal{G}_{\tau}^{\tau'}(z) = \mathcal{G}_{0,\tau}^{\tau'} + z \mathcal{G}_{1,\tau}^{\tau'}$, where $\mathcal{G}_{0,\tau}^{\tau'}$ and $\mathcal{G}_{1,\tau}^{\tau'}$ are given by
%
%
\bqa && \Big( \delta(\tau-\tau'') \partial_{\tau''} + q_{d} \Gamma_{J} \Sigma_{0,\tau}^{\tau''}    \Big) \mathcal{G}_{0,\tau''}^{\tau'}= - \frac{1}{2} \delta(\tau-\tau') , \\ && \Big( \delta(\tau-\tau'') \partial_{\tau''} + q_{d} \Gamma_{J} \Sigma_{0,\tau}^{\tau''}    \Big) \mathcal{G}_{1,\tau''}^{\tau'} = - q_{d} \Gamma_{J} \Sigma_{1,\tau}^{\tau''}    \mathcal{G}_{0,\tau''}^{\tau'} , \eqa
respectively.

Inserting these expansions into Eq. (\ref{HDEFT}) and keeping all terms up to the first order in $z$, we obtain 
\bqa && \mathcal{Z}_{SYK}[\Sigma_{0,\tau}^{\tau'}  , \Pi_{0,\tau}^{\tau'}   , G_{0,\tau}^{\tau'}   ;   z  ] = \int D \Sigma_{1,\tau}^{\tau'}    D \Pi_{1,\tau}^{\tau'}    D G_{1,\tau}^{\tau'}    \int D \eta_{\alpha}(\tau) \nn && \exp\Bigg[ - \int_{0}^{\beta} d \tau \int_{0}^{\beta} d \tau' \Bigg\{ \eta_{\alpha}(\tau) \delta(\tau-\tau') \partial_{\tau'} \eta_{\alpha} (\tau') + q_{d} \Gamma_{J} \Sigma_{0,\tau}^{\tau'}   \Big( \eta_{\alpha}(\tau) \eta_{\alpha}(\tau') - N G_{0,\tau'}^{\tau}   \Big) \nn && \ \ \ \ \ \ \ + N q_{d} \Gamma_{J} \Pi_{0,\tau'}^{\tau}   G_{0,\tau}^{\tau'}    G_{0,\tau'}^{\tau}    + \frac{ N q_{d} \Gamma_{J} }{4} \Pi_{0,\tau'}^{\tau} \Pi_{0,\tau'}^{\tau}      + z q_{d} \Gamma_{J} \Sigma_{1,\tau}^{\tau'}    \Big( \eta_{\alpha}(\tau) \eta_{\alpha}(\tau') - N G_{1,\tau'}^{\tau}    \Big) \Bigg\} \Bigg] \nn &&  \exp\Bigg[ - z N \int_{0}^{\beta} d \tau \int_{0}^{\beta} d \tau' \Bigg\{ q_{d} \Gamma_{J} \Big( \mathcal{G}_{0,\tau'}^{\tau} - G_{0,\tau'}^{\tau}     \Big) \Sigma_{1,\tau}^{\tau'}    - q_{d} \Gamma_{J} \Big( \Sigma_{0,\tau}^{\tau'}   - 2 \Pi_{0,\tau'}^{\tau}   G_{0,\tau}^{\tau'}    \Big) G_{1,\tau'}^{\tau}    \nn && \ \ \ \ \ \ \   + q_{d} \Gamma_{J} \Pi_{1,\tau'}^{\tau} \Big( G_{0,\tau}^{\tau'}    G_{0,\tau'}^{\tau}    + \frac{1}{2} \Pi_{0,\tau'}^{\tau}   \Big) + (q_{d} \Gamma_{J})^{2} \int_{0}^{\beta} d \tau'' \int_{0}^{\beta} d \tau''' \Sigma_{1,\tau}^{\tau'}    \mathcal{G}_{0,\tau'}^{\tau'''} \mathcal{G}_{0,\tau'''}^{\tau''} \Sigma_{1,\tau''}^{\tau}  \nn && \ \ \ \ \ \ \   + 2 q_{d} \Gamma_{J} \Pi_{1,\tau'}^{\tau} G_{1,\tau}^{\tau'}    G_{0,\tau'}^{\tau}    + q_{d} \Gamma_{J} \Pi_{0,\tau'}^{\tau}   G_{1,\tau}^{\tau'}    G_{1,\tau'}^{\tau}    + \frac{ q_{d} \Gamma_{J} }{4} \Pi_{1,\tau'}^{\tau}(z) \Pi_{1,\tau'}^{\tau}(z)  - \frac{1}{2} \ln \Big(- \frac{1}{2} [\mathcal{G}_{0,\tau}^{\tau'}]^{-1} \Big) \Bigg\} \Bigg] , \label{HDEFT_UV_Limit} \eqa 
where the bulk part is given by the order of $z$. Comparing this expression with Eq. (\ref{SYK_Correction}) and Eq. (\ref{Partition_Ft_Saddle}), we observe the following correspondence
\bqa && \Sigma_{1,\tau}^{\tau'}    \leftrightarrow \delta \Sigma_{\tau}^{\tau'} , ~~~~~ \Pi_{1,\tau}^{\tau'}    \leftrightarrow \delta \Pi_{\tau}^{\tau'} , ~~~~~ G_{1,\tau}^{\tau'}    \leftrightarrow \delta G_{\tau}^{\tau'}  , \eqa 
where $\Sigma_{0,\tau}^{\tau'}  $, $\Pi_{0,\tau}^{\tau'}   $, and $G_{0,\tau}^{\tau'}   $ were regarded as background fields in the large $N$ limit, determined self-consistently with the introduction of quantum corrections from $\int D \Sigma_{1,\tau}^{\tau'}    D \Pi_{1,\tau}^{\tau'}    D G_{1,\tau}^{\tau'}   $. In other words, the perturbation structure of Eq. (\ref{HDEFT}) is identical to that of Eq. (\ref{SYK_Correction}) and Eq. (\ref{Partition_Ft_Saddle}).

One may criticize that there exist additional terms in Eq. (\ref{HDEFT_UV_Limit}) that do not appear in Eq. (\ref{SYK_Correction}). In particular, there do not exist Majorana fermion terms in Eq. (\ref{SYK_Correction}), where such fermion degrees of freedom have been integrated out completely. The origin of this difference is that Majorana fermion fields in Eq. (\ref{HDEFT_UV_Limit}) are integrated out only partially in the Wilsonian RG transformation \cite{Emergent_AdS2_BH_RG,RG_Flow_Holography_Monotonicity,Nonperturbative_Wilson_RG_Disorder,Nonperturbative_Wilson_RG,Einstein_Klein_Gordon_RG_Kim,Einstein_Dirac_RG_Kim,RG_GR_Geometry_I_Kim,RG_GR_Geometry_II_Kim,Kondo_Holography_Kim, Kitaev_Entanglement_Entropy_Kim,RG_Holography_First_Kim}. In this respect the infinitesimal number $z$ corresponds to the infinitesimal variation of the cutoff $d \Lambda$, where the Majorana fermion fields between $\Lambda$ and $\Lambda - d \Lambda$ are integrated out and other Majorana fermion degrees of freedom remain untact. Repeating these RG transformations until the IR cutoff $z_{f}$ is reached, we obtain the holographic dual effective field theory Eq. (\ref{HDEFT}), which generalizes the Luttinger-Ward functional approach \cite{LW_Functional_1,LW_Functional_2} to take higher order quantum corrections in the FRG perspective \cite{LW_Functional_FRG}. In other words, the present theoretical framework for the holographic dual effective field theory generalizes our previous one \cite{Emergent_AdS2_BH_RG,RG_Flow_Holography_Monotonicity,Nonperturbative_Wilson_RG,Einstein_Klein_Gordon_RG_Kim,Einstein_Dirac_RG_Kim,RG_GR_Geometry_I_Kim,RG_GR_Geometry_II_Kim, Kondo_Holography_Kim,Kitaev_Entanglement_Entropy_Kim} in a more accurate FRG way. A brute force derivation of this non-perturbative effective field theory has to be performed in the Wilsonian RG transformation near future.

\subsection{Non-perturbative nature of the holographic dual effective field theory}

To verify the non-perturbative nature of the holographic dual effective field theory, we discretize the continuum coordinate $z$ of the extra dimension as follows 
\bqa && \mathcal{Z}_{SYK}(f) = \int \Pi_{k = 1}^{f} D \Sigma_{k,\tau}^{\tau'}  D \Pi_{k,\tau}^{\tau'} D G_{k,\tau}^{\tau'}  \int D \eta_{\alpha}(\tau) \exp\Big[ - \int_{0}^{\beta} d \tau \int_{0}^{\beta} d \tau'  \Big\{ \eta_{\alpha}(\tau) \delta(\tau-\tau') \partial_{\tau'} \eta_{\alpha} (\tau') \nn &&+ q_{d} \Gamma_{J} \Sigma_{f,\tau}^{\tau'}  \Big( \eta_{\alpha}(\tau) \eta_{\alpha}(\tau') - N G_{f,\tau'}^{\tau}  \Big) + N q_{d} \Gamma_{J} \Pi_{0,\tau'}^{\tau}   G_{0,\tau}^{\tau'}    G_{0,\tau'}^{\tau}   + \frac{ N q_{d} \Gamma_{J} }{4} \Pi_{0,\tau'}^{\tau} \Pi_{0,\tau'}^{\tau}      \Big\} \Big]  \nn && \exp\Big[ - N \sum_{k = 1}^{f} \mathcal{S}_{k-1}\Big(\Sigma_{k-1,\tau}^{\tau'} ,\Pi_{k-1,\tau}^{\tau'} ,G_{k-1,\tau}^{\tau'} ; \Sigma_{k,\tau}^{\tau'}  - \Sigma_{k-1,\tau}^{\tau'} ,\Pi_{k,\tau}^{\tau'}  - \Pi_{k-1,\tau}^{\tau'} ,  G_{k,\tau}^{\tau'}  - G_{k-1,\tau}^{\tau'} \Big) \Big] . \label{HDEFT_Discrete} \eqa 
Here, the $k^{th}$ effective action is given by 
\bqa && \mathcal{S}_{k-1}\Big(\Sigma_{k-1,\tau}^{\tau'} ,\Pi_{k-1,\tau}^{\tau'} ,G_{k-1,\tau}^{\tau'} ; \Sigma_{k,\tau}^{\tau'}  - \Sigma_{k-1,\tau}^{\tau'} ,\Pi_{k,\tau}^{\tau'} - \Pi_{k-1,\tau}^{\tau'} ,G_{k,\tau}^{\tau'}  - G_{k-1,\tau}^{\tau'} \Big) \nn && = \Delta z \int_{0}^{\beta} d \tau \int_{0}^{\beta} d \tau' \Bigg\{ q_{d} \Gamma_{J} \mathcal{G}_{k-1,\tau'}^{\tau} \Big( \frac{\Sigma_{k,\tau}^{\tau'}  - \Sigma_{k-1,\tau}^{\tau'} }{\Delta z} \Big) + 2 q_{d} \Gamma_{J} \Pi_{k-1,\tau'}^{\tau}  \Big( \frac{G_{k,\tau}^{\tau'}  - G_{k-1,\tau}^{\tau'} }{\Delta z} \Big) G_{k-1,\tau'}^{\tau}  \nn && + q_{d} \Gamma_{J} \Big( \frac{\Pi_{k,\tau'}^{\tau}  - \Pi_{k-1,\tau'}^{\tau} }{\Delta z} \Big) G_{k-1,\tau}^{\tau'}  G_{k-1,\tau'}^{\tau}  + \frac{ q_{d} \Gamma_{J} }{2} \Big( \frac{\Pi_{k,\tau'}^{\tau}  - \Pi_{k-1,\tau'}^{\tau} }{\Delta z} \Big) \Pi_{k-1,\tau'}^{\tau}  \Bigg\} \nn && + \Delta z \int_{0}^{\beta} d \tau \int_{0}^{\beta} d \tau' \Bigg\{ (q_{d} \Gamma_{J})^{2} \int_{0}^{\beta} d \tau'' \int_{0}^{\beta} d \tau''' \Big( \frac{\Sigma_{k,\tau}^{\tau'}  - \Sigma_{k-1,\tau}^{\tau'} }{\Delta z} \Big) \mathcal{G}_{k-1,\tau'}^{\tau'''} \mathcal{G}_{k-1,\tau'''}^{\tau''} \Big( \frac{\Sigma_{k, \tau''}^{\tau} - \Sigma_{k-1, \tau''}^{\tau}}{\Delta z} \Big) \nn && - q_{d} \Gamma_{J} \Big( \frac{\Sigma_{k,\tau}^{\tau'}  - \Sigma_{k-1,\tau}^{\tau'} }{\Delta z} \Big) \Big( \frac{G_{k,\tau'}^{\tau}  - G_{k-1,\tau'}^{\tau} }{\Delta z} \Big)  + 2 q_{d} \Gamma_{J} \Big( \frac{\Pi_{k,\tau'}^{\tau}  - \Pi_{k-1,\tau'}^{\tau} }{\Delta z} \Big) \Big( \frac{G_{k,\tau}^{\tau'}  - G_{k-1,\tau}^{\tau'} }{\Delta z} \Big) G_{k-1,\tau'}^{\tau}  \nn && + q_{d} \Gamma_{J} \Pi_{k-1,\tau'}^{\tau}  \Big( \frac{G_{k,\tau}^{\tau'}  - G_{k-1,\tau}^{\tau'} }{\Delta z} \Big) \Big( \frac{G_{k,\tau'}^{\tau}  - G_{k-1,\tau'}^{\tau} }{\Delta z} \Big) + \frac{ q_{d} \Gamma_{J} }{4} \Big( \frac{\Pi_{k,\tau'}^{\tau}  - \Pi_{k-1,\tau'}^{\tau} }{\Delta z} \Big)^{2} \Bigg\} \nn && + \Delta z \int_{0}^{\beta} d \tau \int_{0}^{\beta} d \tau' \Bigg\{ - \frac{1}{2} \ln \Big( - \frac{1}{2} [\mathcal{G}_{k-1,\tau}^{\tau'}]^{-1} \Big) \Bigg\} , \eqa
reorganized as the first-order and second-order fluctuation terms as the perturbation series of Eq. (\ref{SYK_Correction}). Then, $k$ corresponds to the RG iteration step, well discussed in Refs.  \cite{Emergent_AdS2_BH_RG,RG_Flow_Holography_Monotonicity,Nonperturbative_Wilson_RG_Disorder,Nonperturbative_Wilson_RG,Einstein_Klein_Gordon_RG_Kim, Einstein_Dirac_RG_Kim,RG_GR_Geometry_I_Kim,RG_GR_Geometry_II_Kim,Kondo_Holography_Kim, Kitaev_Entanglement_Entropy_Kim,RG_Holography_First_Kim}. The Majorana fermion Green's function is given by 
\bqa && \Big( \delta(\tau-\tau'') \partial_{\tau''} + q_{d} \Gamma_{J} \Sigma_{k-1,\tau}^{\tau''}  \Big) \mathcal{G}_{k-1,\tau''}^{\tau'}= - \frac{1}{2} \delta(\tau-\tau') . \eqa

Taking $f = 1$ in the above, we obtain 
\bqa && \mathcal{Z}_{SYK}[\Sigma_{0,\tau}^{\tau'}  , \Pi_{0,\tau}^{\tau'}   , G_{0,\tau}^{\tau'}  ;  1  ] = \int D \Sigma_{1,\tau}^{\tau'}    D \Pi_{1,\tau}^{\tau'}    D G_{1,\tau}^{\tau'}    \int D \eta_{\alpha}(\tau) \nn && \exp\Big[ - \int_{0}^{\beta} d \tau \int_{0}^{\beta} d \tau' \Big\{ \eta_{\alpha}(\tau) \delta(\tau-\tau') \partial_{\tau'} \eta_{\alpha} (\tau') + q_{d} \Gamma_{J} \Sigma_{1,\tau}^{\tau'}    \Big( \eta_{\alpha}(\tau) \eta_{\alpha}(\tau') - N G_{1,\tau'}^{\tau}    \Big) \nn && \ \ \ \ \ \ \ + N q_{d} \Gamma_{J} \Pi_{0,\tau'}^{\tau}   G_{0,\tau}^{\tau'}    G_{0,\tau'}^{\tau}   + \frac{ N q_{d} \Gamma_{J} }{4} \Pi_{0,\tau'}^{\tau} \Pi_{0,\tau'}^{\tau}      \Big\} \Big] \nn && \exp\Big[ - N \mathcal{S}_{0}\Big(\Sigma_{0,\tau}^{\tau'}  ,\Pi_{0,\tau}^{\tau'}   ,G_{0,\tau}^{\tau'}   ;\Sigma_{1,\tau}^{\tau'}    - \Sigma_{0,\tau}^{\tau'}  ,\Pi_{1,\tau}^{\tau'}    - \Pi_{0,\tau}^{\tau'}   ,G_{1,\tau}^{\tau'}    - G_{0,\tau}^{\tau'}   \Big) \Big] . \eqa 
This effective field theory reproduces Eq. (\ref{HDEFT_UV_Limit}) in the previous subsection as expected. In other words, $1/N$ quantum corrections are taken into account self-consistently, performing the path integrals of $\int D \Sigma_{1,\tau}^{\tau'}    D \Pi_{1,\tau}^{\tau'}    D G_{1,\tau}^{\tau'}   $ with $\int D \eta_{\alpha}(\tau)$, as follows 
\bqa \mathcal{Z}_{SYK}[\Sigma_{0,\tau}^{\tau'}  , \Pi_{0,\tau}^{\tau'}   , G_{0,\tau}^{\tau'}  ;  1  ] =&& \exp\Big[ - N \int_{0}^{\beta} d \tau \int_{0}^{\beta} d \tau' \Big\{ - \frac{1}{2} \ln \Big(- \frac{1}{2} [\mathcal{G}_{0,\tau}^{\tau'}]^{-1} \Big)  - q_{d} \Gamma_{J} \Sigma_{0,\tau}^{\tau'}   G_{0,\tau'}^{\tau}  \nn && \ \ \ \ \ \ \  + q_{d} \Gamma_{J} \Pi_{0,\tau'}^{\tau}   G_{0,\tau}^{\tau'}    G_{0,\tau'}^{\tau}    + \frac{ q_{d} \Gamma_{J} }{4} \Pi_{0,\tau'}^{\tau} \Pi_{0,\tau'}^{\tau}      \Big\} \Big]  \nn && \exp\Big[ - \mathcal{V}_{0}\Big(\Sigma_{0,\tau}^{\tau'}  ,\Pi_{0,\tau}^{\tau'}   , G_{0,\tau}^{\tau'}   \Big) - N \Delta \mathcal{V}_{0}\Big(\Sigma_{0,\tau}^{\tau'}  ,\Pi_{0,\tau}^{\tau'}   ,G_{0,\tau}^{\tau'}   \Big) \Big] . \label{One_Loop_Partition_Ft_HDEFT}  \eqa 
Here, $\mathcal{V}_{0}\Big(\Sigma_{0,\tau}^{\tau'}  ,\Pi_{0,\tau}^{\tau'}   , G_{0,\tau}^{\tau'}   \Big)$ represents an effective potential of $\mathcal{O}(N^{0})$, given by logarithms, and $N \Delta \mathcal{V}_{0}\Big(\Sigma_{0,\tau}^{\tau'}  ,\Pi_{0,\tau}^{\tau'}   ,G_{0,\tau}^{\tau'}   \Big)$ denotes a `shift' part of $\mathcal{O}(N^{1})$, given by powers. See appendix section A. 

To introduce both $1/N$ and $1/N^{2}$ corrections, we take $f = 2$ in Eq. (\ref{HDEFT_Discrete}). Then, we obtain 

\bqa && \mathcal{Z}_{SYK}[\Sigma_{0,\tau}^{\tau'}  , \Pi_{0,\tau}^{\tau'}   , G_{0,\tau}^{\tau'}  ;  2   ] = \int D \Sigma_{1,\tau}^{\tau'}    D \Pi_{1,\tau}^{\tau'}    D G_{1,\tau}^{\tau'}    D \Sigma_{2,\tau}^{\tau'}  D \Pi_{2,\tau}^{\tau'}  D G_{2,\tau}^{\tau'}  \int D \eta_{\alpha}(\tau)    \nn &&     \exp\Big[ - \int_{0}^{\beta} d \tau \int_{0}^{\beta} d \tau' \Big\{ \eta_{\alpha}(\tau) \delta(\tau-\tau') \partial_{\tau'} \eta_{\alpha} (\tau') + q_{d} \Gamma_{J}\Sigma_{2,\tau}^{\tau'}  \Big( \eta_{\alpha}(\tau) \eta_{\alpha}(\tau')  - N G_{2,\tau'}^{\tau}  \Big) \nn && \ \ \ \ \ \ \ + N q_{d} \Gamma_{J} \Pi_{0,\tau'}^{\tau}   G_{0,\tau}^{\tau'}    G_{0,\tau'}^{\tau}   + \frac{ N q_{d} \Gamma_{J} }{4} \Pi_{0,\tau'}^{\tau} \Pi_{0,\tau'}^{\tau}      \Big\} \Big] \nn && \exp\Big[ - N \mathcal{S}_{0}\Big(\Sigma_{0,\tau}^{\tau'}  ,\Pi_{0,\tau}^{\tau'}   ,G_{0,\tau}^{\tau'}   ;\Sigma_{1,\tau}^{\tau'}    - \Sigma_{0,\tau}^{\tau'}  ,\Pi_{1,\tau}^{\tau'} - \Pi_{0,\tau}^{\tau'}   ,G_{1,\tau}^{\tau'}    - G_{0,\tau}^{\tau'}   \Big) \nn && \ \ \ \ \ \ \  - N \mathcal{S}_{1}\Big(\Sigma_{1,\tau}^{\tau'}   ,\Pi_{1,\tau}^{\tau'}   ,G_{1,\tau}^{\tau'}   ;\Sigma_{2,\tau}^{\tau'}  - \Sigma_{1,\tau}^{\tau'}   ,\Pi_{2,\tau}^{\tau'}  - \Pi_{1,\tau}^{\tau'}   ,G_{2,\tau}^{\tau'}  - G_{1,\tau}^{\tau'}   \Big) \Big] .  \eqa 

In this expression one can perform the Gaussian integrals of $\int D\Sigma_{2,\tau}^{\tau'}  D \Pi_{2,\tau}^{\tau'}  D G_{2,\tau}^{\tau'} $ to obtain 
\bqa && \mathcal{Z}_{SYK}[\Sigma_{0,\tau}^{\tau'}  , \Pi_{0,\tau}^{\tau'}   , G_{0,\tau}^{\tau'}  ;  2   ] = \int D \Sigma_{1,\tau}^{\tau'}    D \Pi_{1,\tau}^{\tau'}    D G_{1,\tau}^{\tau'}    \int D \eta_{\alpha}(\tau)    \nn &&    \exp\Big[ - \int_{0}^{\beta} d \tau \int_{0}^{\beta} d \tau' \Big\{ \eta_{\alpha}(\tau) \delta(\tau-\tau') \partial_{\tau'} \eta_{\alpha} (\tau') + q_{d} \Gamma_{J} \Sigma_{1,\tau}^{\tau'}    \Big( \eta_{\alpha}(\tau) \eta_{\alpha}(\tau') - N G_{1,\tau'}^{\tau}    \Big)  \nn && \ \ \ \ \ \ \  + N q_{d} \Gamma_{J} \Pi_{0,\tau'}^{\tau}   G_{0,\tau}^{\tau'}    G_{0,\tau'}^{\tau}    + \frac{ N q_{d} \Gamma_{J} }{4} \Pi_{0,\tau'}^{\tau} \Pi_{0,\tau'}^{\tau}      \Big\} \Big] \nn &&  \exp\Big[ - N \mathcal{S}_{0}\Big(\Sigma_{0,\tau}^{\tau'}  ,\Pi_{0,\tau}^{\tau'}   ,G_{0,\tau}^{\tau'}   ;\Sigma_{1,\tau}^{\tau'}    - \Sigma_{0,\tau}^{\tau'}  ,\Pi_{1,\tau}^{\tau'}    - \Pi_{0,\tau}^{\tau'}   ,G_{1,\tau}^{\tau'}    - G_{0,\tau}^{\tau'}   \Big) - \mathcal{V}_{1}\Big(\Sigma_{1,\tau}^{\tau'}   ,\Pi_{1,\tau}^{\tau'}   ,G_{1,\tau}^{\tau'}   \Big)  \nn && \ \ \ \ \ \ \  - N \Delta \mathcal{V}_{1}\Big(\Sigma_{1,\tau}^{\tau'}   ,\Pi_{1,\tau}^{\tau'}   ,G_{1,\tau}^{\tau'}   \Big) \Big] . \eqa 
Naively, the effective potential of $\mathcal{V}_{1}\Big(\Sigma_{1,\tau}^{\tau'}   ,\Pi_{1,\tau}^{\tau'}   ,G_{1,\tau}^{\tau'}   \Big)$ is the order of $\mathcal{O}(N^{0})$, while the shift part of $N \Delta \mathcal{V}_{1}\Big(\Sigma_{1,\tau}^{\tau'}   ,\Pi_{1,\tau}^{\tau'}   ,G_{1,\tau}^{\tau'}   \Big)$ is the order of $\mathcal{O}(N^{1})$. We discuss this issue below in more details.

Unfortunately, one cannot take the path integrals of $\int D \Sigma_{1,\tau}^{\tau'}    D \Pi_{1,\tau}^{\tau'}    D G_{1,\tau}^{\tau'}   $ further. The effective potential $\mathcal{V}_{1}\Big(\Sigma_{1,\tau}^{\tau'}   ,\Pi_{1,\tau}^{\tau'}   ,G_{1,\tau}^{\tau'}   \Big)$ is quite complicated not to allow the path integrals of $\int D \Sigma_{1,\tau}^{\tau'}    D \Pi_{1,\tau}^{\tau'}    D G_{1,\tau}^{\tau'}   $. In this respect the best way would be to consider the saddle-point approximation for these functions. Shifting $\Sigma_{1,\tau}^{\tau'}   $, $\Pi_{1,\tau}^{\tau'}   $, and $G_{1,\tau}^{\tau'}   $ as $\Sigma_{1,\tau}^{\tau'}    + \Sigma_{0,\tau}^{\tau'}  $, $\Pi_{1,\tau}^{\tau'}    + \Pi_{0,\tau}^{\tau'}   $, and $G_{1,\tau}^{\tau'}    + G_{0,\tau}^{\tau'}   $, respectively, and taking the path integral of $\int D \eta_{\alpha}(\tau)$, we obtain  
\bqa && \mathcal{Z}_{SYK}[\Sigma_{0,\tau}^{\tau'}  , \Pi_{0,\tau}^{\tau'}   , G_{0,\tau}^{\tau'}   ; \Sigma_{1,\tau}^{\tau'}   , \Pi_{1,\tau}^{\tau'}   , G_{1,\tau}^{\tau'}   ;   2    ] \nn && = \exp\Big[ - N \int_{0}^{\beta} d \tau \int_{0}^{\beta} d \tau' \Big\{ - \frac{1}{2} \ln \Big(- \frac{1}{2} [\mathcal{G}_{0,\tau}^{\tau'}]^{-1} \Big)  - q_{d} \Gamma_{J} \Sigma_{0,\tau}^{\tau'}   G_{0,\tau'}^{\tau}    + q_{d} \Gamma_{J} \Pi_{0,\tau'}^{\tau}   G_{0,\tau}^{\tau'}    G_{0,\tau'}^{\tau} \nn &&  \ \ \ \ \ \ \ \ \ \ + \frac{ q_{d} \Gamma_{J} }{4} \Pi_{0,\tau'}^{\tau} \Pi_{0,\tau'}^{\tau}      - q_{d} \Gamma_{J} \Sigma_{0,\tau}^{\tau'}   G_{1,\tau'}^{\tau}    - q_{d} \Gamma_{J} \Sigma_{1,\tau}^{\tau'}    \Big( G_{1,\tau'}^{\tau}    + G_{0,\tau'}^{\tau}    \Big) \Big\} \Big] \nn && \ \ \ \exp\Big[ - N \mathcal{S}_{0}\Big(\Sigma_{0,\tau}^{\tau'}  ,\Pi_{0,\tau}^{\tau'}   ,G_{0,\tau}^{\tau'}   ;\Sigma_{1,\tau}^{\tau'}   ,\Pi_{1,\tau}^{\tau'}   ,G_{1,\tau}^{\tau'}   \Big)  - \mathcal{V}_{1}\Big(\Sigma_{1,\tau}^{\tau'}    + \Sigma_{0,\tau}^{\tau'}  ,\Pi_{1,\tau}^{\tau'}    + \Pi_{0,\tau}^{\tau'}   ,G_{1,\tau}^{\tau'}    + G_{0,\tau}^{\tau'}   \Big) \nn && \ \ \ \ \ \ \ \ \ \ - N \Delta \mathcal{V}_{1}\Big(\Sigma_{1,\tau}^{\tau'}    + \Sigma_{0,\tau}^{\tau'}  ,\Pi_{1,\tau}^{\tau'}    + \Pi_{0,\tau}^{\tau'}   ,G_{1,\tau}^{\tau'}    + G_{0,\tau}^{\tau'}   \Big) \Big] . \eqa 
Here, we recall that this expression reproduces Eq. (\ref{One_Loop_Partition_Ft_HDEFT}), neglecting $\mathcal{V}_{1}\Big(\Sigma_{1,\tau}^{\tau'}   ,\Pi_{1,\tau}^{\tau'}   ,G_{1,\tau}^{\tau'}   \Big)$ and $N \Delta \mathcal{V}_{1}\Big(\Sigma_{1,\tau}^{\tau'}   ,\Pi_{1,\tau}^{\tau'}   ,G_{1,\tau}^{\tau'}   \Big)$ and performing the gaussian path integrals of $\int D \Sigma_{1,\tau}^{\tau'}    D \Pi_{1,\tau}^{\tau'}    D G_{1,\tau}^{\tau'}   $ in this case. 

We point out that this mathematical structure is typical in the large $N$ expansion. Here, $\Sigma_{0,\tau}^{\tau'}  $, $\Pi_{0,\tau}^{\tau'}   $, and $G_{0,\tau}^{\tau'}   $ are the order of $\mathcal{O}(N^{0})$ while $\Sigma_{1,\tau}^{\tau'}   $, $\Pi_{1,\tau}^{\tau'}   $, and $G_{1,\tau}^{\tau'}   $ are the order of $\mathcal{O}(N^{-1})$. As a result, $\mathcal{V}_{1}\Big(\Sigma_{1,\tau}^{\tau'}   ,\Pi_{1,\tau}^{\tau'}   ,G_{1,\tau}^{\tau'}   \Big)$ is the order of $\mathcal{O}(N^{-2})$. Actually, we showed our explicit demonstration of such higher-order quantum corrections in ref. \cite{Kondo_Holography_Kim}. Here, we derived an effective holographic dual field theory for the Kondo effect, applying the Wilsonian RG transformation to the Kondo problem. Interestingly, it turns out that the holographic dual effective field theory resums $1/N$, $1/N^{2}$, ... perturbative corrections of the quantum field theory in a large $N$ fashion on one higher dimensional spacetime. As a result, IR divergences are cured by such resummation in the RG-based emergent holographic dual effective field theory. Indeed, the impurity thermodynamics show smooth crossover behaviors across the Kondo temperature. Actually, both the impurity specific heat and spin susceptibility were calculated to show reasonable match with the thermodynamic Bethe ansatz results. In spite of this qualitative match between the holographic dual effective field theory and the thermodynamic Bethe ansatz for the Kondo effect, one should not take the resummation of higher order Feynman diagrams in the holographic dual effective field theory as an exact way. Although quantum fluctuations are resummed in the all loop order certainly, only the partial class of Feynman diagrams are resummed in the holographic dual effective field theory based on the Wilsonian RG transformation. We believe that this aspect has to be investigated more rigorously.

\subsection{Equations of motion}  

Now, we perform the saddle-point analysis in the large $N$ limit. We emphasize that this large $N$ analysis in the holographic dual effective field theory should be distinguished from the field theoretic large $N$ limit in the SYK model since all order expansions are taken into account and resummed through the Wilsonian RG transformation in a nonperturbative way  \cite{Emergent_AdS2_BH_RG,RG_Flow_Holography_Monotonicity,Nonperturbative_Wilson_RG_Disorder,Nonperturbative_Wilson_RG,Einstein_Klein_Gordon_RG_Kim,Einstein_Dirac_RG_Kim,RG_GR_Geometry_I_Kim,RG_GR_Geometry_II_Kim,Kondo_Holography_Kim, Kitaev_Entanglement_Entropy_Kim,RG_Holography_First_Kim}, as discussed in the previous subsection. 

We clean up both the boundary action and the bulk effective action in Eq. (\ref{HDEFT}) to obtain 
\bqa && \mathcal{Z}_{SYK}(z_{f}) = \int D \Sigma_{\tau}^{\tau'}(z)  D \Pi_{\tau}^{\tau'}(z)  DG_{\tau}^{\tau'}(z)  \int D \eta_{\alpha}(\tau) \exp\Bigg[ - \int_{0}^{\beta} d \tau \int_{0}^{\beta} d \tau' \Bigg\{ \eta_{\alpha}(\tau) \delta(\tau-\tau') \partial_{\tau'} \eta_{\alpha} (\tau') \nn && + q_{d} \Gamma_{J} \Sigma_{\tau}^{\tau'}(z_{f})     \Big( \eta_{\alpha}(\tau) \eta_{\alpha}(\tau') - N G_{\tau'}^{\tau}(z_{f})  \Big) + N q_{d} \Gamma_{J} \Pi_{\tau'}^{\tau}(z_{f})    G_{\tau}^{\tau'}(z_{f})   G_{\tau'}^{\tau}(z_{f})  + \frac{ N q_{d} \Gamma_{J} }{4} \Pi_{\tau'}^{\tau}(z_{f}) \Pi_{\tau'}^{\tau}(z_{f})      \Bigg\} \Bigg] \nn && \exp\Bigg[ - N \int_{0}^{z_{f}} d z \int_{0}^{\beta} d \tau \int_{0}^{\beta} d \tau' \Bigg\{ q_{d} \Gamma_{J} \mathcal{G}_{\tau'}^{\tau}(z) \Big(\partial_{z} \Sigma_{\tau}^{\tau'}(z) \Big) - q_{d} \Gamma_{J} \Big(\partial_{z} \Sigma_{\tau}^{\tau'}(z) \Big) \Big(\partial_{z} G_{\tau'}^{\tau}(z)  \Big) \nn && \ \ \ \ \ \ \  + (q_{d} \Gamma_{J})^{2} \int_{0}^{\beta} d \tau'' \int_{0}^{\beta} d \tau''' \Big(\partial_{z} \Sigma_{\tau}^{\tau'}(z) \Big) \mathcal{G}_{\tau'}^{\tau'''}(z) \mathcal{G}_{\tau'''}^{\tau''}(z) \Big(\partial_{z} \Sigma_{\tau''}^{\tau}(z)  \Big) \nn && \ \ \ \ \ \ \  + q_{d} \Gamma_{J} \Big(  \Pi_{\tau'}^{\tau}(z)   - 4G_{\tau}^{\tau'}(z)  G_{\tau'}^{\tau}(z)   \Big) \Big(\partial_{z}G_{\tau}^{\tau'}(z) \Big) \Big(\partial_{z} G_{\tau'}^{\tau}(z)  \Big) \nn && \ \ \ \ \ \ \  + \frac{ q_{d} \Gamma_{J} }{4} \Big(\partial_{z}  \Pi_{\tau'}^{\tau}(z)   + 4 \Big(\partial_{z}G_{\tau}^{\tau'}(z) \Big) G_{\tau'}^{\tau}(z)  \Big)^{2} - \frac{1}{2} \ln \Big( - \frac{1}{2} [\mathcal{G}_{\tau}^{\tau'}(z)]^{-1} \Big) \Bigg\} \Bigg] . \eqa 
We point out that the boundary action is written in terms of all IR degrees of freedom, i.e., $z = z_{f}$, where UV terms are cancelled out by the linear $z-$derivative terms of the bulk effective action through the integration by part.

To consider the Hamiltonian formulation in the next subsection, we shift the $ \Pi_{\tau'}^{\tau}(z)  $ field into
\bqa &&  \Pi_{\tau'}^{\tau}(z)   \longrightarrow  \Pi_{\tau'}^{\tau}(z)   - 2G_{\tau}^{\tau'}(z)  G_{\tau'}^{\tau}(z)   . \eqa

As a result, the above expression of the partition function reads 
\bqa && \mathcal{Z}_{SYK}(z_{f}) = \int D \Sigma_{\tau}^{\tau'}(z)  D \Pi_{\tau}^{\tau'}(z)  DG_{\tau}^{\tau'}(z)  \exp\Bigg[ - N \int_{0}^{\beta} d \tau \int_{0}^{\beta} d \tau' \Bigg\{ - \frac{1}{2} \ln \Big( - \frac{1}{2} [\mathcal{G}_{\tau}^{\tau'}(z_{f})]^{-1} \Big) \nn && - q_{d} \Gamma_{J} \Sigma_{\tau}^{\tau'}(z_{f})     G_{\tau'}^{\tau}(z_{f})  - q_{d} \Gamma_{J} \Big( G_{\tau}^{\tau'}(z_{f})   G_{\tau'}^{\tau}(z_{f})  \Big)^{2} + \frac{ q_{d} \Gamma_{J} }{4} \Pi_{\tau'}^{\tau}(z_{f}) \Pi_{\tau'}^{\tau}(z_{f})      \Bigg\} \Bigg] \nn && \exp\Bigg[ - N \int_{0}^{z_{f}} d z \int_{0}^{\beta} d \tau \int_{0}^{\beta} d \tau' \Bigg\{ q_{d} \Gamma_{J} \mathcal{G}_{\tau'}^{\tau}(z) \Big(\partial_{z} \Sigma_{\tau}^{\tau'}(z) \Big) - q_{d} \Gamma_{J} \Big(\partial_{z} \Sigma_{\tau}^{\tau'}(z) \Big) \Big(\partial_{z} G_{\tau'}^{\tau}(z)  \Big) \nn && \ \ \ \ \ \ \  + (q_{d} \Gamma_{J})^{2} \int_{0}^{\beta} d \tau'' \int_{0}^{\beta} d \tau''' \Big(\partial_{z} \Sigma_{\tau}^{\tau'}(z) \Big) \mathcal{G}_{\tau'}^{\tau'''}(z) \mathcal{G}_{\tau'''}^{\tau''}(z) \Big(\partial_{z} \Sigma_{\tau''}^{\tau}(z)  \Big) \nn && \ \ \ \ \ \ \  + q_{d} \Gamma_{J} \Big(  \Pi_{\tau'}^{\tau}(z)   - 6G_{\tau}^{\tau'}(z)  G_{\tau'}^{\tau}(z)   \Big) \Big(\partial_{z}G_{\tau}^{\tau'}(z) \Big) \Big(\partial_{z} G_{\tau'}^{\tau}(z)  \Big) + \frac{ q_{d} \Gamma_{J} }{4} \Big(\partial_{z}  \Pi_{\tau'}^{\tau}(z)   \Big)^{2}\nn && \ \ \ \ \ \ \   - \frac{1}{2} \ln \Big( - \frac{1}{2} [\mathcal{G}_{\tau}^{\tau'}(z)]^{-1} \Big) \Bigg\} \Bigg] , \label{HDEFT_Shifted_Lagrangian} \eqa 
where $\int D \eta_{\alpha}(\tau)$ has been performed.

Taking variations of the bulk effective action with respect to $\Sigma_{\tau}^{\tau'}(z) $, $G_{\tau}^{\tau'}(z)  $, and $ \Pi_{\tau'}^{\tau}(z)  $, we obtain Lagrange equations of motion,
\bqa && 2 q_{d} \Gamma_{J} \int_{0}^{\beta} d \tau_1 \int_{0}^{\beta} d \tau_2 \mathcal{G}_{\tau}^{\tau_1}(z) \mathcal{G}_{\tau_1}^{\tau_2}(z) \Big(\partial_{z}^{2} \Sigma_{\tau_2}^{\tau'}(z)  \Big) \nn && + 4 (q_{d} \Gamma_{J})^{2} \int_{0}^{\beta} d \tau_1 \int_{0}^{\beta} d \tau_2 \int_{0}^{\beta} d \tau_3 \int_{0}^{\beta} d \tau_4 \Big(\partial_{z} \Sigma_{\tau}^{\tau_1}(z)  \Big) \mathcal{G}_{\tau_1}^{\tau_2}(z) \mathcal{G}_{\tau_2}^{\tau_3}(z) \mathcal{G}_{\tau_3}^{\tau_4}(z) \Big(\partial_{z} \Sigma_{\tau_4}^{\tau'}(z)  \Big) \nn && =  \mathcal{G}_{\tau}^{\tau'}(z) + \partial_{z}^{2}G_{\tau}^{\tau'}(z)  , \label{Eq_Sigma} \eqa 
\bqa && \Big(  \Pi_{\tau'}^{\tau}(z)   - 6G_{\tau}^{\tau'}(z)  G_{\tau'}^{\tau}(z)   \Big) \Big(\partial_{z}^{2}G_{\tau}^{\tau'}(z) \Big)   + \Big( \partial_{z}  \Pi_{\tau'}^{\tau}(z)   - 6G_{\tau}^{\tau'}(z)  \partial_{z} G_{\tau'}^{\tau}(z)   \Big) \Big(\partial_{z}G_{\tau}^{\tau'}(z) \Big)  \nn && = \frac{1}{2} \Big(\partial_{z}^{2} \Sigma_{\tau}^{\tau'}(z) \Big) , \label{Eq_G} \eqa
and
\bqa && \partial_{z}^{2}  \Pi_{\tau'}^{\tau}(z)   = 2 \Big(\partial_{z}G_{\tau}^{\tau'}(z) \Big) \Big(\partial_{z} G_{\tau'}^{\tau}(z)  \Big), \label{Eq_Pi} \eqa 
respectively. To figure out these coupled differential equations, one may neglect $\partial_{z}$ in these differential equations. Then, it is easy to realize that the resulting coupled algebraic equations are similar to those of the Luttinger-Ward functional approach for the SYK model.

\subsection{Boundary conditions} 

To solve the three nonlinearly coupled second-order differential equations (\ref{Eq_Sigma}), (\ref{Eq_G}), and (\ref{Eq_Pi}), we need three types of UV and IR boundary conditions. These boundary conditions are given by an effective  boundary action. To determine the effective  boundary action, we introduce the Hamiltonian formulation as 
\bqa && \mathcal{Z}_{SYK}(z_{f}) = \int D \Sigma_{\tau}^{\tau'}(z)  D \Pi_{\Sigma,\tau}^{\tau'}(z)   D \Pi_{\tau}^{\tau'}(z)  D \Pi_{\Pi,\tau}^{\tau'}(z)   DG_{\tau}^{\tau'}(z)  D \Pi_{G,\tau}^{\tau'}(z)   \nn && \bar{\mathcal{Z}}_{SYK}[\Sigma_{\tau}^{\tau'}(z_{f})    ,\Pi_{\tau}^{\tau'}(z_{f})  ,G_{\tau}^{\tau'}(z_{f})  ] \exp\Bigg[ - N q_{d} \Gamma_{J} \int_{0}^{z_{f}} d z \int_{0}^{\beta} d \tau \int_{0}^{\beta} d \tau' \Bigg\{ \Big( \Pi_{\Sigma,\tau'}^{\tau}(z)   + \mathcal{G}_{\tau'}^{\tau}(z)\Big) \Big(\partial_{z} \Sigma_{\tau}^{\tau'}(z) \Big) \nn && - \frac{1}{4} \int_{0}^{\beta} d \tau'' \Pi_{\Sigma,\tau}^{\tau'}(z)   \Bigg( q_{d} \Gamma_{J} \int_{0}^{\beta} d \tau''' \mathcal{G}_{\tau'}^{\tau'''}(z) \mathcal{G}_{\tau'''}^{\tau''}(z) - \frac{1}{4} \Big( \Pi_{\tau'}^{\tau''}(z)   - 6G_{\tau''}^{\tau'}(z)   G_{\tau'}^{\tau''}(z)   \Big)^{-1} \Bigg)^{-1} \Pi_{\Sigma,\tau''}^{\tau}(z)   \nn && + \Pi_{G,\tau'}^{\tau}(z)   \Big\{\frac{1}{2}\partial_{z} \Sigma_{\tau}^{\tau'}(z)  -  \Big(  \Pi_{\tau'}^{\tau}(z)   - 6G_{\tau}^{\tau'}(z)  G_{\tau'}^{\tau}(z)   \Big) \partial_{z}G_{\tau}^{\tau'}(z) \Big\}  - \frac{1}{4} \Pi_{G,\tau}^{\tau'}(z)   \Big(  \Pi_{\tau'}^{\tau}(z)   - 6G_{\tau}^{\tau'}(z)  G_{\tau'}^{\tau}(z)   \Big) \Pi_{G,\tau'}^{\tau}(z)  \nn && + \Pi_{\Pi,\tau}^{\tau'}(z)   \Big(\partial_{z}  \Pi_{\tau'}^{\tau}(z)   \Big) - [\Pi_{\Pi,\tau}^{\tau'}(z)   ]^{2}  - \frac{1}{2 q_{d} \Gamma_{J}} \ln \Big( - \frac{1}{2} [\mathcal{G}_{\tau}^{\tau'}(z)]^{-1} \Big) \Bigg\} \Bigg] , \eqa 
where the IR boundary partition function is given by
\bqa && \bar{\mathcal{Z}}_{SYK}[\Sigma_{\tau}^{\tau'}(z_{f})    ,\Pi_{\tau}^{\tau'}(z_{f})  ,G_{\tau}^{\tau'}(z_{f})  ] = \exp\Bigg[ - N \int_{0}^{\beta} d \tau \int_{0}^{\beta} d \tau' \Bigg\{ - \frac{1}{2} \ln \Big( - \frac{1}{2} [\mathcal{G}_{\tau}^{\tau'}(z_{f})]^{-1} \Big) \nn && - q_{d} \Gamma_{J} \Sigma_{\tau}^{\tau'}(z_{f})     G_{\tau'}^{\tau}(z_{f})  - q_{d} \Gamma_{J} \Big( G_{\tau}^{\tau'}(z_{f})   G_{\tau'}^{\tau}(z_{f})  \Big)^{2} + \frac{ q_{d} \Gamma_{J} }{4} \Pi_{\tau'}^{\tau}(z_{f}) \Pi_{\tau'}^{\tau}(z_{f})      \Bigg\} \Bigg] . \eqa
Here, $\Pi_{\Sigma,\tau}^{\tau'}(z)  $, $\Pi_{\Pi,\tau}^{\tau'}(z)  $, and $\Pi_{G,\tau}^{\tau'}(z)  $ are the momentum of $\Sigma_{\tau}^{\tau'}(z) $, $\Pi_{\tau}^{\tau'}(z) $, and $G_{\tau}^{\tau'}(z)  $, respectively. 
These momenta are given by one sector of the Hamilton equation of motion, 
\bqa && \Pi_{\Sigma,\tau}^{\tau'}(z)   = 2 \int_{0}^{\beta} d \tau''' \Bigg( q_{d} \Gamma_{J} \int_{0}^{\beta} d \tau'' \mathcal{G}_{\tau}^{\tau''}(z) \mathcal{G}_{\tau''}^{\tau'''}(z)  - \frac{1}{4} \Big( \Pi_{\tau}^{\tau'''}(z)   - 6 G_{\tau'''}^{\tau}(z)   G_{\tau}^{\tau'''}(z)   \Big)^{-1} \Bigg) \partial_{z} \Sigma_{\tau'''}^{\tau'}(z)  , 
%
%
\label{Eq_Pi_Sigma} \eqa 
\bqa && \Pi_{G,\tau}^{\tau'}(z)   = - 2 \partial_{z}G_{\tau}^{\tau'}(z)  +  \int_{0}^{\beta} d \tau'' \Big( \Pi_{\tau}^{\tau''}(z)  - 6 G_{\tau}^{\tau''}(z)   G_{\tau''}^{\tau}(z)   \Big)^{-1} \Big(\partial_{z} \Sigma_{\tau''}^{\tau'}(z)  \Big) , \label{Eq_Pi_G} \eqa 
and 
\bqa && \Pi_{\Pi,\tau}^{\tau'}(z)   = \frac{1}{2} \partial_{z} \Pi_{\tau}^{\tau'}(z)  , \label{Eq_Pi_Pi} \eqa 
respectively, which will be utilized to determine the boundary conditions below. Performing the Gaussian integrals for these momenta, we reproduce the Lagrangian formulation, Eq. (\ref{HDEFT_Shifted_Lagrangian}).

The above Hamiltonian formulation gives rise to the effective boundary action, 
\bqa && \mathcal{S}_{eff} = \mathcal{S}_{IR}^{Bdry} + \mathcal{S}_{IR}^{Bulk} + \mathcal{S}_{UV}^{Bulk} , \label{BC_EFT} \eqa
where 
\bqa && \mathcal{S}_{IR}^{Bdry} = N q_{d} \Gamma_{J} \int_{0}^{\beta} d \tau \int_{0}^{\beta} d \tau' \Bigg\{ - \frac{1}{2} \ln \Big( - \frac{1}{2 q_{d} \Gamma_{J}} [\mathcal{G}_{\tau}^{\tau'}(z_{f})]^{-1} \Big) - \Sigma_{\tau}^{\tau'}(z_{f})     G_{\tau'}^{\tau}(z_{f})  \nn && \ \ \ \ \ \ \ \ \ \ - \Big( G_{\tau}^{\tau'}(z_{f})   G_{\tau'}^{\tau}(z_{f})  \Big)^{2} + \frac{ 1 }{4} \Pi_{\tau'}^{\tau}(z_{f}) \Pi_{\tau'}^{\tau}(z_{f})      \Bigg\} , \\ && \mathcal{S}_{IR}^{Bulk} = N q_{d} \Gamma_{J} \int_{0}^{\beta} d \tau \int_{0}^{\beta} d \tau' \Bigg\{ \Big( \Pi_{\Sigma,\tau'}^{\tau}(z_{f})   + \frac{1}{2}\Pi_{G,\tau'}^{\tau}(z_{f})   + \mathcal{G}_{\tau'}^{\tau}(z_{f}) \Big) \Sigma_{\tau}^{\tau'}(z_{f})     \nn && \ \ \ \ \ \ \ \ \ \ -  \Big( \Pi_{\tau'}^{\tau}(z_{f})    - 6 G_{\tau}^{\tau'}(z_{f})   G_{\tau'}^{\tau}(z_{f})  \Big) \Pi_{G,\tau'}^{\tau}(z_{f})   G_{\tau}^{\tau'}(z_{f})   + \Pi_{\Pi,\tau}^{\tau'}(z_{f})   \Pi_{\tau'}^{\tau}(z_{f})    \Bigg\} , \\ && \mathcal{S}_{UV}^{Bulk} = N q_{d} \Gamma_{J} \int_{0}^{\beta} d \tau \int_{0}^{\beta} d \tau' \Bigg\{ - \Big( \Pi_{\Sigma,\tau'}^{\tau}(0)   +\frac{1}{2} \Pi_{G,\tau'}^{\tau}(0)   + \mathcal{G}_{\tau'}^{\tau}(0) \Big) \Sigma_{\tau}^{\tau'}(0)   \nn && \ \ \ \ \ \ \ \ \ \ + \Big( \Pi_{\tau'}^{\tau}(0)  - 6 G_{\tau}^{\tau'}(0)  G_{\tau'}^{\tau}(0)  \Big) \Pi_{G,\tau'}^{\tau}(0)   G_{\tau}^{\tau'}(0)  - \Pi_{\Pi,\tau}^{\tau'}(0)   \Pi_{\tau'}^{\tau}(0)  \Bigg\} . \eqa 
Here, $\mathcal{S}_{IR}^{Bulk} + \mathcal{S}_{UV}^{Bulk}$ comes from the $z-$linear derivative terms of the bulk effective action through the integration by part, where most bulk terms do not contribute, resulting from the equations of motion.

Taking variations of the boundary effective action with respect to $\Sigma_{\tau}^{\tau'}(z_{f})    $, $G_{\tau}^{\tau'}(z_{f})  $, and $\Pi_{\tau}^{\tau'}(z_{f})  $, we obtain 
\bqa && \Pi_{\Sigma,\tau}^{\tau'}(z_{f})   +\frac{1}{2} \Pi_{G,\tau}^{\tau'}(z_{f})   + \mathcal{G}_{\tau}^{\tau'}(z_{f}) + 2 q_{d} \Gamma_{J} \int_{0}^{\beta} d \tau'' \mathcal{G}_{\tau'}^{\tau''}(z_{f}) \mathcal{G}_{\tau''}^{\tau'}(z_{f}) \Sigma_{\tau'}^{\tau}(z_{f})   \nn && = G_{\tau}^{\tau'}(z_{f})   + \mathcal{G}_{\tau}^{\tau'}(z_{f}) , \eqa 
\bqa && - \Big( \Pi_{\tau'}^{\tau}(z_{f})    - 18 G_{\tau}^{\tau'}(z_{f})   G_{\tau'}^{\tau}(z_{f})  \Big) \Pi_{G,\tau}^{\tau'}(z_{f})   = \Sigma_{\tau}^{\tau'}(z_{f})     + 4 G_{\tau}^{\tau'}(z_{f})   [G_{\tau'}^{\tau}(z_{f}) ]^{2} , \eqa 
and
\bqa && \Pi_{\Pi,\tau}^{\tau'}(z_{f})   = - \frac{1}{2} \Pi_{\tau}^{\tau'}(z_{f})   + \frac{1}{2} \Pi_{G,\tau}^{\tau'}(z_{f})   G_{\tau'}^{\tau}(z_{f})  , \eqa 
respectively. Recalling Eqs. (\ref{Eq_Pi_Sigma}), (\ref{Eq_Pi_G}), and (\ref{Eq_Pi_Pi}), we determine three types of the IR boundary conditions as follows 
\bqa && 2 q_{d} \Gamma_{J} \int_{0}^{\beta} d \tau_1 \int_{0}^{\beta} d \tau_2 \int_{0}^{\beta} d \tau_3 \Big( \Pi_{\tau}^{\tau_1}(z_{f})   - 6 G_{\tau}^{\tau_1}(z_{f})   G_{\tau_1}^{\tau}(z_{f})   \Big) \mathcal{G}_{\tau_1}^{\tau_2}(z_{f}) \mathcal{G}_{\tau_2}^{\tau_3}(z_{f}) \Big(\partial_{z_{f}} \Sigma_{\tau_3}^{\tau'}(z_{f})  \Big) \nn && - \int_{0}^{\beta} d \tau_1 \Big( \Pi_{\tau}^{\tau_1}(z_{f})   - 6 G_{\tau}^{\tau_1}(z_{f})   G_{\tau_1}^{\tau}(z_{f})   \Big) \Bigg( \partial_{z_{f}} G_{\tau_1}^{\tau'}(z_{f})   - 2 q_{d} \Gamma_{J} \int_{0}^{\beta} d \tau_2 \int_{0}^{\beta} d \tau_3 \mathcal{G}_{\tau_1}^{\tau_2}(z_{f}) \mathcal{G}_{\tau_2}^{\tau_3}(z_{f}) \Sigma_{\tau_3}^{\tau'}(z_{f})   \Bigg) \nn && = \int_{0}^{\beta} d \tau_1 \Big( \Pi_{\tau}^{\tau_1}(z_{f})   - 6 G_{\tau}^{\tau_1}(z_{f})   G_{\tau_1}^{\tau}(z_{f})   \Big) G_{\tau_1}^{\tau'}(z_{f})   , \label {IR_BC_Sigma} \eqa 
\bqa && 2 \Big( \Pi_{\tau'}^{\tau}(z_{f})    - 18 G_{\tau}^{\tau'}(z_{f})   G_{\tau'}^{\tau}(z_{f})  \Big) \Big( \partial_{z_{f}} G_{\tau}^{\tau'}(z_{f})   \Big) \nn && -\frac{1}{2} \Big( \Pi_{\tau'}^{\tau}(z_{f})    - 18 G_{\tau}^{\tau'}(z_{f})   G_{\tau'}^{\tau}(z_{f})  \Big) \int_{0}^{\beta} d \tau'' \Big( \Pi_{\tau}^{\tau''}(z_{f})   - 6 G_{\tau}^{\tau''}(z_{f})   G_{\tau''}^{\tau}(z_{f})   \Big)^{-1} \Big(\partial_{z_{f}} \Sigma_{\tau''}^{\tau'}(z_{f})  \Big) \nn && = \Sigma_{\tau}^{\tau'}(z_{f})     + 4 G_{\tau}^{\tau'}(z_{f})   [G_{\tau'}^{\tau}(z_{f}) ]^{2} , \label {IR_BC_G} \eqa 
and
\bqa && \int_{0}^{\beta} d \tau'' \Big( \Pi_{\tau}^{\tau''}(z_{f})   - 6 G_{\tau}^{\tau''}(z_{f})   G_{\tau''}^{\tau}(z_{f})   \Big) \Bigg( \partial_{z_{f}} \Pi_{\tau''}^{\tau'}(z_{f})   + \Pi_{\tau''}^{\tau'}(z_{f})   +2 G_{\tau''}^{\tau'}(z_{f})   \partial_{z_{f}} G_{\tau'}^{\tau''}(z_{f})   \Bigg) \nn && =  G_{\tau}^{\tau'}(z_{f})   \Big(\partial_{z_{f}} \Sigma_{\tau'}^{\tau}(z_{f})  \Big) . \label {IR_BC_Pi} \eqa 

Taking variations of the boundary effective action with respect to $\Sigma_{\tau}^{\tau'}(0)  $, $G_{\tau}^{\tau'}(0) $, and $\Pi_{\tau}^{\tau'}(0)  $, we obtain 
\bqa && \Pi_{\Sigma,\tau}^{\tau'}(0)    + \frac{1}{2}\Pi_{G,\tau}^{\tau'}(0)  + \mathcal{G}_{\tau}^{\tau'}(0) + 2 q_{d} \Gamma_{J} \int_{0}^{\beta} d \tau_1 \int_{0}^{\beta} d \tau_2 \mathcal{G}_{\tau}^{\tau_1}(0) \mathcal{G}_{\tau_1}^{\tau_2}(0) \Sigma_{\tau_2}^{\tau'}(0)   = 0 , \\ && \Big( \Pi_{\tau'}^{\tau}(0)  - 18 G_{\tau}^{\tau'}(0)  G_{\tau'}^{\tau}(0)  \Big) \Pi_{G,\tau}^{\tau'}(0)  = 0 , \\ && \Pi_{\Pi,\tau}^{\tau'}(0)   = \frac{1}{2} \Pi_{G,\tau'}^{\tau}(0)   G_{\tau}^{\tau'}(0)  . \eqa 
Considering $\Pi_{\tau'}^{\tau}(0)  \not= 18 G_{\tau}^{\tau'}(0)  G_{\tau'}^{\tau}(0) $, we obtain
\bqa && \Pi_{\Sigma,\tau}^{\tau'}(0)    + \mathcal{G}_{\tau}^{\tau'}(0) + 2 q_{d} \Gamma_{J} \int_{0}^{\beta} d \tau_1 \int_{0}^{\beta} d \tau_2 \mathcal{G}_{\tau}^{\tau_1}(0) \mathcal{G}_{\tau_1}^{\tau_2}(0) \Sigma_{\tau_2}^{\tau'}(0)   = 0 , \\ && \Pi_{G,\tau}^{\tau'}(0)  = 0 , \\ && \Pi_{\Pi,\tau}^{\tau'}(0)   = 0 . \eqa
Similarly to the case of the IR boundary conditions, we obtain three UV boundary conditions,
\bqa && 2 q_{d} \Gamma_{J} \int_{0}^{\beta} d \tau_1 \int_{0}^{\beta} d \tau_2 \int_{0}^{\beta} d \tau_3 \Big( \Pi_{\tau}^{\tau_1}(0)   - 6 G_{\tau_1}^{\tau}(0)   G_{\tau}^{\tau_1}(0)   \Big) \mathcal{G}_{\tau_1}^{\tau_2}(0) \mathcal{G}_{\tau_2}^{\tau_3}(0) [\partial_{z} \Sigma_{\tau_3}^{\tau'}(z)  ]_{z = 0} \nn && + \int_{0}^{\beta} d \tau_1 \Big( \Pi_{\tau}^{\tau_1}(0)   - 6 G_{\tau_1}^{\tau}(0)   G_{\tau}^{\tau_1}(0)   \Big) \Bigg( \mathcal{G}_{\tau_1}^{\tau'}(0)  + 2 q_{d} \Gamma_{J} \int_{0}^{\beta} d \tau_2 \int_{0}^{\beta} d \tau_3 \mathcal{G}_{\tau_1}^{\tau_2}(0) \mathcal{G}_{\tau_2}^{\tau_3}(0) \Sigma_{\tau_3}^{\tau'}(0)   \Bigg) \nn && = \frac{1}{2} [\partial_{z} \Sigma_{\tau}^{\tau'}(z) ]_{z = 0} , \label {UV_BC_Sigma} \eqa 
\bqa && 2\int_{0}^{\beta} d \tau'' \Big( \Pi_{\tau}^{\tau''}(0)   - 6 G_{\tau}^{\tau''}(0)   G_{\tau''}^{\tau}(0)   \Big) [\partial_{z}G_{\tau''}^{\tau'}(z)  ]_{z = 0} =  [\partial_{z} \Sigma_{\tau}^{\tau'}(z) ]_{z = 0} , \label {UV_BC_G} \eqa 
and
\bqa && [\partial_{z} \Pi_{\tau}^{\tau'}(z) ]_{z = 0} = 0 , \label {UV_BC_Pi} \eqa 
respectively.

\section{Conformal ansatz}

\subsection{Conformal vacuum solution} 

We are ready to solve Eqs. (\ref{Eq_Sigma}), (\ref{Eq_G}), and (\ref{Eq_Pi}) with their IR boundary conditions (\ref{IR_BC_Sigma}), (\ref{IR_BC_G}), and (\ref{IR_BC_Pi}) and UV boundary ones (\ref{UV_BC_Sigma}), (\ref{UV_BC_G}), and (\ref{UV_BC_Pi}). However, these equations are quite complicated and still not easy to solve. It is natural to start from the conformal ansatz as follows
\bqa && \Sigma_{\tau}^{\tau'}(z)  = \mbox{sgn}(\tau-\tau') \frac{\mathcal{C}_{\Sigma}(z)}{|\tau-\tau'|^{\Delta_{\Sigma}(z)}} , \\ &&G_{\tau}^{\tau'}(z)  = \mbox{sgn}(\tau-\tau') \frac{\mathcal{C}_{G}(z)}{|\tau-\tau'|^{\Delta_{G}(z)}} , \\ && \Pi_{\tau}^{\tau'}(z)  = \frac{\mathcal{C}_{\Pi}(z)}{|\tau-\tau'|^{\Delta_{\Pi}(z)}} . \eqa  

Let's first consider the UV boundary $z = 0$, which corresponds to the case that there are no quantum corrections to the field theoretic large $N$ solution of the SYK model. In this case we know the answer, which gives rise to a finite entropy contribution even at zero temperature \cite{SYK_Model_I,SYK_Model_II,SYK_Model_III,SYK_Model_IV,SYK_Model_V}. Now, we consider the $z \rightarrow 0$ limit, where the field theoretic $1/N$ quantum corrections are taken into account self-consistently. As a result, the conformal invariance breaks down, where the density of states vanishes \cite{SYK_Model_II,SYK_Schwarzian_I,SYK_Schwarzian_II,SYK_Schwarzian_III,SYK_Schwarzian_IV,SYK_Schwarzian_V,SYK_Schwarzian_VI} to exhibit a pseudogap behavior as discussed in the introduction.

Within the above ansatz, we catch the breakdown of the conformal invariance by the RG flows of $\mathcal{C}_{\Sigma}(z)$, $\mathcal{C}_{G}(z)$, and $\mathcal{C}_{\Pi}(z)$. Here, the critical exponents of $\Delta_{\Sigma}(z)$, $\Delta_{G}(z)$, and $\Delta_{\Pi}(z)$ turn out not to $RG-flow$. In other words, they do not depend on $z$. 

One may introduce this conformal vacuum solution into Eqs. (\ref{Eq_Sigma}), (\ref{Eq_G}), and (\ref{Eq_Pi}) with their IR boundary conditions (\ref{IR_BC_Sigma}), (\ref{IR_BC_G}), and (\ref{IR_BC_Pi}) and UV boundary ones (\ref{UV_BC_Sigma}), (\ref{UV_BC_G}), and (\ref{UV_BC_Pi}). Here, we start from the first, considering an effective action in terms of such RG-flow variables as $\mathcal{C}_{\Sigma}(z)$, $\mathcal{C}_{G}(z)$, and $\mathcal{C}_{\Pi}(z)$. Introducing this conformal ansatz into the holographic dual effective field theory Eq. (\ref{HDEFT}), we obtain the following expression
\bqa && \mathcal{Z}_{SYK}(z_{f}) = \int D \mathcal{C}_{\Sigma}(z) D \mathcal{C}_{\Pi}(z) D \mathcal{C}_{G}(z)\nn && \exp\Bigg[ - N \int_{-\infty}^{\infty} d \omega \Bigg\{ - \frac{1}{2} \ln \Big( q_{d} \Gamma_{J} \mbox{sgn}(\omega) |\omega|^{\Delta_{\Sigma} - 1} F(\Delta_{\Sigma}) \mathcal{C}_{\Sigma}(z_{f}) \Big)  - q_{d} \Gamma_{J} |\omega|^{\Delta_{\Sigma} + \Delta_{G} - 2} F(\Delta_{\Sigma}) F(\Delta_{G}) \mathcal{C}_{\Sigma}(z_{f}) \mathcal{C}_{G}(z_{f}) \nn && \ \ \ \ \ \ \  + q_{d} \Gamma_{J} |\omega|^{\Delta_{\Pi} + 2 \Delta_{G} - 2} F(\Delta_{\Pi}) F(2 \Delta_{G}) \mathcal{C}_{\Pi}(0) \mathcal{C}_{G}^{2}(0)  +\frac{ q_{d} \Gamma_{J} }{4} |\omega|^{2 \Delta_{\Pi} - 2} F^{2}(\Delta_{\Pi}) \mathcal{C}_{\Pi}^{2}(0) \Bigg\} \Bigg]\nn &&  \exp\Bigg[ - N \int_{0}^{z_{f}} d z \int_{-\infty}^{\infty} d \omega \Bigg\{ 2 q_{d} \Gamma_{J} |\omega|^{\Delta_{\Pi} + 2 \Delta_{G} - 2} F(\Delta_{\Pi}) F(2 \Delta_{G}) \mathcal{C}_{\Pi}(z) \mathcal{C}_{G}(z) [\partial_{z} \mathcal{C}_{G}(z)] \nn && \ \ \ \ \ \ \   + q_{d} \Gamma_{J} |\omega|^{\Delta_{\Pi} + 2 \Delta_{G} - 2} F(\Delta_{\Pi}) F(2 \Delta_{G}) \mathcal{C}_{G}^{2}(z) [\partial_{z} \mathcal{C}_{\Pi}(z)] + \frac{ q_{d} \Gamma_{J} }{2} |\omega|^{2 \Delta_{\Pi} - 2} F^{2}(\Delta_{\Pi}) \mathcal{C}_{\Pi}(z) [\partial_{z} \mathcal{C}_{\Pi}(z)] \nn && \ \ \ \ \ \ \   - \frac{1}{2} [\partial_{z} \ln \mathcal{C}_{\Sigma}(z)] + \frac{1}{4} [\partial_{z} \ln \mathcal{C}_{\Sigma}(z)]^{2} - q_{d} \Gamma_{J} |\omega|^{\Delta_{\Sigma} + \Delta_{G} - 2} F(\Delta_{\Sigma}) F(\Delta_{G}) [\partial_{z} \mathcal{C}_{\Sigma}(z)] [\partial_{z} \mathcal{C}_{G}(z)] \nn && \ \ \ \ \ \ \  + 2 q_{d} \Gamma_{J} |\omega|^{\Delta_{\Pi} + 2 \Delta_{G} - 2} F(\Delta_{\Pi}) F(2 \Delta_{G}) [\partial_{z} \mathcal{C}_{\Pi}(z)] \mathcal{C}_{G}(z) [\partial_{z} \mathcal{C}_{G}(z)] + \frac{ q_{d} \Gamma_{J} }{4} |\omega|^{2 \Delta_{\Pi} - 2} F^{2}(\Delta_{\Pi}) [\partial_{z} \mathcal{C}_{\Pi}(z)]^{2} \nn && \ \ \ \ \ \ \ + q_{d} \Gamma_{J} |\omega|^{\Delta_{\Pi} + 2 \Delta_{G} - 2} F(\Delta_{\Pi}) F(2 \Delta_{G}) \mathcal{C}_{\Pi}(z) [\partial_{z} \mathcal{C}_{G}(z)]^{2} - \frac{1}{2} \ln \Big( q_{d} \Gamma_{J} \mbox{sgn}(\omega) |\omega|^{\Delta_{\Sigma} - 1} F(\Delta_{\Sigma}) \mathcal{C}_{\Sigma}(z) \Big) \Bigg\} \Bigg] . \eqa
Here, the bulk effective action was organized as the first and second order derivatives with respect to $z$. We also performed the Fourier transformations for the collective bi-local fields as follows
\bqa && \Sigma(i \omega,z) = \int_{-\infty}^{\infty} d \tau e^{i \omega \tau} \Sigma(\tau,z) = F(\Delta_{\Sigma}) \mathcal{C}_{\Sigma}(z) \mbox{sgn}(\omega) |\omega|^{\Delta_{\Sigma} - 1} , \\ && G(i \omega,z) = \int_{-\infty}^{\infty} d \tau e^{i \omega \tau} G(\tau,z)  = F(\Delta_{G}) \mathcal{C}_{G}(z) \mbox{sgn}(\omega) |\omega|^{\Delta_{G} - 1} , \\ && \Pi(i \omega,z) = \int_{-\infty}^{\infty} d \tau e^{i \omega \tau} \Pi(\tau,z) = F(\Delta_{\Pi}) \mathcal{C}_{\Pi}(z) |\omega|^{\Delta_{\Pi} - 1} , \eqa
where 
\bqa && F(y) = \int_{-\infty}^{\infty} d x \frac{\mbox{sgn}(x) e^{i x}}{|x|^{y}} = i 2^{1 - y} \sqrt{\pi} \frac{\Gamma(1 - \frac{1}{2} y)}{\Gamma(\frac{1}{2} + \frac{1}{2} y)} . \eqa

A natural procedure is to minimize the effective action functional, taking variations with respect to $\mathcal{C}_{\Sigma}(z)$, $\mathcal{C}_{G}(z)$, and $\mathcal{C}_{\Pi}(z)$. The resulting Lagrange equations of motion for $\mathcal{C}_{\Sigma}(z)$, $\mathcal{C}_{G}(z)$, and $\mathcal{C}_{\Pi}(z)$ have to be satisfied, regardless of the frequency. In this respect the frequency dependence in the effective action should disappear. As a result, we have three constraints for the three critical exponents as follows
\bqa && \Delta_{\Sigma} + \Delta_{G} - 2 = 0 , \\ && \Delta_{\Pi} + 2 \Delta_{G} - 2 = 0 , \\ && \Delta_{\Pi} - 1 = 0 . \eqa
These equations give 
\bqa && \Delta_{\Sigma} = \frac{3}{2} , ~~~~~  \Delta_{G} = \frac{1}{2} , ~~~~~ \Delta_{\Pi} = 1 , \eqa
consistent with the saddle-point solution in the large $N$ limit. As a result, the above expression reads
\bqa && \mathcal{Z}_{SYK}(z_{f}) = \exp\Big[ - N \int_{-\Lambda_{\omega}}^{\Lambda_{\omega}} d \omega \Big\{ - \frac{1}{2} \ln \Big( q_{d} \Gamma_{J} \mbox{sgn}(\omega) \sqrt{|\omega|} F(\Delta_{\Sigma}) \mathcal{C}_{\Sigma}(z_{f}) \Big) - q_{d} \Gamma_{J} F(\Delta_{\Sigma}) F(\Delta_{G}) \mathcal{C}_{\Sigma}(z_{f}) \mathcal{C}_{G}(z_{f}) \nn && + q_{d} \Gamma_{J} F(\Delta_{\Pi}) F(2 \Delta_{G}) \mathcal{C}_{\Pi}(0) \mathcal{C}_{G}^{2}(0) + \frac{ q_{d} \Gamma_{J} }{4} F^{2}(\Delta_{\Pi}) \mathcal{C}_{\Pi}^{2}(0) \Big\} \Big] \nn && \exp\Big[ - N \int_{0}^{z_{f}} d z \int_{-\Lambda_{\omega}}^{\Lambda_{\omega}} d \omega \Big\{ 2 q_{d} \Gamma_{J} F(\Delta_{\Pi}) F(2 \Delta_{G}) \mathcal{C}_{\Pi}(z) \mathcal{C}_{G}(z) [\partial_{z} \mathcal{C}_{G}(z)]  + q_{d} \Gamma_{J} F(\Delta_{\Pi}) F(2 \Delta_{G}) \mathcal{C}_{G}^{2}(z) [\partial_{z} \mathcal{C}_{\Pi}(z)] \nn && \ \ \ \ \ \ \  + \frac{ q_{d} \Gamma_{J} }{2} F^{2}(\Delta_{\Pi}) \mathcal{C}_{\Pi}(z) [\partial_{z} \mathcal{C}_{\Pi}(z)]  - \frac{1}{2} [\partial_{z} \ln \mathcal{C}_{\Sigma}(z)] + \frac{1}{4} [\partial_{z} \ln \mathcal{C}_{\Sigma}(z)]^{2} - q_{d} \Gamma_{J} F(\Delta_{\Sigma}) F(\Delta_{G}) [\partial_{z} \mathcal{C}_{\Sigma}(z)] [\partial_{z} \mathcal{C}_{G}(z)] \nn && \ \ \ \ \ \ \  + 2 q_{d} \Gamma_{J} F(\Delta_{\Pi}) F(2 \Delta_{G}) [\partial_{z} \mathcal{C}_{\Pi}(z)] \mathcal{C}_{G}(z) [\partial_{z} \mathcal{C}_{G}(z)]\nn && \ \ \ \ \ \ \   + q_{d} \Gamma_{J} F(\Delta_{\Pi}) F(2 \Delta_{G}) \mathcal{C}_{\Pi}(z) [\partial_{z} \mathcal{C}_{G}(z)]^{2} + \frac{ q_{d} \Gamma_{J} }{4} F^{2}(\Delta_{\Pi}) [\partial_{z} \mathcal{C}_{\Pi}(z)]^{2}  - \frac{1}{2} \ln \Big( q_{d} \Gamma_{J} \mbox{sgn}(\omega) \sqrt{|\omega|} F(\Delta_{\Sigma}) \mathcal{C}_{\Sigma}(z) \Big) \Big\} \Big] ,\nn \eqa
where the frequency cutoff $\Lambda_{\omega}$ has been introduced to regulate the divergence in a formal expression. The frequency dependence does not arise in $\mathcal{C}_{\Sigma}(z)$, $\mathcal{C}_{G}(z)$, and $\mathcal{C}_{\Pi}(z)$ although $\mbox{sgn}(\omega) \sqrt{|\omega|}$ is shown inside the $log$.

It is straightforward to obtain the following Lagrange equations of motion
\bqa && \frac{\partial_{z}^{2} \mathcal{C}_{\Sigma}(z)}{\mathcal{C}_{\Sigma}^{2}(z)} - \frac{[\partial_{z} \mathcal{C}_{\Sigma}(z)]^{2}}{\mathcal{C}_{\Sigma}^{3}(z)} - 2 q_{d} \Gamma_{J} F(\Delta_{\Sigma}) F(\Delta_{G}) [\partial_{z}^{2} \mathcal{C}_{G}(z)] + \frac{1}{\mathcal{C}_{\Sigma}(z)} = 0 , \label{CIA_Self_Energy} \\ && \mathcal{C}_{\Pi}(z) [\partial_{z}^{2} \mathcal{C}_{G}(z)] + [\partial_{z} \mathcal{C}_{\Pi}(z)] [\partial_{z} \mathcal{C}_{G}(z)] + [\partial_{z}^{2} \mathcal{C}_{\Pi}(z)] \mathcal{C}_{G}(z) - \frac{1}{2} \frac{F(\Delta_{\Sigma}) F(\Delta_{G})}{F(\Delta_{\Pi}) F(2 \Delta_{G}) } [\partial_{z}^{2} \mathcal{C}_{\Sigma}(z)] = 0 , \label{CIA_Green_Function} \\ && \partial_{z}^{2} \mathcal{C}_{\Pi}(z) + 4 \frac{F(2 \Delta_{G})}{ F(\Delta_{\Pi})} \mathcal{C}_{G}(z) [\partial_{z}^{2} \mathcal{C}_{G}(z)] + 2 \frac{F(2 \Delta_{G})}{ F(\Delta_{\Pi})} [\partial_{z} \mathcal{C}_{G}(z)]^{2} = 0 . \label{CIA_Polarization_Function} \eqa
We point out that these expressions are not the same as those by inserting the conformal ansatz into Eqs. (\ref{Eq_Sigma}), (\ref{Eq_G}), and (\ref{Eq_Pi}). This will be clarified below.

To find both the UV and IR boundary conditions consistent with these equations of motion, it is convenient to consider the Hamiltonian formulation. The above expression of the partition function gives rise to the following effective Lagrangian in the Hamiltonian formulation
\bqa  L_{eff} = && - \frac{1}{2} \ln \Big( F(\Delta_{\Sigma}) \mathcal{C}_{\Sigma}(z_{f}) \Big) - q_{d} \Gamma_{J} F(\Delta_{\Sigma}) F(\Delta_{G}) \mathcal{C}_{\Sigma}(z_{f}) \mathcal{C}_{G}(z_{f}) + q_{d} \Gamma_{J} F(\Delta_{\Pi}) F(2 \Delta_{G}) \mathcal{C}_{\Pi}(z_{f}) \mathcal{C}_{G}^{2}(z_{f}) \nn && + \frac{ q_{d} \Gamma_{J} }{4} F^{2}(\Delta_{\Pi}) \mathcal{C}_{\Pi}^{2}(z_{f})  - \frac{1}{2} \ln \mathcal{C}_{\Sigma}(z_{f}) + \frac{1}{2} \ln \mathcal{C}_{\Sigma}(0) \nn && + \int_{0}^{z_{f}} d z \Bigg\{ \Pi_{\mathcal{C}_{\Sigma}}(z) \partial_{z} \mathcal{C}_{\Sigma}(z) - \Bigg(\frac{1}{\mathcal{C}_{\Sigma}^{2}(z)} - \frac{q_{d} \Gamma_{J} \frac{F^{2}(\Delta_{\Sigma}) F^{2}(\Delta_{G})}{ F(\Delta_{\Pi}) F(2 \Delta_{G})}}{\mathcal{C}_{\Pi}(z) - 4 \frac{F(2 \Delta_{G})}{F(\Delta_{\Pi})} \mathcal{C}_{G}^{2}(z)} \Bigg)^{-1} \Pi_{\mathcal{C}_{\Sigma}}^{2}(z) \nn && + \Pi_{\mathcal{C}_{G}}(z) \Bigg(\partial_{z} \mathcal{C}_{G}(z) - \frac{1}{2} \frac{\frac{F(\Delta_{\Sigma}) F(\Delta_{G})}{ F(\Delta_{\Pi}) F(2 \Delta_{G})}}{\mathcal{C}_{\Pi}(z) - 4 \frac{F(2 \Delta_{G})}{F(\Delta_{\Pi})} \mathcal{C}_{G}^{2}(z)} \partial_{z} \mathcal{C}_{\Sigma}(z) \Bigg)  + \Pi_{\mathcal{C}_{\Pi}}(z) \Bigg(\partial_{z} \mathcal{C}_{\Pi}(z) + 4 \frac{ F(2 \Delta_{G})}{F(\Delta_{\Pi})} \mathcal{C}_{G}(z) [\partial_{z} \mathcal{C}_{G}(z)] \Bigg)\nn && - \frac{1}{4 q_{d} \Gamma_{J}} \frac{1}{F(\Delta_{\Pi}) F(2 \Delta_{G})} \Bigg( \mathcal{C}_{\Pi}(z) - 4 \frac{F(2 \Delta_{G})}{F(\Delta_{\Pi})} \mathcal{C}_{G}^{2}(z) \Bigg)^{-1} \Pi_{\mathcal{C}_{G}}^{2}(z) - \frac{ 1 }{q_{d} \Gamma_{J}} \frac{1}{F^{2}(\Delta_{\Pi})} \Pi_{\mathcal{C}_{\Pi}(z)}^{2}(z) - \frac{1}{2} \ln \Big( F(\Delta_{\Sigma}) \mathcal{C}_{\Sigma}(z) \Big) \Bigg\} ,\nn \eqa
where the frequency dependence inside the $log$ has been droped out. $\Pi_{\mathcal{C}_{\Sigma}}(z)$, $\Pi_{\mathcal{C}_{\Pi}}(z)$, and $\Pi_{\mathcal{C}_{G}}(z)$ are the momentum to $\mathcal{C}_{\Sigma}(z)$, $\mathcal{C}_{\Pi}(z)$, and $\mathcal{C}_{G}(z)$, respectively. They are given by the Hamiltonian equation of motion as follows
\bqa && \Pi_{\mathcal{C}_{\Sigma}}(z) = \frac{1}{2} \Bigg(\frac{1}{\mathcal{C}_{\Sigma}^{2}(z)} - \frac{q_{d} \Gamma_{J} \frac{F^{2}(\Delta_{\Sigma}) F^{2}(\Delta_{G})}{ F(\Delta_{\Pi}) F(2 \Delta_{G})}}{\mathcal{C}_{\Pi}(z) - 4 \frac{F(2 \Delta_{G})}{F(\Delta_{\Pi})} \mathcal{C}_{G}^{2}(z)} \Bigg) \partial_{z} \mathcal{C}_{\Sigma}(z) , \\ && \Pi_{\mathcal{C}_{G}}(z) = 2 q_{d} \Gamma_{J} F(\Delta_{\Pi}) F(2 \Delta_{G}) \Bigg( \mathcal{C}_{\Pi}(z) - 4 \frac{F(2 \Delta_{G})}{F(\Delta_{\Pi})} \mathcal{C}_{G}^{2}(z) \Bigg) \Bigg(\partial_{z} \mathcal{C}_{G}(z) - \frac{1}{2} \frac{\frac{F(\Delta_{\Sigma}) F(\Delta_{G})}{ F(\Delta_{\Pi}) F(2 \Delta_{G})}}{\mathcal{C}_{\Pi}(z) - 4 \frac{F(2 \Delta_{G})}{F(\Delta_{\Pi})} \mathcal{C}_{G}^{2}(z)} \partial_{z} \mathcal{C}_{\Sigma}(z) \Bigg) , \\ && \Pi_{\mathcal{C}_{\Pi}}(z) = \frac{q_{d} \Gamma_{J}}{ 2 } F^{2}(\Delta_{\Pi}) \Bigg(\partial_{z} \mathcal{C}_{\Pi}(z) + 4 \frac{ F(2 \Delta_{G})}{F(\Delta_{\Pi})} \mathcal{C}_{G}(z) [\partial_{z} \mathcal{C}_{G}(z)] \Bigg) . \eqa 

It is straightforward to find the boundary effective Lagrangian from the above effective Lagrangian as 
\bqa  L_{bdry.} = &&- \ln \mathcal{C}_{\Sigma}(z_{f}) - q_{d} \Gamma_{J} F(\Delta_{\Sigma}) F(\Delta_{G}) \mathcal{C}_{\Sigma}(z_{f}) \mathcal{C}_{G}(z_{f}) + q_{d} \Gamma_{J} F(\Delta_{\Pi}) F(2 \Delta_{G}) \mathcal{C}_{\Pi}(z_{f}) \mathcal{C}_{G}^{2}(z_{f}) \nn && + \frac{ q_{d} \Gamma_{J} }{4} F^{2}(\Delta_{\Pi}) \mathcal{C}_{\Pi}^{2}(z_{f}) + \Pi_{\mathcal{C}_{\Sigma}}(z_{f}) \mathcal{C}_{\Sigma}(z_{f}) \nn && + \Pi_{\mathcal{C}_{G}}(z_{f}) \Bigg( \mathcal{C}_{G}(z_{f}) - \frac{1}{2} \frac{\frac{F(\Delta_{\Sigma}) F(\Delta_{G})}{ F(\Delta_{\Pi}) F(2 \Delta_{G})}}{\mathcal{C}_{\Pi}(z_{f}) - 4 \frac{F(2 \Delta_{G})}{F(\Delta_{\Pi})} \mathcal{C}_{G}^{2}(z_{f})} \mathcal{C}_{\Sigma}(z_{f}) \Bigg) + \Pi_{\mathcal{C}_{\Pi}}(z_{f}) \Bigg( \mathcal{C}_{\Pi}(z_{f}) + 2 \frac{ F(2 \Delta_{G})}{F(\Delta_{\Pi})} \mathcal{C}_{G}^{2}(z_{f}) \Bigg) \nn && + \frac{1}{2} \ln \mathcal{C}_{\Sigma}(0) - \Pi_{\mathcal{C}_{\Sigma}}(0) \mathcal{C}_{\Sigma}(0) - \Pi_{\mathcal{C}_{G}}(0) \Bigg( \mathcal{C}_{G}(0) - \frac{1}{2} \frac{\frac{F(\Delta_{\Sigma}) F(\Delta_{G})}{ F(\Delta_{\Pi}) F(2 \Delta_{G})}}{\mathcal{C}_{\Pi}(0) - 4 \frac{F(2 \Delta_{G})}{F(\Delta_{\Pi})} \mathcal{C}_{G}^{2}(0)} \mathcal{C}_{\Sigma}(0) \Bigg)\nn && - \Pi_{\mathcal{C}_{\Pi}}(0) \Bigg( \mathcal{C}_{\Pi}(0) + 2 \frac{ F(2 \Delta_{G})}{F(\Delta_{\Pi})} \mathcal{C}_{G}^{2}(0) \Bigg) . \eqa
Following the same strategy as discussed in the previous section, we obtain the IR boundary conditions as
\bqa && \partial_{z_{f}} \mathcal{C}_{\Sigma}(z_{f}) - 2 \mathcal{C}_{\Sigma}(z_{f}) - 2 q_{d} \Gamma_{J} F(\Delta_{\Sigma}) F(\Delta_{G}) \Big( \mathcal{C}_{G}(z_{f}) + \partial_{z_{f}} \mathcal{C}_{G}(z_{f}) \Big) \mathcal{C}_{\Sigma}^{2}(z_{f}) = 0 , \label{Conformal_IR_BC_Sigma}  \\ && \Bigg( \mathcal{C}_{\Pi}(z_{f}) - 4 \mathcal{C}_{G}^{2}(z_{f}) - 4 \frac{F(\Delta_{\Sigma}) F(\Delta_{G})}{ F^{2}(\Delta_{\Pi})} \frac{\mathcal{C}_{G}(z_{f}) \mathcal{C}_{\Sigma}(z_{f})}{\mathcal{C}_{\Pi}(z_{f}) - 4 \mathcal{C}_{G}^{2}(z_{f})} \Bigg) \Bigg(\partial_{z_{f}} \mathcal{C}_{G}(z_{f}) - \frac{1}{2} \frac{\frac{F(\Delta_{\Sigma}) F(\Delta_{G})}{ F(\Delta_{\Pi}) F(2 \Delta_{G})}}{\mathcal{C}_{\Pi}(z_{f}) - 4 \mathcal{C}_{G}^{2}(z_{f})} \partial_{z_{f}} \mathcal{C}_{\Sigma}(z_{f}) \Bigg) \nn && - \frac{1}{2} \frac{F(\Delta_{\Sigma}) F(\Delta_{G})}{F(\Delta_{\Pi}) F(2 \Delta_{G})} \mathcal{C}_{\Sigma}(z_{f}) + \mathcal{C}_{\Pi}(z_{f}) \mathcal{C}_{G}(z_{f}) + \mathcal{C}_{G}(z_{f}) \Big(\partial_{z_{f}} \mathcal{C}_{\Pi}(z_{f}) + 4 \mathcal{C}_{G}(z_{f}) \partial_{z_{f}} \mathcal{C}_{G}(z_{f}) \Big) = 0 , \label{Conformal_IR_BC_G} \\ && \partial_{z_{f}} \mathcal{C}_{\Pi}(z_{f}) + 4 \mathcal{C}_{G}(z_{f}) \partial_{z_{f}} \mathcal{C}_{G}(z_{f}) + \mathcal{C}_{\Pi}(z_{f}) + 2 \mathcal{C}_{G}^{2}(z_{f}) \nn && + 2 \frac{F(\Delta_{\Sigma}) F(\Delta_{G})}{F^{2}(\Delta_{\Pi})} \mathcal{C}_{\Sigma}(z_{f}) \Bigg(\partial_{z_{f}} \mathcal{C}_{G}(z_{f}) - \frac{1}{2} \frac{\frac{F(\Delta_{\Sigma}) F(\Delta_{G})}{ F(\Delta_{\Pi}) F(2 \Delta_{G})}}{\mathcal{C}_{\Pi}(z_{f}) - 4 \mathcal{C}_{G}^{2}(z_{f})} \partial_{z_{f}} \mathcal{C}_{\Sigma}(z_{f}) \Bigg) = 0 , \label{Conformal_IR_BC_Pi} \eqa 
and the UV ones as
\bqa && \Bigg(\frac{1}{\mathcal{C}_{\Sigma}(0)} - q_{d} \Gamma_{J} \frac{F^{2}(\Delta_{\Sigma}) F^{2}(\Delta_{G})}{ F(\Delta_{\Pi}) F(2 \Delta_{G})} \frac{\mathcal{C}_{\Sigma}(0)}{\mathcal{C}_{\Pi}(0) - 4 \mathcal{C}_{G}^{2}(0)} \Bigg) [\partial_{z} \mathcal{C}_{\Sigma}(z)]_{z = 0} = 1 , \label{Conformal_UV_BC_Sigma} \\ && [\partial_{z} \mathcal{C}_{G}(z)]_{z = 0} - \frac{1}{2} \frac{\frac{F(\Delta_{\Sigma}) F(\Delta_{G})}{ F(\Delta_{\Pi}) F(2 \Delta_{G})}}{\mathcal{C}_{\Pi}(0) - 4 \mathcal{C}_{G}^{2}(0)} [\partial_{z} \mathcal{C}_{\Sigma}(z)]_{z = 0} = 0 , \label{Conformal_UV_BC_G} \\ && [\partial_{z} \mathcal{C}_{\Pi}(z)]_{z = 0} + 4 \mathcal{C}_{G}(0) [\partial_{z} \mathcal{C}_{G}(z)]_{z = 0} = 0 . \label{Conformal_UV_BC_Pi}   \eqa

These equations of motion with the IR and UV boundary conditions are not the same as those discussed in the previous section. Of course, they are related by a simple transformation. Integrating over all the momentum fields in the above effective Lagrangian, we obtain the following expression
\bqa L_{eff} = &&- \frac{1}{2} \ln \Big( F(\Delta_{\Sigma}) \mathcal{C}_{\Sigma}(z_{f}) \Big) - q_{d} \Gamma_{J} F(\Delta_{\Sigma}) F(\Delta_{G}) \mathcal{C}_{\Sigma}(z_{f}) \mathcal{C}_{G}(z_{f}) + q_{d} \Gamma_{J} F(\Delta_{\Pi}) F(2 \Delta_{G}) \mathcal{C}_{\Pi}(z_{f}) \mathcal{C}_{G}^{2}(z_{f}) \nn && + \frac{ q_{d} \Gamma_{J} }{4} F^{2}(\Delta_{\Pi}) \mathcal{C}_{\Pi}^{2}(z_{f})  - \frac{1}{2} \ln \mathcal{C}_{\Sigma}(z_{f}) + \frac{1}{2} \ln \mathcal{C}_{\Sigma}(0) \nn && + \int_{0}^{z_{f}} d z \Bigg\{ \frac{1}{4} \Bigg(\frac{1}{\mathcal{C}_{\Sigma}^{2}(z)} - \frac{q_{d} \Gamma_{J} \frac{F^{2}(\Delta_{\Sigma}) F^{2}(\Delta_{G})}{ F(\Delta_{\Pi}) F(2 \Delta_{G})}}{\mathcal{C}_{\Pi}(z) - 4 \frac{F(2 \Delta_{G})}{F(\Delta_{\Pi})} \mathcal{C}_{G}^{2}(z)} \Bigg) [\partial_{z} \mathcal{C}_{\Sigma}(z)]^{2} \nn && +q_{d} \Gamma_{J} F(\Delta_{\Pi}) F(2 \Delta_{G}) \Bigg( \mathcal{C}_{\Pi}(z) - 4 \frac{F(2 \Delta_{G})}{F(\Delta_{\Pi})} \mathcal{C}_{G}^{2}(z) \Bigg) \Bigg(\partial_{z} \mathcal{C}_{G}(z) - \frac{1}{2} \frac{\frac{F(\Delta_{\Sigma}) F(\Delta_{G})}{ F(\Delta_{\Pi}) F(2 \Delta_{G})}}{\mathcal{C}_{\Pi}(z) - 4 \frac{F(2 \Delta_{G})}{F(\Delta_{\Pi})} \mathcal{C}_{G}^{2}(z)} \partial_{z} \mathcal{C}_{\Sigma}(z) \Bigg)^{2} \nn && + \frac{ q_{d} \Gamma_{J} }{4} F^{2}(\Delta_{\Pi}) \Bigg(\partial_{z} \mathcal{C}_{\Pi}(z) + 4 \frac{ F(2 \Delta_{G})}{F(\Delta_{\Pi})} \mathcal{C}_{G}(z) [\partial_{z} \mathcal{C}_{G}(z)] \Bigg)^{2} - \frac{1}{2} \ln \Big( F(\Delta_{\Sigma}) \mathcal{C}_{\Sigma}(z) \Big) \Bigg\} . \eqa
Then, we consider a shift transformation for $\mathcal{C}_{\Pi}(z)$ as
\bqa && \mathcal{C}_{\Pi}(z) \longrightarrow \mathcal{C}_{\Pi}(z) - 2 \frac{F(2 \Delta_{G})}{F(\Delta_{\Pi})} \mathcal{C}_{G}^{2}(z) . \eqa
As a result, we obtain
%
%
\bqa  L_{eff} = &&- \ln \Big( F(\Delta_{\Sigma}) \mathcal{C}_{\Sigma}(z_{f}) \Big) - q_{d} \Gamma_{J} F(\Delta_{\Sigma}) F(\Delta_{G}) \mathcal{C}_{\Sigma}(z_{f}) \mathcal{C}_{G}(z_{f}) - q_{d} \Gamma_{J} F^{2}(2 \Delta_{G}) \mathcal{C}_{G}^{4}(z_{f})\nn && + \frac{ q_{d} \Gamma_{J} }{4} F^{2}(\Delta_{\Pi}) \mathcal{C}_{\Pi}^{2}(z_{f}) + \frac{1}{2} \ln \mathcal{C}_{\Sigma}(0) \nn && + \int_{0}^{z_{f}} d z \Big\{ \frac{1}{4} \frac{1}{\mathcal{C}_{\Sigma}^{2}(z)} [\partial_{z} \mathcal{C}_{\Sigma}(z)]^{2} - q_{d} \Gamma_{J} F(\Delta_{\Sigma}) F(\Delta_{G}) [\partial_{z} \mathcal{C}_{G}(z)] [\partial_{z} \mathcal{C}_{\Sigma}(z)] \nn && + q_{d} \Gamma_{J} F(\Delta_{\Pi}) F(2 \Delta_{G}) \Big( \mathcal{C}_{\Pi}(z) - 6 \frac{F(2 \Delta_{G})}{F(\Delta_{\Pi})} \mathcal{C}_{G}^{2}(z) \Big) [\partial_{z} \mathcal{C}_{G}(z)]^{2} + \frac{ q_{d} \Gamma_{J} }{4} F^{2}(\Delta_{\Pi}) [\partial_{z} \mathcal{C}_{\Pi}(z)]^{2} \nn && - \frac{1}{2} \ln \Big( F(\Delta_{\Sigma}) \mathcal{C}_{\Sigma}(z) \Big) \Big\} . \eqa
This is the effective Lagrangian of the previous section within the conformal ansatz.

The resulting equations of motion are given by
\bqa && \mathcal{C}_{\Sigma}(z) \partial_{z}^{2} \mathcal{C}_{\Sigma}(z) - [\partial_{z} \mathcal{C}_{\Sigma}(z)]^{2} + \mathcal{C}_{\Sigma}^{2}(z) = 2 q_{d} \Gamma_{J} F(\Delta_{\Sigma}) F(\Delta_{G}) \mathcal{C}_{\Sigma}^{3}(z) \partial_{z}^{2} \mathcal{C}_{G}(z) , \label{C_Sigma_Old} \\ && \Big( \mathcal{C}_{\Pi}(z) - 6 \mathcal{C}_{G}^{2}(z) \Big) \partial_{z}^{2} \mathcal{C}_{G}(z) +  \Big( \partial_{z} \mathcal{C}_{\Pi}(z) - 6 \mathcal{C}_{G}(z) \partial_{z} \mathcal{C}_{G}(z) \Big) \partial_{z} \mathcal{C}_{G}(z) = \frac{1}{2} \frac{ F(\Delta_{\Sigma}) F(\Delta_{G})}{F(\Delta_{\Pi}) F(2 \Delta_{G})} \partial_{z}^{2} \mathcal{C}_{\Sigma}(z) , \label{C_G_Old} \\ && \partial_{z}^{2} \mathcal{C}_{\Pi}(z) = 2 [\partial_{z} \mathcal{C}_{G}(z)]^{2} . \label{C_Pi_Old} \eqa
These equations can be found from those of the present section by taking the shift transformation. One can find both the IR and UV boundary conditions in the same fashion, not shown here.

\subsection{RG flows based on the matching method} 

In this subsection, we show important details on how to solve the previously derived nonlinearly intertwined second order coupled differential equations because we believe that this technical aspect is essential for general applications of the present emergent dual holographic theoretical framework to various strongly coupled systems. Readers who are not interested in this methodology can skip this technical part and may start from Fig. 2 to see the solution.

\subsubsection{Asymptotic solutions near UV and IR boundaries}

For convenience, we introduce $X(\zeta) = \mathcal{C}_{\Sigma}(\zeta)$, $Y(\zeta) = \mathcal{C}_{G}(\zeta)$, and $Z(\zeta) = \mathcal{C}_{\Pi}(\zeta)$, where $\zeta = \frac{z}{\Lambda}$ is a `normalized' coordinate of the extra dimension. Here, $\Lambda = z_{f}$ is the cutoff scale. UV and IR boundary conditions are given by $x_0=X(0)$, $y_0=Y(0)$, $z_0=Z(0)$ and $x_f=X(1)$, $y_f=Y(1)$, $z_f=Z(1)$, respectively, which satisfy their UV and IR boundary equations discussed in the previous subsection. Then, the bulk equations of motion given by (\ref{CIA_Self_Energy})--(\ref{CIA_Polarization_Function}) or `equivalently', Eqs. (\ref{Eq_Sigma})--(\ref{Eq_Pi}) with the near conformal-solution ansatz can be rewritten as follows
	\bqa &&    X''(\zeta )=-\frac{2 \left(4 Y(\zeta )^2 X'(\zeta )^2-Z(\zeta ) X'(\zeta )^2+2 \alpha  X(\zeta )^3 Y(\zeta ) Y'(\zeta )^2-\alpha  X(\zeta )^3 Y'(\zeta ) Z'(\zeta )-4 \Lambda ^2 X(\zeta )^2 Y(\zeta )^2+\Lambda ^2 X(\zeta )^2 Z(\zeta )\right)}{X(\zeta ) \left(\alpha  a_0 X(\zeta )^2-8 Y(\zeta )^2+2 Z(\zeta )\right)},\nn  \label{eq:46} \\
	 && Y''(\zeta )=\frac{a_0 X'(\zeta )^2-a_0 \Lambda ^2    X(\zeta )^2+4 X(\zeta ) Y(\zeta ) Y'(\zeta )^2-2 X(\zeta ) Y'(\zeta ) Z'(\zeta )}{X(\zeta ) \left(\alpha  a_0 X(\zeta )^2-8 Y(\zeta )^2+2 Z(\zeta )\right)} ,  \label{eq:47} \\ 
	 &&  Z''(\zeta )=-\frac{2 \left(2 a_0 Y(\zeta ) X'(\zeta )^2+\alpha  a_0 X(\zeta )^3 Y'(\zeta )^2-2 a_0 \Lambda ^2 X(\zeta )^2 Y(\zeta )-4 X(\zeta ) Y(\zeta ) Y'(\zeta ) Z'(\zeta )+2 X(\zeta ) Z(\zeta ) Y'(\zeta )^2\right)}{X(\zeta ) \left(\alpha  a_0 X(\zeta )^2-8 Y(\zeta )^2+2 Z(\zeta )\right)}.  \label{eq:48} 
	 \eqa
Here, the $'$ symbol denotes a derivative with respect to the argument. We also introduced $\alpha= - 2 q_{d} \Gamma_{J} F(\Delta_{\Sigma}) F(\Delta_{G}) $ and $a_0=  \frac{F(\Delta_{\Sigma}) F(\Delta_{G})}{F(\Delta_{\Pi}) F(2 \Delta_{G}) }$ based on the near conformal-solution ansatz.

To solve these coupled nonlinear differential equations, we resort to the following recursion equations
	\bqa &&    X^{(n+1)^{''}} =-\frac{2 \left( 2 \alpha  X^{(n)^3} Y^{(n)} Y^{(n)'^2} -\alpha  X^{(n)^3} Y^{(n)'}  Z^{(n)'}-4 \Lambda ^2 X^{(n)^2} Y^{(n)^2} +\Lambda ^2 X^{(n)^2}  Z^{(n)} +\left(4 Y^{(n)^2}   -Z^{(n)} \right)X^{(n+1)'^2} \right)}{X^{(n)}  \left(\alpha  a_0 X^{(n)^2} -8 Y^{(n)^2} +2 Z^{(n)} \right)},\nn  \label{eq:49}\\
	&& Y^{(n+1)^{''}} =\frac{a_0 X^{(n)'^2}-a_0 \Lambda ^2    X^{(n)^2}-2 X^{(n)} Z^{(n)'} Y^{(n+1)^{'}} + 4 X^{(n) } Y^{(n) } Y^{(n+1)'^2}  }{X^{(n)}  \left(\alpha  a_0 X^{(n)^2} -8 Y^{(n)^2} +2 Z^{(n)} \right)} ,  \label{eq:50}\\
	 &&  Z^{(n+1)^{''}} =-\frac{2 \left(2 a_0 Y^{(n) } X^{(n)'^2}+\alpha  a_0 X^{(n)^3} Y^{(n)'^2}-2 a_0 \Lambda ^2 X^{(n)^2} Y^{(n)}+2 X^{(n) } Z^{(n) } Y^{(n)'^2} -4 X^{(n) } Y^{(n) } Y^{(n)' } Z^{(n+1)^{'}} \right)}{X^{(n)}  \left(\alpha  a_0 X^{(n)^2} -8 Y^{(n)^2} +2 Z^{(n)} \right)} . \label{eq:51}
	 \eqa
These second-order coupled differential equations are supported by the UV boundary conditions,
 \bqa
X^{(0)}=\mathit{x}_0, \;\; Y^{(0)}=\mathit{y}_0, \;\; Z^{(0)}=\mathit{z}_0,\;\; 
X^{(0)'}=\mathit{x}_{\text{id}}, \;\; Y^{(0)'}=\mathit{y}_{\text{id}}, \;\; Z^{(0)'}=\mathit{z}_{\text{id}} , \label{eq:52_1}
 \eqa
and the IR boundary ones, 
 \bqa
 X^{(0)}=\mathit{x}_f, \;\; Y^{(0)}=\mathit{y}_f, \;\; Z^{(0)}=\mathit{z}_f,\;\; 
 X^{(0)'}=\mathit{x}_{\text{fd}}, \;\; Y^{(0)'}=\mathit{y}_{\text{fd}}, \;\; Z^{(0)'}=\mathit{z}_{\text{fd}} , \label{eq:52_2}
 \eqa
respectively. These UV and IR boundary conditions have to satisfy the UV and IR boundary equations, respectively, in addition to the matching condition, to be clarified below.

Previously, we used this recursion-equation technique to solve complex coupled nonlinear differential equations \cite{Emergent_AdS2_BH_RG}. The strategy is to find a solution when $n = 0$ and to show that the $n = 0$ solution near UV (and IR) is an asymptotically nice solution. The $n=0$ solutions of Eqs. (\ref{eq:49})-- (\ref{eq:51}) near the UV boundary are 
 	\bqa && X_{uv}(\zeta )=\mathit{x}_0-\mathcal{A}_1 \log \left(\cosh \left(\mathcal{A}_2 \zeta \right)-\mathcal{A}_3 \sinh \left(\mathcal{A}_2 \zeta \right)\right) ,  \label{eq:52_3}\\
 	&&Y_{uv}(\zeta )=\mathcal{A}_6 \left(-\log \left(\mathcal{A}_5 \sinh \left(\mathcal{A}_4 \zeta \right)+\cosh \left(\mathcal{A}_4 \zeta \right)\right)\right)+\frac{\zeta  \mathit{z}_{\text{id}}}{4 \mathit{y}_0}+\mathit{y}_0,  \label{eq:53}\\
 	&&  Z_{uv}(\zeta )=\mathcal{A}_7 \zeta +\mathcal{A}_8 \left(1-e^{\mathcal{A}_9 \zeta }\right)+\mathit{z}_0 , \label{eq:54}
 	\eqa
where all coefficients are given by
 	\bqa && \mathcal{A}_1=\frac{\mathit{x}_0 \left(\alpha  a_0 \mathit{x}_0^2-8 \mathit{y}_0^2+2 \mathit{z}_0\right)}{2 \left(\mathit{z}_0-4 \mathit{y}_0^2\right)}, ~~~~~ \mathcal{A}_2=\frac{2 \sqrt{\mathit{x}_0^2 \left(\mathit{z}_0-4 \mathit{y}_0^2\right) \left(2 \alpha  \mathit{x}_0 \mathit{y}_0 \mathit{y}_{\text{id}}^2-\alpha  \mathit{x}_0 \mathit{y}_{\text{id}} \mathit{z}_{\text{id}}-4 \Lambda ^2 \mathit{y}_0^2+\Lambda ^2 \mathit{z}_0\right)}}{\mathit{x}_0 \left(\alpha  a_0 \mathit{x}_0^2-8 \mathit{y}_0^2+2 \mathit{z}_0\right)}, \nonumber\\
 	&& \mathcal{A}_3=\mathit{x}_{\text{id}} \sqrt{\frac{\mathit{z}_0-4 \mathit{y}_0^2}{\mathit{x}_0^2 \left(2 \alpha  \mathit{x}_0 \mathit{y}_0 \mathit{y}_{\text{id}}^2-\alpha  \mathit{x}_0 \mathit{y}_{\text{id}} \mathit{z}_{\text{id}}-4 \Lambda ^2 \mathit{y}_0^2+\Lambda ^2 \mathit{z}_0\right)}}, ~~~~~ \mathcal{A}_4=\sqrt{\frac{4 a_0 \mathit{y}_0 \left(\Lambda ^2 \mathit{x}_0^2-\mathit{x}_{\text{id}}^2\right)+\mathit{x}_0 \mathit{z}_{\text{id}}^2}{\mathit{x}_0 \left(\alpha  a_0 \mathit{x}_0^2-8 \mathit{y}_0^2+2 \mathit{z}_0\right){}^2}}, \nonumber\\
 	&& \mathcal{A}_5=\frac{\mathit{z}_{\text{id}}-4 \mathit{y}_0 \mathit{y}_{\text{id}}}{\left(\alpha  a_0 \mathit{x}_0^2-8 \mathit{y}_0^2+2 \mathit{z}_0\right) \sqrt{\frac{4 a_0 \mathit{y}_0 \left(\Lambda ^2 \mathit{x}_0^2-\mathit{x}_{\text{id}}^2\right)+\mathit{x}_0 \mathit{z}_{\text{id}}^2}{\mathit{x}_0 \left(\alpha  a_0 \mathit{x}_0^2-8 \mathit{y}_0^2+2 \mathit{z}_0\right){}^2}}}, ~~~ \mathcal{A}_6=\frac{\alpha  a_0 \mathit{x}_0^2-8 \mathit{y}_0^2+2 \mathit{z}_0}{4 \mathit{y}_0}, \nonumber\\
  && \mathcal{A}_7=\frac{a_0 \left(\alpha  \mathit{x}_0^3 \mathit{y}_{\text{id}}^2+2 \mathit{y}_0 \mathit{x}_{\text{id}}^2-2 \Lambda ^2 \mathit{x}_0^2 \mathit{y}_0\right)+2 \mathit{x}_0 \mathit{z}_0 \mathit{y}_{\text{id}}^2}{4 \mathit{x}_0 \mathit{y}_0 \mathit{y}_{\text{id}}},\nonumber\\
 	&& \mathcal{A}_8=\frac{\left(\alpha  a_0 \mathit{x}_0^2-8 \mathit{y}_0^2+2 \mathit{z}_0\right) \left(a_0 \left(\alpha  \mathit{x}_0^3 \mathit{y}_{\text{id}}^2+2 \mathit{y}_0 \mathit{x}_{\text{id}}^2-2 \Lambda ^2 \mathit{x}_0^2 \mathit{y}_0\right)+2 \mathit{x}_0 \mathit{y}_{\text{id}} \left(\mathit{z}_0 \mathit{y}_{\text{id}}-2 \mathit{y}_0 \mathit{z}_{\text{id}}\right)\right)}{32 \mathit{x}_0 \mathit{y}_0^2 \mathit{y}_{\text{id}}^2}, \nonumber\\
  &&\mathcal{A}_9=\frac{8 \mathit{y}_0 \mathit{y}_{\text{id}}}{\alpha  a_0 \mathit{x}_0^2-8 \mathit{y}_0^2+2 \mathit{z}_0}. \nonumber
 	\eqa
The $n=0$ solutions of Eqs. (\ref{eq:49})-- (\ref{eq:51}) near the IR boundary are 
	\bqa && X_{ir}(\zeta )=\mathit{x}_f-\mathcal{B}_1 \log \left(\mathcal{B}_3 \sinh \left((1-\zeta ) \mathcal{B}_2\right)+\cosh \left((1-\zeta ) \mathcal{B}_2\right)\right),  \label{eq:55}\\
	&&Y_{ir}(\zeta )=-\frac{(1-\zeta ) \mathit{z}_{\text{fd}}}{4 \mathit{y}_f}+\mathit{y}_f+\mathcal{B}_4 \left(-\log \left(\cosh \left((1-\zeta ) \mathcal{B}_5\right)-\mathcal{B}_6 \sinh \left((1-\zeta ) \mathcal{B}_5\right)\right)\right),  \label{eq:56}\\
	&&  Z_{ir}(\zeta )=\mathit{z}_f+(1-\zeta ) \mathcal{B}_7+\mathcal{B}_8 \left(1-e^{(1-\zeta ) \left(-\mathcal{B}_9\right)}\right), \label{eq:57}
	\eqa
where all coefficients are given by
	\bqa && \mathcal{B}_1=\frac{\mathit{x}_f \left(\alpha  a_0 \mathit{x}_f^2-8 \mathit{y}_f^2+2 \mathit{z}_f\right)}{2 \left(4 \mathit{y}_f^2-\mathit{z}_f\right)}, ~~~~~ \mathcal{B}_2=\frac{2 \sqrt{\mathit{x}_f^2 \left(4 \mathit{y}_f^2-\mathit{z}_f\right) \left(2 \alpha  \mathit{x}_f \mathit{y}_f \mathit{y}_{\text{fd}}^2-\alpha  \mathit{x}_f \mathit{y}_{\text{fd}} \mathit{z}_{\text{fd}}-4 \Lambda ^2 \mathit{y}_f^2+\Lambda ^2 \mathit{z}_f\right)}}{\mathit{x}_f \left(\alpha  a_0 \mathit{x}_f^2-8 \mathit{y}_f^2+2 \mathit{z}_f\right)}, \nonumber\\
	&&\mathcal{B}_3=\frac{\mathit{x}_{\text{fd}} \sqrt{4 \mathit{y}_f^2-\mathit{z}_f}}{\sqrt{\mathit{x}_f^2 \left(2 \alpha  \mathit{x}_f \mathit{y}_f \mathit{y}_{\text{fd}}^2-\alpha  \mathit{x}_f \mathit{y}_{\text{fd}} \mathit{z}_{\text{fd}}-4 \Lambda ^2 \mathit{y}_f^2+\Lambda ^2 \mathit{z}_f\right)}}, ~~~~~ \mathcal{B}_4=\frac{\alpha  a_0 \mathit{x}_f^2-8 \mathit{y}_f^2+2 \mathit{z}_f}{4 \mathit{y}_f}, \nonumber\\
	&& \mathcal{B}_5=\sqrt{\frac{4 a_0 \mathit{y}_f \left(\Lambda ^2 \mathit{x}_f^2-\mathit{x}_{\text{fd}}^2\right)+\mathit{x}_f \mathit{z}_{\text{fd}}^2}{\mathit{x}_f \left(\alpha  a_0 \mathit{x}_f^2-8 \mathit{y}_f^2+2 \mathit{z}_f\right){}^2}}, ~~~ \mathcal{B}_6=\frac{\left(\mathit{z}_{\text{fd}}-4 \mathit{y}_f \mathit{y}_{\text{fd}}\right) \sqrt{\frac{\mathit{x}_f \left(\alpha  a_0 \mathit{x}_f^2-8 \mathit{y}_f^2+2 \mathit{z}_f\right){}^2}{a_0 \mathit{y}_f \left(\Lambda ^2 \mathit{x}_f^2-\mathit{x}_{\text{fd}}^2\right)+\mathit{x}_f \mathit{z}_{\text{fd}}^2}}}{2 \left(\alpha  a_0 \mathit{x}_f^2-8 \mathit{y}_f^2+2 \mathit{z}_f\right)}, \nonumber\\
 && \mathcal{B}_7=\frac{a_0 \left(2 \mathit{y}_f \left(\Lambda ^2 \mathit{x}_f^2-\mathit{x}_{\text{fd}}^2\right)-\alpha  \mathit{x}_f^3 \mathit{y}_{\text{fd}}^2\right)-2 \mathit{x}_f \mathit{z}_f \mathit{y}_{\text{fd}}^2}{4 \mathit{x}_f \mathit{y}_f \mathit{y}_{\text{fd}}},\nonumber\\
&& \mathcal{B}_8=\frac{\left(\alpha  a_0 \mathit{x}_f^2-8 \mathit{y}_f^2+2 \mathit{z}_f\right) \left(a_0 \left(\alpha  \mathit{x}_f^3 \mathit{y}_{\text{fd}}^2+2 \mathit{y}_f \mathit{x}_{\text{fd}}^2-2 \Lambda ^2 \mathit{x}_f^2 \mathit{y}_f\right)+2 \mathit{x}_f \mathit{y}_{\text{fd}} \left(\mathit{z}_f \mathit{y}_{\text{fd}}-2 \mathit{y}_f \mathit{z}_{\text{fd}}\right)\right)}{32 \mathit{x}_f \mathit{y}_f^2 \mathit{y}_{\text{fd}}^2}, \nonumber\\
 && \mathcal{B}_9=\frac{8 \mathit{y}_f \mathit{y}_{\text{fd}}}{\alpha  a_0 \mathit{x}_f^2-8 \mathit{y}_f^2+2 \mathit{z}_f}. \nonumber
	\eqa

\begin{figure}[h]
	\centering
	\includegraphics[width=0.6\linewidth]{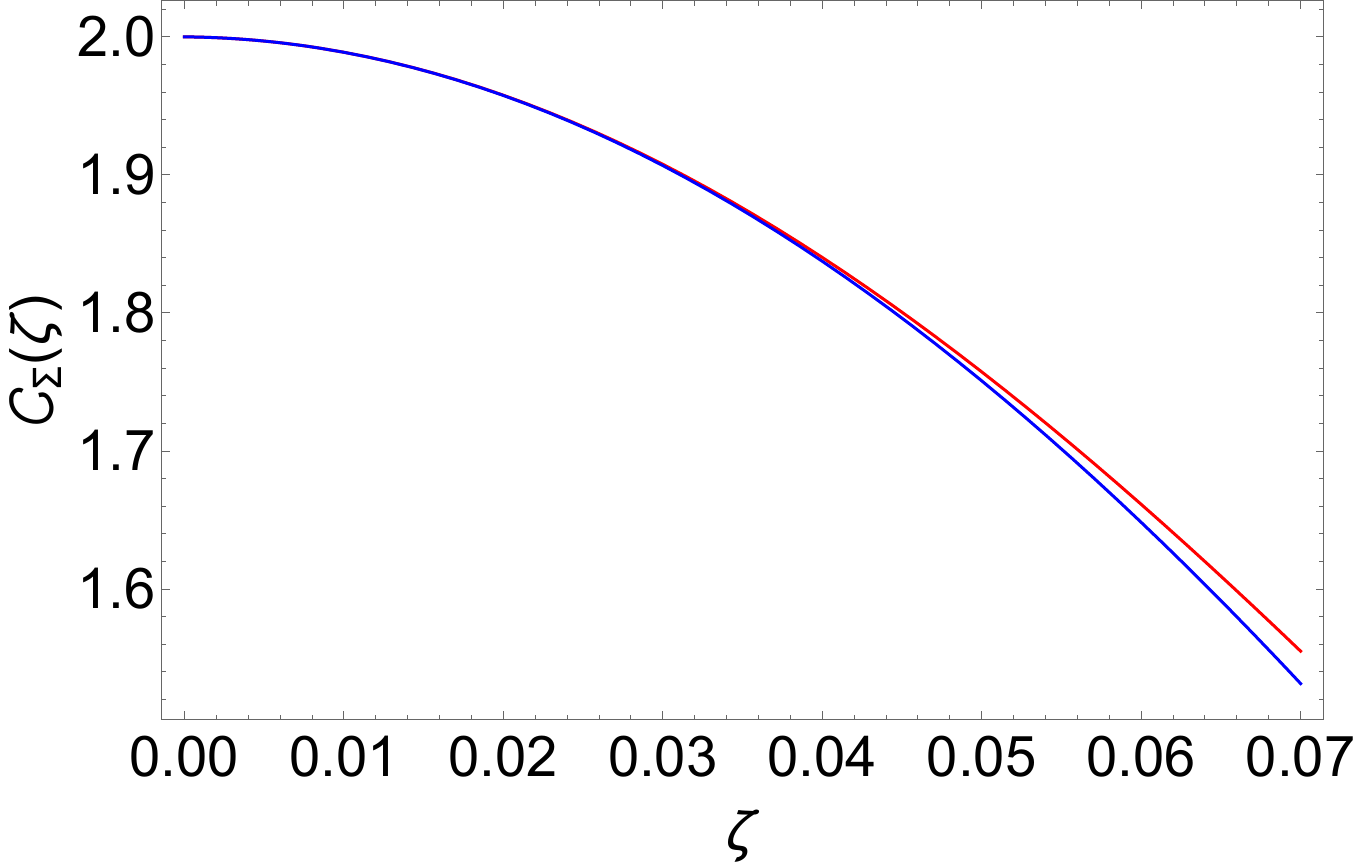}
	\caption{ Comparison between the iteration solution of Eq. (\ref{eq:49}) and the solution of the full differential equation (\ref{eq:46}) near the UV boundary $\zeta \sim 0$. }
	\label{num}
\end{figure}

To justify this solution, we compare it with the solution of the full differential equation (\ref{eq:46}) near the UV boundary $\zeta \sim 0$. The blue curve in Fig.~\ref{num} is the plot of Eq. (\ref{eq:52_1}), obtained by the iteration method. The red curve in Fig.~\ref{num} is the numerical result of Eq. (\ref{eq:46}). Here, we check out this comparison with some arbitrary UV boundary values, $\mathit{x}_0=2$, $\mathit{y}_0=0.2$, $\mathit{z}_0=-0.3$, $\mathit{x}_{\text{id}}=-0.1$, $\mathit{y}_{\text{id}}=2$, $\mathit{z}_{\text{id}}=3$, $\alpha =0.1$, $\Lambda =10$, and $a_0=-0.001$. As $\zeta \rightarrow 0$, two curves become identical. One can also confirm that such two methods give idential solutions in the IR limit $\zeta \rightarrow 1$, not shown here.

The UV boundary conditions given by Eqs. (\ref{Conformal_UV_BC_Sigma})--(\ref{Conformal_UV_BC_Pi}) result in
%
%
\bqa
\mathit{x}_{\text{id}}=-\Lambda\frac{2 \mathit{x}_0 \left(4 \mathit{y}_0^2-\mathit{z}_0\right)}{\alpha  a_0 \mathit{x}_0^2-8 \mathit{y}_0^2+2 \mathit{z}_0}, ~~~~~ \mathit{y}_{\text{id}} = \Lambda \frac{a_0 \mathit{x}_0}{\alpha  a_0 \mathit{x}_0^2-8 \mathit{y}_0^2+2 \mathit{z}_0}, ~~~~~ \mathit{z}_{\text{id}} = - \Lambda \frac{4 a_0 \mathit{x}_0 \mathit{y}_0}{\alpha  a_0 \mathit{x}_0^2-8 \mathit{y}_0^2+2 \mathit{z}_0} . \nonumber
\eqa
%
%
Introducing these results into Eqs. (\ref{eq:52_3})--(\ref{eq:54}), we obtain the UV solution as   
	\bqa && X_{uv}(\zeta )= \mathit{x}_0-F_0 \log \left(F_2 \sinh \left(\zeta  F_1 \Lambda \right)+\cosh \left(\zeta  F_1 \Lambda \right)\right),  \label{eq:58}\\
	&&Y_{uv}(\zeta )=\mathit{y}_0-\Lambda  F_3  \zeta    -F_4 \log \left(\cosh \left(\zeta  F_5 \Lambda \right)-F_6 \sinh \left(\zeta  F_5 \Lambda \right)\right)  ,  \label{eq:59}\\
	&&  Z_{uv}(\zeta )=   \mathit{z}_0+ F_7 \Lambda  \zeta -F_8 \left(1-e^{\zeta  F_9 \Lambda }\right) , \label{eq:60}
	\eqa
where all coefficients are given by 
	\bqa && F_0=\frac{\mathit{x}_0 \left(\alpha  a_0 \mathit{x}_0^2-8 \mathit{y}_0^2+2 \mathit{z}_0\right)}{2 \left(\mathit{z}_0-4 \mathit{y}_0^2\right)}, ~~~ F_1=\frac{2 \sqrt{\mathit{x}_0^2 \left(\mathit{z}_0-4 \mathit{y}_0^2\right) \left(\frac{6 \alpha  a_0^2 \mathit{x}_0^3 \mathit{y}_0}{\left(\alpha  a_0 \mathit{x}_0^2-8 \mathit{y}_0^2+2 \mathit{z}_0\right)^2}-4 \mathit{y}_0^2+\mathit{z}_0\right)}}{\mathit{x}_0 \left(\alpha  a_0 \mathit{x}_0^2-8 \mathit{y}_0^2+2 \mathit{z}_0\right)},\nonumber\\
 && F_2=\frac{\left(2 \mathit{x}_0 \left(4 \mathit{y}_0^2-\mathit{z}_0\right)\right) \sqrt{\frac{\mathit{z}_0-4 \mathit{y}_0^2}{\mathit{x}_0^2 \left(\frac{6 \alpha  a_0^2 \mathit{x}_0^3 \mathit{y}_0}{\left(\alpha  a_0 \mathit{x}_0^2-8 \mathit{y}_0^2+2 \mathit{z}_0\right)^2}-4 \mathit{y}_0^2+\mathit{z}_0\right)}}}{\alpha  a_0 \mathit{x}_0^2-8 \mathit{y}_0^2+2 \mathit{z}_0}, \nonumber\\
	&&   F_3=\frac{a_0 \mathit{x}_0}{\alpha  a_0 \mathit{x}_0^2-8 \mathit{y}_0^2+2 \mathit{z}_0}, ~~~ F_4=\frac{\alpha  a_0 \mathit{x}_0^2-8 \mathit{y}_0^2+2 \mathit{z}_0}{4 \mathit{y}_0}, ~~~ F_5=\frac{2 \sqrt{a_0^2 \mathit{x}_0^2 \mathit{y}_0 \left(\alpha ^2 a_0 \mathit{x}_0^3+4 \left(-4 \alpha  \mathit{x}_0 \mathit{y}_0^2+\alpha  \mathit{x}_0 \mathit{z}_0+\mathit{y}_0\right)\right)}}{\left(\alpha  a_0 \mathit{x}_0^2-8 \mathit{y}_0^2+2 \mathit{z}_0\right)^2}, \nonumber\\
	&&   F_6=4 a_0 \mathit{x}_0 \mathit{y}_0 \sqrt{\frac{1}{a_0^2 \mathit{x}_0^2 \mathit{y}_0 \left(\alpha ^2 a_0 \mathit{x}_0^3+4 \left(-4 \alpha  \mathit{x}_0 \mathit{y}_0^2+\alpha  \mathit{x}_0 \mathit{z}_0+\mathit{y}_0\right)\right)}}, \nonumber \\
 && F_7=\frac{a_0 \mathit{x}_0 \left(\alpha  a_0 \mathit{x}_0^2 \left(2 \alpha  \mathit{x}_0 \mathit{y}_0-1\right)-2 \left(4 \alpha  \mathit{x}_0 \mathit{y}_0 \left(4 \mathit{y}_0^2-\mathit{z}_0\right)+\mathit{z}_0\right)\right)}{4 \mathit{y}_0 \left(-\alpha  a_0 \mathit{x}_0^2+8 \mathit{y}_0^2-2 \mathit{z}_0\right)},\nonumber\\
	&&  F_8=\frac{\left(-\alpha  a_0 \mathit{x}_0^2+8 \mathit{y}_0^2-2 \mathit{z}_0\right) \left(\alpha  a_0 \mathit{x}_0^2 \left(1-2 \alpha  \mathit{x}_0 \mathit{y}_0\right)+2 \left(16 \alpha  \mathit{x}_0 \mathit{y}_0^3-4 \alpha  \mathit{x}_0 \mathit{y}_0 \mathit{z}_0+8 \mathit{y}_0^2+\mathit{z}_0\right)\right)}{32 \mathit{y}_0^2}, ~~~ F_9=\frac{8 a_0 \mathit{x}_0 \mathit{y}_0}{\left(\alpha  a_0 \mathit{x}_0^2-8 \mathit{y}_0^2+2 \mathit{z}_0\right)^2} . \nonumber
	\eqa

The IR boundary conditions given by Eqs. (\ref{Conformal_IR_BC_Sigma})--(\ref{Conformal_IR_BC_Pi}) result in
\begin{footnotesize}
\bqa
\mathit{x}_{\text{fd}} &=& \Lambda \frac{\mathit{x}_f \left(4 \mathit{y}_f^2-\mathit{z}_f\right) \left(a_0 \mathit{x}_f \left(16 \alpha  \mathit{x}_f \mathit{y}_f^4-4 \alpha  \mathit{x}_f \mathit{y}_f^2 \left(\mathit{z}_f+3\right)+\alpha  \mathit{x}_f \mathit{z}_f+8 \mathit{y}_f \left(\mathit{z}_f+2\right)-32 \mathit{y}_f^3\right)+2 \left(4 \mathit{y}_f^2-\mathit{z}_f\right) \left(2 \mathit{y}_f^2 \left(\alpha  \mathit{x}_f \mathit{y}_f-4\right)+\mathit{z}_f \left(2-\alpha  \mathit{x}_f \mathit{y}_f\right)\right)\right)}{\left(2 a_0 \mathit{x}_f \mathit{y}_f \left(4 \mathit{y}_f^2-\mathit{z}_f-2\right)+\left(\mathit{z}_f-4 \mathit{y}_f^2\right)^2\right) \left(\alpha  a_0 \mathit{x}_f^2-8 \mathit{y}_f^2+2 \mathit{z}_f\right)},\nn \label{xfd} \\
\mathit{y}_{\text{fd}} &=& \Lambda \frac{-2 a_0^2 \mathit{x}_f^2 \mathit{y}_f \left(4 \mathit{y}_f^2-\mathit{z}_f-2\right) \left(\alpha  \mathit{x}_f \mathit{y}_f-2\right)-a_0 \mathit{x}_f \left(\mathit{z}_f-4 \mathit{y}_f^2\right)^2 \left(\alpha  \mathit{x}_f \mathit{y}_f-3\right)+4 \mathit{y}_f^3 \left(\mathit{z}_f-4 \mathit{y}_f^2\right)^2}{\left(2 a_0 \mathit{x}_f \mathit{y}_f \left(4 \mathit{y}_f^2-\mathit{z}_f-2\right)+\left(\mathit{z}_f-4 \mathit{y}_f^2\right)^2\right) \left(\alpha  a_0 \mathit{x}_f^2-8 \mathit{y}_f^2+2 \mathit{z}_f\right)}, \label{yfd} \\
\mathit{z}_{\text{fd}} &=& \Lambda \frac{a_0 A \mathit{x}_f \left(4 \mathit{y}_f^2-\mathit{z}_f\right)+2 a_0^2 B \mathit{x}_f^2+\alpha  a_0^3 \mathit{x}_f^4 \left(4 \mathit{y}_f^2-\mathit{z}_f\right)+2 \mathit{z}_f \left(2 \mathit{y}_f^2-\mathit{z}_f\right) \left(\mathit{z}_f-4 \mathit{y}_f^2\right){}^2}{\left(2 a_0 \mathit{x}_f \mathit{y}_f \left(4 \mathit{y}_f^2-\mathit{z}_f-2\right)+\left(\mathit{z}_f-4 \mathit{y}_f^2\right){}^2\right) \left(\alpha  a_0 \mathit{x}_f^2-8 \mathit{y}_f^2+2 \mathit{z}_f\right)} , \label{zfd}
\eqa
\end{footnotesize} 
where $A=8 \mathit{y}_f^3 \left(\alpha  \mathit{x}_f \mathit{y}_f-8\right)+2 \mathit{y}_f \mathit{z}_f \left(-3 \alpha  \mathit{x}_f \mathit{y}_f+8 \mathit{y}_f^2+2\right)+\mathit{z}_f^2 \left(\alpha  \mathit{x}_f-4 \mathit{y}_f\right)$ and $B=16 \alpha  \mathit{x}_f \mathit{y}_f^5-4 \alpha  \mathit{x}_f \mathit{y}_f^3 \left(2 \mathit{z}_f+1\right)+\alpha  \mathit{x}_f \mathit{y}_f \mathit{z}_f \left(\mathit{z}_f+2\right)+16 \mathit{y}_f^2 \left(\mathit{z}_f+1\right)-48 \mathit{y}_f^4-\mathit{z}_f^2$.
Introducing these results into Eqs. (\ref{eq:55})--(\ref{eq:57}), we obtain the IR solution, not shown here.

\subsubsection{Matching method}

To solve these three coupled nonlinear second-order differential equations, we consider the matching method as follows:
\begin{enumerate}
	\item
	$X(\zeta)$, $Y(\zeta)$, and $Z(\zeta)$: We say $X(\zeta)$ ($Y(\zeta)$ and $Z(\zeta)$) near UV boundary as $X_{uv}(\zeta)$ ($Y_{uv}(\zeta)$ and $Z_{uv}(\zeta)$) and $X(\zeta)$ ($Y(\zeta)$ and $Z(\zeta)$)  near IR boundary as $X_{ir}(\zeta)$ ($Y_{ir}(\zeta)$ and $Z_{ir}(\zeta)$). Then, we consider the matching condition at $x_{1}$ ($x_{2}$ and $x_{3}$), given by \\
 (1) $X_{uv}(x_1)=X_{ir}(x_1)$ ($Y_{uv}(x_2) = Y_{ir}(x_2)$ and $Z_{uv}(x_3) = Z_{ir}(x_3)$),\\
 (2) $\partial_{\zeta}X_{uv}(\zeta) \Big|_{\zeta = x_1} = \partial_{\zeta}X_{ir}(\zeta) \Big|_{\zeta= x_1}$ $\Big($ $\partial_{\zeta}Y_{uv}(\zeta) \Big|_{\zeta = x_2} = \partial_{\zeta}Y_{ir}(\zeta) \Big|_{\zeta = x_2}$ and $\partial_{\zeta}Z_{uv}(\zeta) \Big|_{\zeta = x_3} = \partial_{\zeta}Z_{ir}(\zeta) \Big|_{\zeta = x_3}$ $\Big)$ , and \\ (3) $\partial_{\zeta}^2 X_{uv}(\zeta) \Big|_{\zeta = x_1} = \partial_{\zeta}^2 X_{ir}(\zeta) \Big|_{\zeta= x_1}$ $\Big($ $\partial_{\zeta}^2 Y_{uv}(\zeta) \Big|_{\zeta = x_2} = \partial_{\zeta}^2 Y_{ir}(\zeta) \Big|_{\zeta = x_2}$ and $\partial_{\zeta}^2 Z_{uv}(\zeta) \Big|_{\zeta = x_3} = \partial_{\zeta}^2 Z_{ir}(\zeta) \Big|_{\zeta = x_3}$ $\Big)$.  
	\item
	Solving these three equations, we obtain $x_1$ ($x_2$ and $x_3$) and determine $x_0$ ($y_0$ and $z_0$) and  $x_f$ ($y_f$ and $z_f$) in a self-consistent way. 
		\item
In general, we find that the matching point occurs near either UV or IR boundary. Counting the power of $\Lambda$ in the four constrained conditions, $\partial_{\zeta}X_{uv}(\zeta) \Big|_{\zeta = x_1} = \partial_{\zeta}X_{ir}(\zeta) \Big|_{\zeta= x_1}$ and $\partial_{\zeta}^2 X_{uv}(\zeta) \Big|_{\zeta = x_1} = \partial_{\zeta}^2 X_{ir}(\zeta) \Big|_{\zeta= x_1}$  $\Big($ $\partial_{\zeta}Y_{uv}(\zeta) \Big|_{\zeta = x_2} = \partial_{\zeta}Y_{ir}(\zeta) \Big|_{\zeta = x_2}$ and $\partial_{\zeta}^2 Y_{uv}(\zeta) \Big|_{\zeta = x_2} = \partial_{\zeta}^2 Y_{ir}(\zeta) \Big|_{\zeta = x_2}$ $\Big)$, we obtain $\sim (z_f-4 y_f^2) \Lambda^3$ in the leading order and $\sim (z_0-4 y_0^2) \Lambda^2$ in the subleading one. As a result, we find $z_0 = 4 y_0^2$ and $z_f = 4 y_f^2$. These equations simplify the procedure significantly.
	\item
	After tedious and lengthy calculations, the three matching points are given by $x_1=\frac{\gamma _1}{\Lambda }+\frac{\gamma _2}{\Lambda^2 }$, $x_2=\frac{\gamma _3}{\Lambda^{3/2} }$, and $x_3=1-\frac{\gamma _4}{\Lambda }- \frac{\gamma _5}{\Lambda^2 }$, respectively. Here, we obtain $\gamma _i \sim \mathcal{O}(1)$ to be specified below.
\end{enumerate}

Taking into account $z_0= 4 y_0^2$ and $z_f= 4 y_f^2$, we simplify both UV and IR solutions (Eqs. (\ref{eq:55})--(\ref{eq:57}) \& Eqs. (\ref{eq:58})--(\ref{eq:60})). The UV solution is 
	\bqa && X_{uv}(\zeta )= \mathit{x}_0-\frac{6  \Lambda ^2 \mathit{y}_0}{\alpha ^2 a_0 \mathit{x}_0^2} \zeta ^2 ,  \label{eq:61}\\
	&&Y_{uv}(\zeta )= \mathit{y}_0-\frac{\zeta  \Lambda }{\alpha  \mathit{x}_0} -\frac{ \alpha  a_0 \mathit{x}_0^2 }{4 \mathit{y}_0} \log (\cosh (\zeta  T)-R \sinh (\zeta  T))  ,  \label{eq:62}\\
	&&  Z_{uv}(\zeta )=   \frac{\alpha  a_0 \mathit{x}_0^2 \left(24 \mathit{y}_0^2-\alpha  a_0 \mathit{x}_0^2 \left(2 \alpha  \mathit{x}_0 \mathit{y}_0-1\right)\right)  }{32 \mathit{y}_0^2}\left(1-e^{\frac{8 \zeta  \Lambda  \mathit{y}_0}{\alpha ^2 a_0 \mathit{x}_0^3}}\right)+\frac{a_0 \zeta  \Lambda  \mathit{x}_0 \left(1-2 \alpha  \mathit{x}_0 \mathit{y}_0\right)}{4 \mathit{y}_0}+\frac{2 \zeta  \Lambda  \mathit{y}_0}{\alpha  \mathit{x}_0}+4 \mathit{y}_0^2 , \label{eq:63}
	\eqa
where
	\bqa T=\frac{2 \Lambda  \sqrt{a_0^2 \mathit{x}_0^2 \mathit{y}_0 \left(\alpha ^2 a_0 \mathit{x}_0^3+4 \mathit{y}_0\right)}}{\alpha ^2 a_0^2 \mathit{x}_0^4}, ~~~~~ R=4 a_0 \mathit{x}_0 \mathit{y}_0 \sqrt{\frac{1}{\alpha ^2 a_0^3 \mathit{x}_0^5 \mathit{y}_0+4 a_0^2 \mathit{x}_0^2 \mathit{y}_0^2}} . \nonumber
	\eqa
The IR solution is 
	\bqa && X_{ir}(\zeta ) \approx \frac{   12 \mathit{y}_f \left(\alpha  \Lambda  \mathfrak{X}_d \mathit{y}_f-2\right) }{\alpha ^2 a_0 \mathfrak{X}_d^2}(1-\zeta )^2+\Lambda  \mathfrak{X}_d,  \label{eq:64}\\
	&&Y_{ir}(\zeta ) \approx \mathit{y}_f +\frac{ \alpha  \Lambda  \mathfrak{X}_d \mathit{y}_f+4 }{2 \alpha  \mathfrak{X}_d}(1-\zeta ) -\frac{ \alpha  a_0 \Lambda ^2 \mathfrak{X}_d^2 }{4 \mathit{y}_f} \log (\cosh ((1-\zeta ) V)-W \sinh ((1-\zeta ) V))  ,  \label{eq:65}\\
	&&  Z_{ir}(\zeta ) \approx G \left(1-\exp \left(\frac{  8 \mathit{y}_f \left(\alpha  \Lambda  \mathfrak{X}_d \mathit{y}_f-2\right) }{\alpha ^2 a_0 \Lambda ^2 \mathfrak{X}_d^3}(1-\zeta ) \right)\right)+4 \mathit{y}_f^2+(1-\zeta ) Q .  \label{eq:66}
	\eqa
Here, we introduced $\mathit{x}_f=\Lambda  \mathfrak{X}_d$ for convenience. In these expressions, $V$, $W$, $Q$, and $G$ are given by
	\bqa V&=&2 \sqrt{\frac{\mathit{y}_f \left(\alpha ^2 a_0 \Lambda ^3 \mathfrak{X}_d^3+\alpha ^2 \Lambda ^2 \mathfrak{X}_d^2 \mathit{y}_f^3+8 \alpha  \Lambda  \mathfrak{X}_d \mathit{y}_f^2+16 \mathit{y}_f\right)}{\alpha ^4 a_0^2 \Lambda ^4 \mathfrak{X}_d^6}}, \nonumber\\ W&=&\frac{\mathit{y}_f \left(\alpha  \Lambda  \mathfrak{X}_d \mathit{y}_f-8\right) \sqrt{\frac{\alpha ^4 a_0^2 \Lambda ^4 \mathfrak{X}_d^6}{\alpha ^2 a_0 \Lambda ^3 \mathfrak{X}_d^3 \mathit{y}_f+4 \alpha ^2 \Lambda ^2 \mathfrak{X}_d^2 \mathit{y}_f^4+32 \alpha  \Lambda  \mathfrak{X}_d \mathit{y}_f^3+64 \mathit{y}_f^2}}}{\alpha ^2 a_0 \Lambda ^2 \mathfrak{X}_d^3},\nonumber\\
	Q &=&\frac{a_0 \Lambda ^2 \mathfrak{X}_d \left(\alpha  \Lambda  \mathfrak{X}_d \mathit{y}_f \left(\alpha  \Lambda  \mathfrak{X}_d \mathit{y}_f-6\right)+4\right)}{4 \mathit{y}_f \left(\alpha  \Lambda  \mathfrak{X}_d \mathit{y}_f-2\right)}+2 \mathit{y}_f \left(\Lambda  \mathit{y}_f-\frac{2}{\alpha  \mathfrak{X}_d}\right), \nonumber\\
  G &=&\frac{\alpha  a_0 \Lambda ^2 \mathfrak{X}_d^2 \left(\frac{\alpha  a_0 \Lambda ^2 \mathfrak{X}_d^2 \left(\alpha  \Lambda  \mathfrak{X}_d \mathit{y}_f \left(\alpha  \Lambda  \mathfrak{X}_d \mathit{y}_f-6\right)+4\right)}{\mathit{y}_f^2}-48 \alpha  \Lambda  \mathfrak{X}_d \mathit{y}_f+96\right)}{32 \left(\alpha  \Lambda  \mathfrak{X}_d \mathit{y}_f-2\right)^2} , \nonumber
	\eqa
respectively. 

After lengthy but straightforward computations at sufficiently large $\Lambda$, we obtain the UV solution
	\bqa &&  \mathit{x}_0=x_p+\frac{\epsilon _0}{\Lambda }, ~~~~~ \mathit{y}_0=\frac{1}{8} \left(\sqrt{\alpha ^4 a_0^2 x_0^6-8 \alpha  a_0 x_0^2}+\alpha ^2 a_0 x_0^3\right), ~~~~~ \mathit{z}_0=4 \mathit{y}_0^2, \label{eq:67} \\
	&& x_p=\frac{2 \sqrt[4]{3}}{\alpha ^{3/4} \sqrt[4]{-a_0}}-\frac{\sqrt{41722 \sqrt{42}+484599}}{   4 \sqrt{\Lambda } 3^{3/4} \sqrt{34}   \alpha ^{3/4} \sqrt[4]{-a_0} }, \\ && \epsilon _0=\frac{3 \alpha ^{13/2} a_0^3 x_0^9+10 \sqrt{3} \sqrt{-a_0} \sqrt{\alpha  a_0 x_0^2 \left(\alpha ^3 a_0 x_0^4-8\right)}+20 \sqrt{3} \alpha ^2 \left(-a_0\right)^{3/2} x_0^3+3 \alpha ^{7/2} a_0^2 x_0^5 \left(\alpha  x_0 \sqrt{\alpha  a_0 x_0^2 \left(\alpha ^3 a_0 x_0^4-8\right)}+8\right)}{12 \sqrt{\alpha } a_0 \left(3 \alpha ^3 a_0 x_0^4+8\right)},\nn \label{eq:68}
	\eqa
and the IR solution
	\bqa 
	 && \mathit{x}_f=\Lambda  \mathfrak{X}_d, ~~~~~ \mathit{z}_f=4 \mathit{y}_f^2, ~~~~~ \mathit{y}_f=\frac{\sqrt[4]{-a_0}}{2 \sqrt[4]{3} \sqrt[4]{\alpha }}+\frac{\delta _1}{\Lambda }, \label{eq:69} \\
	&& \mathfrak{X}_d=\frac{\sqrt{-a_0} \left(-12 \alpha ^4 \mathit{x}_0^4 \mathit{y}_f^6+4 \alpha ^2 \mathit{x}_0^2 \mathit{y}_0 \mathit{y}_f^3 \left(\alpha  \mathit{x}_0 \mathit{y}_f-6\right)+\mathit{y}_0^2 \left(\alpha ^2 \mathit{x}_0^2 \mathit{y}_f^2+4 \alpha  \mathit{x}_0 \mathit{y}_f-12\right)\right)}{\Lambda ^2 \left(16 \sqrt{3} \alpha ^{3/2} \mathit{y}_0^2 \mathit{y}_f^3\right)}+\frac{2 \sqrt{3} \mathit{y}_f}{\sqrt{\alpha } \sqrt{-a_0}}+\frac{-\frac{\alpha  \mathit{x}_0^2 \mathit{y}_f^2}{\mathit{y}_0}-\frac{1}{\alpha  \mathit{y}_f}+\frac{\mathit{x}_0}{2}}{\Lambda }, \label{eq:70}\\
	&&   \delta _1=\frac{a_0 \left(18 \alpha ^2 \sqrt{-a_0} x_0^2 y_f^3-9 \alpha  \sqrt{-a_0} x_0 \mathit{y}_0 y_f+\mathit{y}_0 \left(19 \sqrt{-a_0}+140 \sqrt{3} \sqrt{\alpha } y_f^2\right)\right)+54 \alpha  \mathit{y}_0 y_f^3 \left(8 \sqrt{-a_0} y_f-8 \sqrt{-a_0} \mathit{y}_0+\sqrt{3} \sqrt{\alpha } y_f^3\right)}{36 \left(\sqrt{3} \sqrt{\alpha } a_0 \mathit{y}_0 y_f-12 \alpha  \sqrt{-a_0} \mathit{y}_0 y_f^3\right)} ,\nn \label{eq:71}
	\eqa
respectively. These two solutions meet at
	\bqa
	&& x_1=\frac{\gamma _1}{\Lambda }+ \frac{\gamma _2}{\Lambda ^2}, ~~~~~ x_2=\frac{\gamma _3}{\Lambda ^{3/2}}, ~~~~~ x_3=1-\frac{\gamma _4}{\Lambda  }-\frac{\gamma _5}{\Lambda ^2}, \label{eq:72}\\
	&& \gamma _1=\frac{2 \alpha  \mathit{x}_0^2 \mathit{y}_f^2}{\mathit{y}_0 \mathfrak{X}_d}, ~~~~~ \gamma _2=-\frac{4 \left(\alpha ^2 \mathit{x}_0^4 \mathit{y}_f^4+\mathit{x}_0^2 \mathit{y}_0 \mathit{y}_f\right)}{\mathit{y}_0^2 \mathfrak{X}_d^2}, \nonumber\\
 &&\gamma _3=-\frac{a_0 \alpha ^2 x_0^{7/2}}{24} \sqrt{\frac{6 a_0 k y_f^2-81 \sqrt{\alpha } \sqrt{-a_0} \mathit{y}_0 y_f^5 \left(\alpha  x_0^2 \left(33 y_f-32 \mathit{y}_0\right)+32 \epsilon _0\right)-\sqrt{3} a_0^2 x_0^2 \mathit{y}_0}{x_0^2 \mathit{y}_0^2 y_f^3 \left(12 \sqrt{\alpha } \sqrt{-a_0} y_f^2-\sqrt{3} a_0\right) \left(3 \alpha ^2 a_0 x_0^3-36 \mathit{y}_0\right)}} , \label{eq:73} \\
	&& \gamma _4=1, ~~~~~ \gamma _5=20-\frac{\sqrt{3} \epsilon _0 \left(\sqrt{\alpha  a_0 x_0^2 \left(\alpha ^3 a_0 x_0^4-8\right)}+2 \alpha ^2 a_0 x_0^3\right)}{\sqrt{\alpha } \sqrt{-a_0} x_0^2}, \label{eq:74}\\
	&& k= 18 \sqrt{-a_0} \alpha ^{5/2} x_0^4 y_f^3-9 \sqrt{-a_0} \alpha ^{3/2} x_0^3 \mathit{y}_0 y_f-16 \sqrt{\alpha } \sqrt{-a_0} x_0^2 \mathit{y}_0+4 \sqrt{3} \mathit{y}_0 y_f \left(2 \alpha  x_0^2 \left(8 y_f-9 \mathit{y}_0\right)+9 \epsilon _0\right). \label{eq:75}
	\eqa 

\begin{figure}[!htb] 
		
	\subfigure[]
	{ \includegraphics[width=0.42\linewidth]{sigma1.pdf}}  
	\subfigure[]
	{ \includegraphics[width=0.4\linewidth]{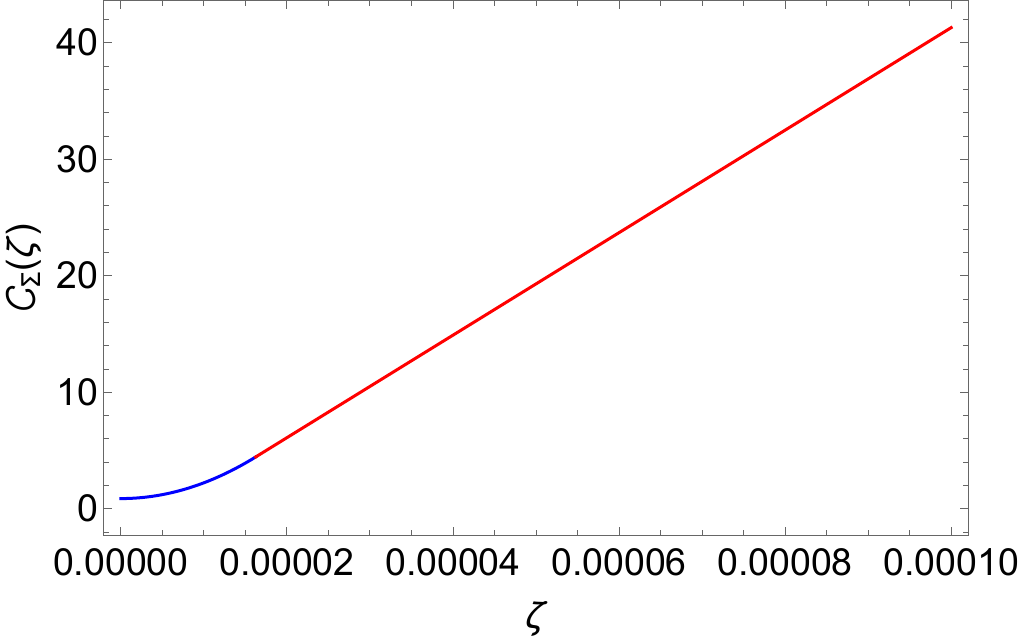}}  
	\caption{ RG flow of $\mathcal{C}_{\Sigma}(\zeta)$. (a) $\mathcal{C}_{\Sigma}(\zeta)$ vs. $\zeta$ overall. (b) $\mathcal{C}_{\Sigma}(\zeta)$ vs. $\zeta$ zoomed near UV. Here, we set $a_0=-0.01$, $\Lambda =500000$, and $\alpha =20$.} 
	\label{sigma}
\end{figure}

\begin{figure}[!htb] 
		
	\subfigure[]
	{ \includegraphics[width=0.4\linewidth]{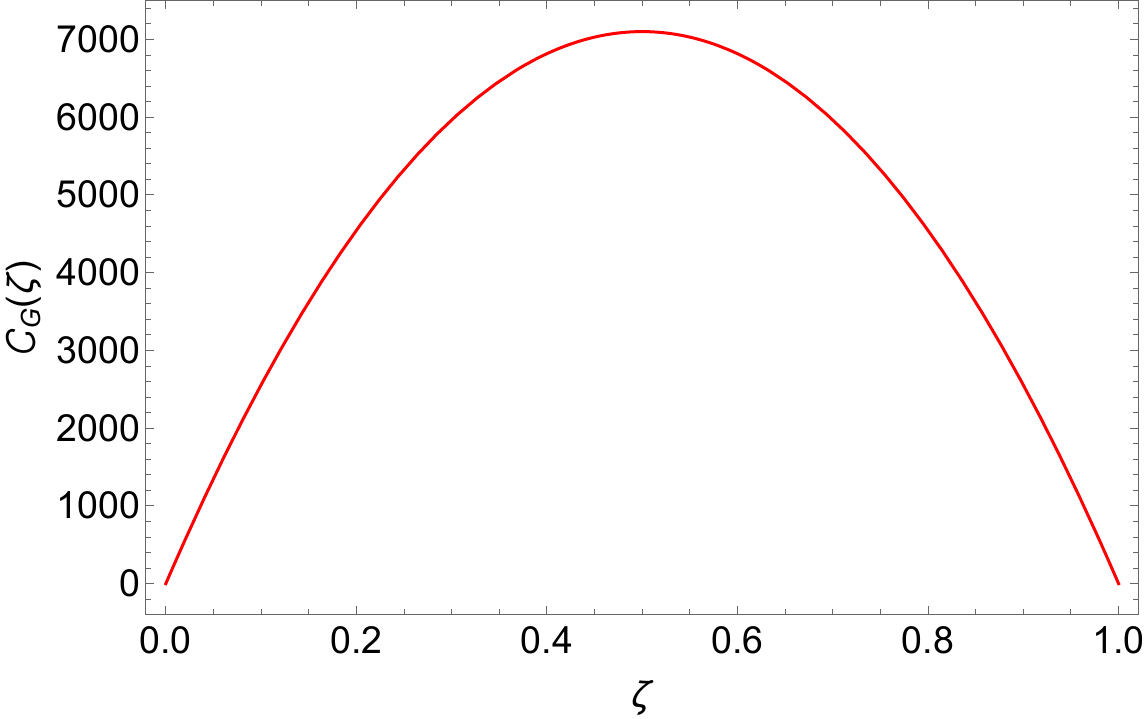}}  
	\subfigure[]
	{ \includegraphics[width=0.4\linewidth]{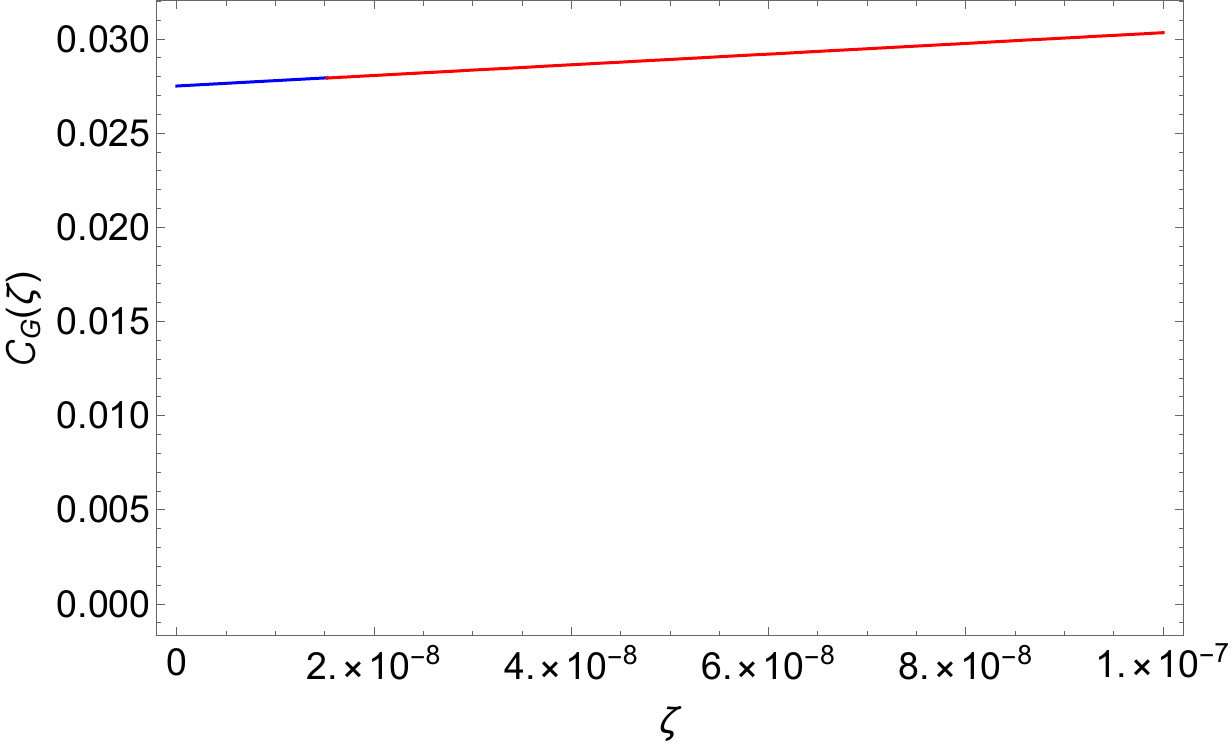}}  
	\caption{ RG flow of $\mathcal{C}_{G}(\zeta)$. (a) $\mathcal{C}_{G}(\zeta)$ vs. $\zeta$ overall. (b) $\mathcal{C}_{G}(\zeta)$ vs. $\zeta$ zoomed near UV. Here, we set $a_0=-0.01$, $\Lambda =500000$, and $\alpha =20$.} 
	\label{sigma1}
\end{figure}

\begin{figure}[!htb] 
 		
 	\subfigure[]
 	{ \includegraphics[width=0.38\linewidth]{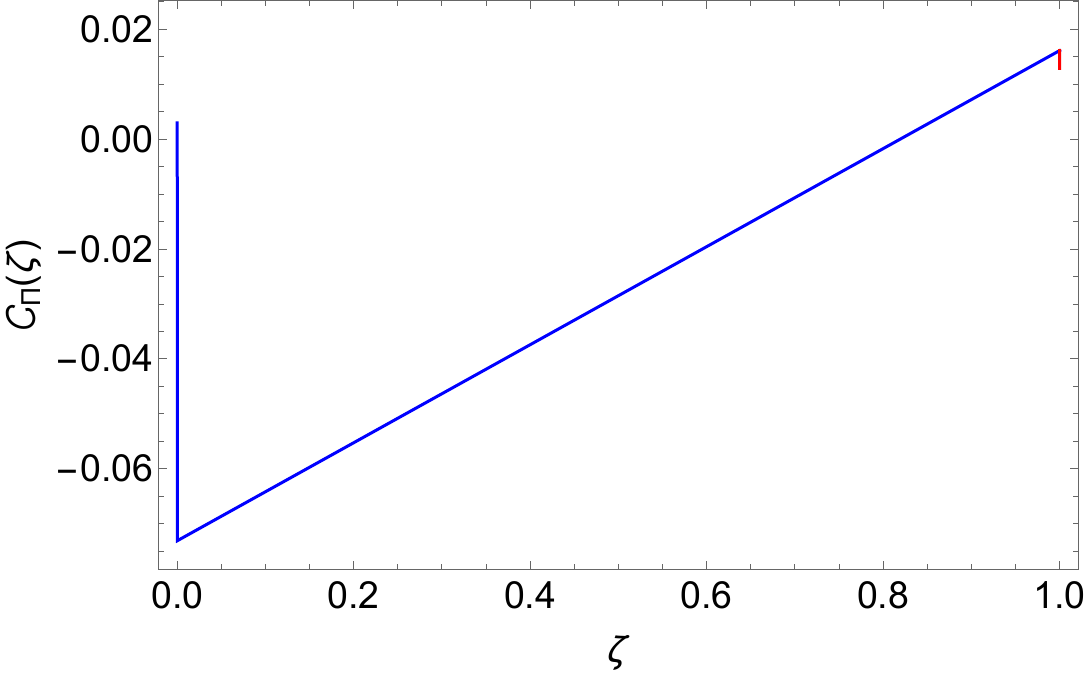}}  
 	\subfigure[]
 	{ \includegraphics[width=0.4\linewidth]{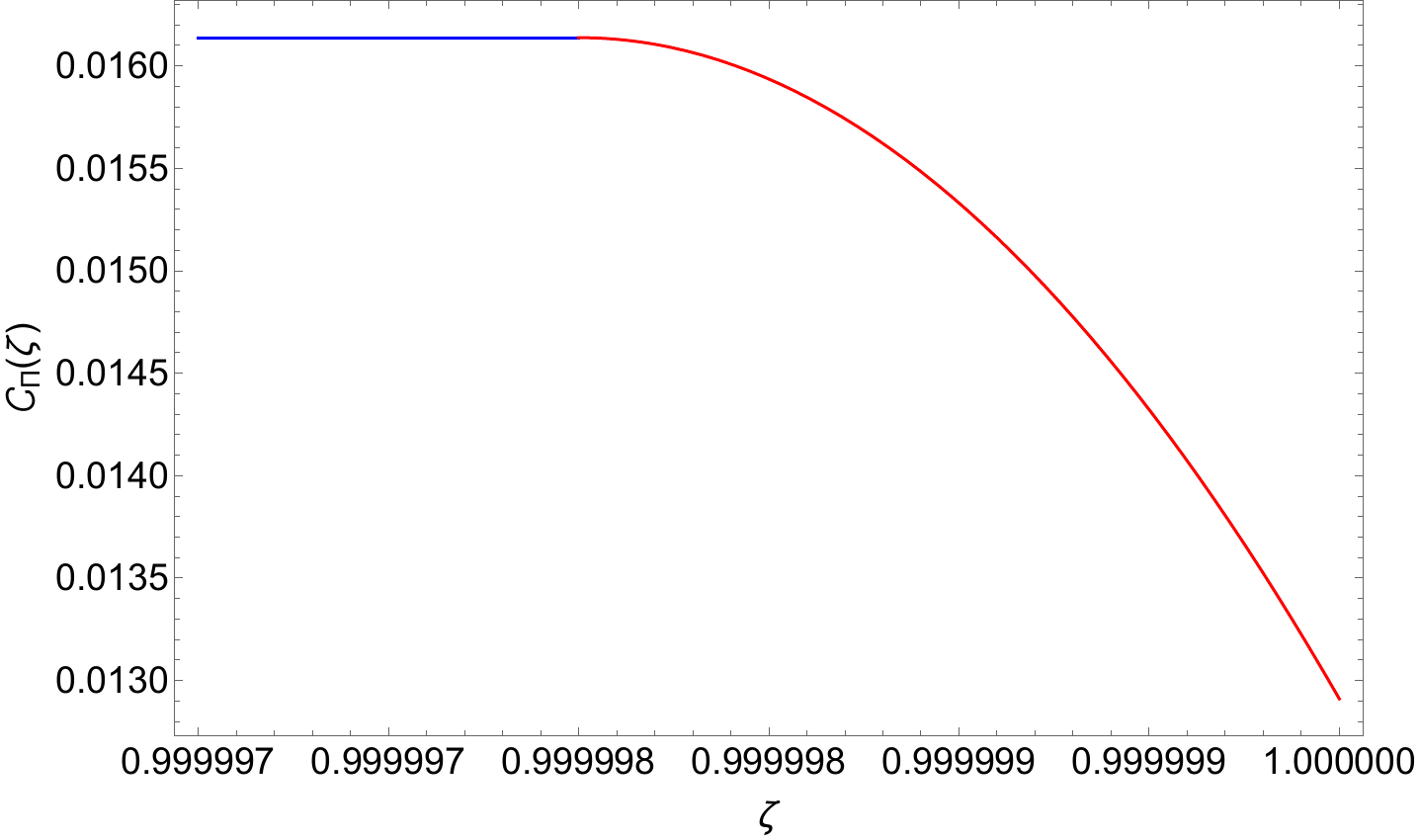}}  
 	\caption{ RG flow of $\mathcal{C}_{\Pi}(\zeta)$. (a) $\mathcal{C}_{\Pi}(\zeta)$ vs. $\zeta$ overall. (b) $\mathcal{C}_{\Pi}(\zeta)$ vs. $\zeta$ zoomed near IR. Here, we set $a_0=-0.01$, $\Lambda =500000$, and $\alpha =20$.} 
 	\label{sigma2}
 \end{figure}

Finally, we are ready to discuss the RG flows of $\mathcal{C}_{\Sigma}(\zeta)$, $\mathcal{C}_{G}(\zeta)$, and $\mathcal{C}_{\Pi}(\zeta)$. The blue curve in Fig.~\ref{sigma} (b) is a plot of the UV solution Eq. (\ref{eq:62}) while the red one is a plot of the IR solution Eq. (\ref{eq:64}). The matching point is obtained by Eq. (\ref{eq:72}), which occurs near the UV boundary. Similarly, the RG flow of $\mathcal{C}_{G}(\zeta)$ is given in Fig.~\ref{sigma1}, where the matching point is realized near the UV boundary. The RG flow of $\mathcal{C}_{\Pi}(\zeta)$ is given in Fig.~\ref{sigma2}. In this case, the matching point appears near the IR boundary. It is interesting to observe $\mathcal{C}_{\Sigma}(1) \sim \mathcal{O}(\Lambda)$ while $\mathcal{C}_{G}(1) \sim \mathcal{O}(1)$ and $\mathcal{C}_{\Pi}(1) \sim \mathcal{O}(1)$. In other words, the renormalized self-energy diverges at IR, which indicates that the conformal solution becomes unstable and an excitation gap is formed. As a result, we obtain an insulating behavior, which will be more clarified in the density of states below.

\subsection{Thermodynamics}

To verify the holographic dual effective field theory for an SYK model, we investigate thermodynamics. In particular, we show that the finite density of states due to the finite residual entropy in the field theoretic large $N$ limit vanishes in the large $N$ limit of the holographic dual effective field theory.  

It is straightforward to generalize the conformal ansatz at zero temperature to that at finite temperatures, given by \cite{SYK_Model_I,SYK_Model_II,SYK_Model_III,SYK_Model_IV,SYK_Model_V}
\bqa && \Sigma(\tau'-\tau,z) = \mathcal{C}_{\Sigma}(z) \mbox{sgn}(\tau-\tau') \Big|\frac{\pi \beta^{-1}}{\sin [\pi \beta^{-1} (\tau-\tau')]}\Big|^{\Delta_{\Sigma}} , \nn && G_{\tau'}^{\tau}(z)   = \mathcal{C}_{G}(z) \mbox{sgn}(\tau-\tau') \Big|\frac{\pi \beta^{-1}}{\sin [\pi \beta^{-1} (\tau-\tau')]}\Big|^{\Delta_{G}} , \nn &&  \Pi_{\tau'}^{\tau}(z)   = \mathcal{C}_{\Pi}(z) \Big|\frac{\pi \beta^{-1}}{\sin [\pi \beta^{-1} (\tau-\tau')]}\Big|^{\Delta_{\Pi}} . \eqa
Inserting this finite-temperature ansatz into Eqs. (\ref{Eq_Sigma}), (\ref{Eq_G}), and (\ref{Eq_Pi}), we obtain the same differential equations (\ref{C_Sigma_Old}), (\ref{C_G_Old}), and (\ref{C_Pi_Old}) for the RG flows of $\mathcal{C}_{\Sigma}(z)$, $\mathcal{C}_{G}(z)$, and $\mathcal{C}_{\Pi}(z)$ and the same UV and IR boundary conditions as those at zero temperature, where the conformal dimensions of $\Delta_{\Sigma} = \frac{3}{2}$, $\Delta_{G} = \frac{1}{2}$, and $\Delta_{\Pi} = 1$ remain unchanged. In appendix B, we confirm this statement based on our explicit calculations.

To find the free energy, we perform the Fourier transformations for all the collective bi-local fields as follows
\bqa && \Sigma(i \omega_{n},z) = \int_{0}^{\beta} d \tau e^{i \omega_{n} \tau} \Sigma(\tau,z) = \mbox{sgn}(\omega_{n}) F(i\omega_{n};\beta,\Delta_{\Sigma}) \mathcal{C}_{\Sigma}(z) , \\ && G(i \omega_{n},z) = \int_{0}^{\beta} d \tau e^{i \omega_{n} \tau} G(\tau,z) = \mbox{sgn}(\omega_{n}) F(i\omega_{n};\beta,\Delta_{G}) \mathcal{C}_{G}(z) , \\ && \Pi(i \omega_{n},z) = \int_{0}^{\beta} d \tau e^{i \omega_{n} \tau} \Pi(\tau,z) = F(i\omega_{n};\beta,\Delta_{\Pi}) \mathcal{C}_{\Pi}(z) , \eqa
where
\bqa && F(i\omega_{n};\beta,y) = \int_{0}^{\beta} d \tau e^{i \omega_{n} \tau} \mbox{sgn}(\omega_{n} \tau) \Big|\frac{\pi \beta^{-1}}{\sin [\pi \beta^{-1} \tau]}\Big|^{y} . \eqa
%
%
Here, $\omega_{n} = (2 n + 1) \pi / \beta$ ($\omega_{n} = 2 n \pi / \beta$) is the fermionic (bosonic) Matsubara frequency with integer $n$.

%
%

Introducing these Fourier-transformed representations into Eq. (\ref{HDEFT}), we obtain
\bqa && \mathcal{Z}_{SYK}(z_{f}) = \int D \mathcal{C}_{\Sigma}(z) D \mathcal{C}_{\Pi}(z) D \mathcal{C}_{G}(z) \nn && \exp\Bigg[ - N \sum_{i \omega_{n}} \Bigg\{ - \frac{1}{2} \ln \Big( q_{d} \Gamma_{J} \mbox{sgn}(\omega_{n}) F(i\omega_{n};\beta,\Delta_{\Sigma}) \mathcal{C}_{\Sigma}(z_{f}) \Big)  - q_{d} \Gamma_{J} F(i\omega_{n};\beta,\Delta_{\Sigma}) F(i\omega_{n};\beta,\Delta_{G}) \mathcal{C}_{\Sigma}(z_{f}) \mathcal{C}_{G}(z_{f})\nn && + q_{d} \Gamma_{J} F(i\omega_{n};\beta,\Delta_{\Pi}) F(i\omega_{n};\beta,2 \Delta_{G}) \mathcal{C}_{\Pi}(0) \mathcal{C}_{G}^{2}(0)  + \frac{ q_{d} \Gamma_{J} }{4} F^{2}(i\omega_{n};\beta,\Delta_{\Pi}) \mathcal{C}_{\Pi}^{2}(0) \Bigg\} \Bigg] \nn && \exp\Bigg[ - N \sum_{i \omega_{n}}  \int_{0}^{z_{f}} d z \Bigg\{ 2 q_{d} \Gamma_{J} F(i\omega_{n};\beta,\Delta_{\Pi}) F(i\omega_{n};\beta,2 \Delta_{G}) \mathcal{C}_{\Pi}(z) \mathcal{C}_{G}(z) [\partial_{z} \mathcal{C}_{G}(z)] \nn && + q_{d} \Gamma_{J} F(i\omega_{n};\beta,\Delta_{\Pi}) F(i\omega_{n};\beta,2 \Delta_{G}) \mathcal{C}_{G}^{2}(z) [\partial_{z} \mathcal{C}_{\Pi}(z)]  + \frac{ q_{d} \Gamma_{J} }{2} F^{2}(i\omega_{n};\beta,\Delta_{\Pi}) \mathcal{C}_{\Pi}(z) [\partial_{z} \mathcal{C}_{\Pi}(z)]\nn && - \frac{1}{2} [\partial_{z} \ln \mathcal{C}_{\Sigma}(z)] + \frac{1}{4} [\partial_{z} \ln \mathcal{C}_{\Sigma}(z)]^{2} - q_{d} \Gamma_{J} F(i\omega_{n};\beta,\Delta_{\Sigma}) F(i\omega_{n};\beta,\Delta_{G}) [\partial_{z} \mathcal{C}_{\Sigma}(z)] [\partial_{z} \mathcal{C}_{G}(z)] \nn && + 2 q_{d} \Gamma_{J} F(i\omega_{n};\beta,\Delta_{\Pi}) F(i\omega_{n};\beta,2 \Delta_{G}) [\partial_{z} \mathcal{C}_{\Pi}(z)] \mathcal{C}_{G}(z) [\partial_{z} \mathcal{C}_{G}(z)] + q_{d} \Gamma_{J} F(i\omega_{n};\beta,\Delta_{\Pi}) F(i\omega_{n};\beta,2 \Delta_{G}) \mathcal{C}_{\Pi}(z) [\partial_{z} \mathcal{C}_{G}(z)]^{2} \nn && + \frac{ q_{d} \Gamma_{J} }{4} F^{2}(i\omega_{n};\beta,\Delta_{\Pi}) [\partial_{z} \mathcal{C}_{\Pi}(z)]^{2} - \frac{1}{2} \ln \Big( q_{d} \Gamma_{J} \mbox{sgn}(\omega_{n}) F(i\omega_{n};\beta,\Delta_{\Sigma}) \mathcal{C}_{\Sigma}(z) \Big) \Bigg\} \Bigg] . \eqa
As discussed before, we extremize this effective action to take variations with respect to $\mathcal{C}_{\Sigma}(z)$, $\mathcal{C}_{\Pi}(z)$, and $\mathcal{C}_{G}(z)$. To make the resulting three coupled equations be satisfied, regardless of the frequency, we enforce three constraint equations, which fix the critical exponents. As discussed in appendix B, such critical exponents remain unchanged.

To find the onshell effective action, we derive the Hamiltonian formulation from the above expression as 
\bqa && \mathcal{Z}_{SYK}(z_{f}) = \int D \mathcal{C}_{\Sigma}(z) D \Pi_{\mathcal{C}_{\Sigma}}(z) D \mathcal{C}_{\Pi}(z) D  \Pi_{\mathcal{C}_{\Pi}}(z) D \mathcal{C}_{G}(z) D \Pi_{\mathcal{C}_{G}}(z)\nn && \exp\Bigg[ - N \sum_{i \omega_{n}} \Bigg\{ - \frac{1}{2} \ln \Big( q_{d} \Gamma_{J} \mbox{sgn}(\omega_{n}) F(i\omega_{n};\beta,\Delta_{\Sigma}) \mathcal{C}_{\Sigma}(z_{f}) \Big)  - q_{d} \Gamma_{J} F(i\omega_{n};\beta,\Delta_{\Sigma}) F(i\omega_{n};\beta,\Delta_{G}) \mathcal{C}_{\Sigma}(z_{f}) \mathcal{C}_{G}(z_{f}) \nn && + q_{d} \Gamma_{J} F(i\omega_{n};\beta,\Delta_{\Pi}) F(i\omega_{n};\beta,2 \Delta_{G}) \mathcal{C}_{\Pi}(z_{f}) \mathcal{C}_{G}^{2}(z_{f}) + \frac{ q_{d} \Gamma_{J} }{4} F^{2}(i\omega_{n};\beta,\Delta_{\Pi}) \mathcal{C}_{\Pi}^{2}(z_{f}) - \frac{1}{2} \ln \mathcal{C}_{\Sigma}(z_{f}) + \frac{1}{2} \ln \mathcal{C}_{\Sigma}(0) \Bigg\} \Bigg] \nn && \exp\Bigg[ - N \sum_{i \omega_{n}}  \int_{0}^{z_{f}} d z \Bigg\{ \Pi_{\mathcal{C}_{\Sigma}}(z) \partial_{z} \mathcal{C}_{\Sigma}(z) - \Bigg(\frac{1}{\mathcal{C}_{\Sigma}^{2}(z)} - \frac{q_{d} \Gamma_{J} \frac{F^{2}(i\omega_{n};\beta,\Delta_{\Sigma}) F^{2}(i\omega_{n};\beta,\Delta_{G})}{ F(i\omega_{n};\beta,\Delta_{\Pi}) F(i\omega_{n};\beta,2 \Delta_{G})}}{\mathcal{C}_{\Pi}(z) - 4 \frac{F(i\omega_{n};\beta,2 \Delta_{G})}{F(i\omega_{n};\beta,\Delta_{\Pi})} \mathcal{C}_{G}^{2}(z)} \Bigg)^{-1} \Pi_{\mathcal{C}_{\Sigma}}^{2}(z) \nn && + \Pi_{\mathcal{C}_{G}}(z) \Bigg(\partial_{z} \mathcal{C}_{G}(z) - \frac{1}{2} \frac{\frac{F(i\omega_{n};\beta,\Delta_{\Sigma}) F(i\omega_{n};\beta,\Delta_{G})}{ F(i\omega_{n};\beta,\Delta_{\Pi}) F(i\omega_{n};\beta,2 \Delta_{G})}}{\mathcal{C}_{\Pi}(z) - 4 \frac{F(i\omega_{n};\beta,2 \Delta_{G})}{F(i\omega_{n};\beta,\Delta_{\Pi})} \mathcal{C}_{G}^{2}(z)} \partial_{z} \mathcal{C}_{\Sigma}(z) \Bigg) - \frac{1}{4 q_{d} \Gamma_{J}} \frac{\Bigg( \mathcal{C}_{\Pi}(z) - 4 \frac{F(i\omega_{n};\beta,2 \Delta_{G})}{F(i\omega_{n};\beta,\Delta_{\Pi})} \mathcal{C}_{G}^{2}(z) \Bigg)^{-1}}{F(i\omega_{n};\beta,\Delta_{\Pi}) F(i\omega_{n};\beta,2 \Delta_{G})} \Pi_{\mathcal{C}_{G}}^{2}(z) \nn && + \Pi_{\mathcal{C}_{\Pi}}(z) \Bigg(\partial_{z} \mathcal{C}_{\Pi}(z) + 4 \frac{ F(i\omega_{n};\beta,2 \Delta_{G})}{F(i\omega_{n};\beta,\Delta_{\Pi})} \mathcal{C}_{G}(z) [\partial_{z} \mathcal{C}_{G}(z)] \Bigg) - \frac{ 1 }{q_{d} \Gamma_{J}} \frac{1}{F^{2}(i\omega_{n};\beta,\Delta_{\Pi})} \Pi_{\mathcal{C}_{\Pi}(z)}^{2}(z) \nn &&  - \frac{1}{2} \ln \Big( q_{d} \Gamma_{J} F(i\omega_{n};\beta,\Delta_{\Sigma}) \mathcal{C}_{\Sigma}(z) \Big) \Bigg\} \Bigg] ,  \eqa
where boundary action parts were cleaned up as discussed before. Most bulk terms do not contribute to the onshell free energy in the large $N-$limit of this holographic dual effective field theory, where equations of motion are taken into account. As a result, we obtain
\bqa && F_{eff}(\beta) = \frac{N}{\beta} \sum_{i \omega_{n}} \Bigg\{ - \frac{1}{2} \ln \Big( q_{d} \Gamma_{J} \mbox{sgn}(\omega_{n}) F(i\omega_{n};\beta,\Delta_{\Sigma}) \mathcal{C}_{\Sigma}(z_{f}) \Big) - q_{d} \Gamma_{J} F(i\omega_{n};\beta,\Delta_{\Sigma}) F(i\omega_{n};\beta,\Delta_{G}) \mathcal{C}_{\Sigma}(z_{f}) \mathcal{C}_{G}(z_{f}) \nn && + q_{d} \Gamma_{J} F(i\omega_{n};\beta,\Delta_{\Pi}) F(i\omega_{n};\beta,2 \Delta_{G}) \mathcal{C}_{\Pi}(z_{f}) \mathcal{C}_{G}^{2}(z_{f}) + \frac{ q_{d} \Gamma_{J} }{4} F^{2}(i\omega_{n};\beta,\Delta_{\Pi}) \mathcal{C}_{\Pi}^{2}(z_{f}) - \frac{1}{2} \ln \mathcal{C}_{\Sigma}(z_{f}) + \frac{1}{2} \ln \mathcal{C}_{\Sigma}(0) \Bigg\} \nn && + \frac{N}{\beta} \sum_{i \omega_{n}} \Bigg\{ \Pi_{\mathcal{C}_{\Sigma}}(z_{f}) \mathcal{C}_{\Sigma}(z_{f}) + \Pi_{\mathcal{C}_{\Pi}}(z_{f}) \Bigg( \mathcal{C}_{\Pi}(z_{f}) + 2 \frac{ F(i\omega_{n};\beta,2 \Delta_{G})}{F(i\omega_{n};\beta,\Delta_{\Pi})} \mathcal{C}_{G}^{2}(z_{f}) \Bigg)\nn && + \Pi_{\mathcal{C}_{G}}(z_{f}) \Bigg( \mathcal{C}_{G}(z_{f}) - \frac{1}{2} \frac{\frac{F(i\omega_{n};\beta,\Delta_{\Sigma}) F(i\omega_{n};\beta,\Delta_{G})}{ F(i\omega_{n};\beta,\Delta_{\Pi}) F(i\omega_{n};\beta,2 \Delta_{G})}}{\mathcal{C}_{\Pi}(z_{f}) - 4 \frac{F(i\omega_{n};\beta,2 \Delta_{G})}{F(i\omega_{n};\beta,\Delta_{\Pi})} \mathcal{C}_{G}^{2}(z_{f})} \mathcal{C}_{\Sigma}(z_{f}) \Bigg)  - \Pi_{\mathcal{C}_{\Sigma}}(0) \mathcal{C}_{\Sigma}(0) \nn && - \Pi_{\mathcal{C}_{\Pi}}(0) \Bigg( \mathcal{C}_{\Pi}(0) + 2 \frac{ F(i\omega_{n};\beta,2 \Delta_{G})}{F(i\omega_{n};\beta,\Delta_{\Pi})} \mathcal{C}_{G}^{2}(0) \Bigg) - \Pi_{\mathcal{C}_{G}}(0) \Bigg( \mathcal{C}_{G}(0) - \frac{1}{2} \frac{\frac{F(i\omega_{n};\beta,\Delta_{\Sigma}) F(i\omega_{n};\beta,\Delta_{G})}{ F(i\omega_{n};\beta,\Delta_{\Pi}) F(i\omega_{n};\beta,2 \Delta_{G})}}{\mathcal{C}_{\Pi}(0) - 4 \frac{F(i\omega_{n};\beta,2 \Delta_{G})}{F(i\omega_{n};\beta,\Delta_{\Pi})} \mathcal{C}_{G}^{2}(0)} \mathcal{C}_{\Sigma}(0) \Bigg) \Bigg\} . \eqa

The three momentums, $\Pi_{\mathcal{C}_{\Sigma}}(z_{f})$, $\Pi_{\mathcal{C}_{\Pi}}(z_{f})$, and $\Pi_{\mathcal{C}_{G}}(z_{f})$ ($\Pi_{\mathcal{C}_{\Sigma}}(0)$, $\Pi_{\mathcal{C}_{\Pi}}(0)$, and $\Pi_{\mathcal{C}_{G}}(0)$) can be represented by their conjugate fields, respectively. Recalling the IR boundary conditions,
\bqa && \Pi_{\mathcal{C}_{\Sigma}}(z_{f}) = \frac{1}{2} \Bigg(\frac{1}{\mathcal{C}_{\Sigma}^{2}(z_{f})} - \frac{q_{d} \Gamma_{J} \frac{F^{2}(i\omega_{n};\beta,\Delta_{\Sigma}) F^{2}(i\omega_{n};\beta,\Delta_{G})}{ F(i\omega_{n};\beta,\Delta_{\Pi}) F(i\omega_{n};\beta,2 \Delta_{G})}}{\mathcal{C}_{\Pi}(z_{f}) - 4 \frac{F(i\omega_{n};\beta,2 \Delta_{G})}{F(i\omega_{n};\beta,\Delta_{\Pi})} \mathcal{C}_{G}^{2}(z_{f})} \Bigg) \partial_{z_{f}} \mathcal{C}_{\Sigma}(z_{f}) , \\ && \Pi_{\mathcal{C}_{\Pi}}(z_{f}) = \frac{q_{d} \Gamma_{J}}{ 2 } F^{2}(i\omega_{n};\beta,\Delta_{\Pi}) \Bigg(\partial_{z_{f}} \mathcal{C}_{\Pi}(z_{f}) + 4 \frac{ F(i\omega_{n};\beta,2 \Delta_{G})}{F(i\omega_{n};\beta,\Delta_{\Pi})} \mathcal{C}_{G}(z_{f}) [\partial_{z_{f}} \mathcal{C}_{G}(z_{f})] \Bigg) , \\ && \Pi_{\mathcal{C}_{G}}(z_{f}) = 2 q_{d} \Gamma_{J} F(i\omega_{n};\beta,\Delta_{\Pi}) F(i\omega_{n};\beta,2 \Delta_{G}) \Bigg( \mathcal{C}_{\Pi}(z_{f}) - 4 \frac{F(i\omega_{n};\beta,2 \Delta_{G})}{F(i\omega_{n};\beta,\Delta_{\Pi})} \mathcal{C}_{G}^{2}(z_{f}) \Bigg) \nn && \quad \quad \quad \quad \quad \times \Bigg(\partial_{z_{f}} \mathcal{C}_{G}(z_{f}) - \frac{1}{2} \frac{\frac{F(i\omega_{n};\beta,\Delta_{\Sigma}) F(i\omega_{n};\beta,\Delta_{G})}{ F(i\omega_{n};\beta,\Delta_{\Pi}) F(i\omega_{n};\beta,2 \Delta_{G})}}{\mathcal{C}_{\Pi}(z_{f}) - 4 \frac{F(i\omega_{n};\beta,2 \Delta_{G})}{F(i\omega_{n};\beta,\Delta_{\Pi})} \mathcal{C}_{G}^{2}(z_{f})} \partial_{z_{f}} \mathcal{C}_{\Sigma}(z_{f}) \Bigg) , \eqa 
and the UV ones,
\bqa && \Pi_{\mathcal{C}_{\Pi}}(0) = 0 , ~~~~~ \Pi_{\mathcal{C}_{G}}(0) = 0 , ~~~~~ \Pi_{\mathcal{C}_{\Sigma}}(0) = \frac{1}{2} \frac{1}{\mathcal{C}_{\Sigma}(0)} , \eqa
we obtain the onshell effective free energy functional as follows
\bqa && F_{eff}(\beta) = \frac{N}{\beta} \sum_{i \omega_{n}} \Bigg\{ - \frac{1}{2} \ln \Big( q_{d} \Gamma_{J} \mbox{sgn}(\omega_{n}) F(i\omega_{n};\beta,\Delta_{\Sigma}) \mathcal{C}_{\Sigma}(z_{f}) \Big) - q_{d} \Gamma_{J} F(i\omega_{n};\beta,\Delta_{\Sigma}) F(i\omega_{n};\beta,\Delta_{G}) \mathcal{C}_{\Sigma}(z_{f}) \mathcal{C}_{G}(z_{f}) \nn && + q_{d} \Gamma_{J} F(i\omega_{n};\beta,\Delta_{\Pi}) F(i\omega_{n};\beta,2 \Delta_{G}) \mathcal{C}_{\Pi}(z_{f}) \mathcal{C}_{G}^{2}(z_{f}) + \frac{ q_{d} \Gamma_{J} }{4} F^{2}(i\omega_{n};\beta,\Delta_{\Pi}) \mathcal{C}_{\Pi}^{2}(z_{f}) - \frac{1}{2} \ln \mathcal{C}_{\Sigma}(z_{f}) + \frac{1}{2} \ln \mathcal{C}_{\Sigma}(0) \Bigg\} \nn && + \frac{N}{\beta} \sum_{i \omega_{n}} \Bigg\{ \frac{1}{2} \Bigg(\frac{1}{\mathcal{C}_{\Sigma}^{2}(z_{f})} - \frac{q_{d} \Gamma_{J} \frac{F^{2}(i\omega_{n};\beta,\Delta_{\Sigma}) F^{2}(i\omega_{n};\beta,\Delta_{G})}{ F(i\omega_{n};\beta,\Delta_{\Pi}) F(i\omega_{n};\beta,2 \Delta_{G})}}{\mathcal{C}_{\Pi}(z_{f}) - 4 \frac{F(i\omega_{n};\beta,2 \Delta_{G})}{F(i\omega_{n};\beta,\Delta_{\Pi})} \mathcal{C}_{G}^{2}(z_{f})} \Bigg) \mathcal{C}_{\Sigma}(z_{f}) [\partial_{z_{f}} \mathcal{C}_{\Sigma}(z_{f})] \nn && + \frac{q_{d} \Gamma_{J}}{ 2 } F^{2}(i\omega_{n};\beta,\Delta_{\Pi}) \Bigg( \mathcal{C}_{\Pi}(z_{f}) + 2 \frac{ F(i\omega_{n};\beta,2 \Delta_{G})}{F(i\omega_{n};\beta,\Delta_{\Pi})} \mathcal{C}_{G}^{2}(z_{f}) \Bigg) \Bigg(\partial_{z_{f}} \mathcal{C}_{\Pi}(z_{f}) + 4 \frac{ F(i\omega_{n};\beta,2 \Delta_{G})}{F(i\omega_{n};\beta,\Delta_{\Pi})} \mathcal{C}_{G}(z_{f}) [\partial_{z_{f}} \mathcal{C}_{G}(z_{f})] \Bigg) \nn && + 2 q_{d} \Gamma_{J} F(i\omega_{n};\beta,\Delta_{\Pi}) F(i\omega_{n};\beta,2 \Delta_{G}) \Bigg( \mathcal{C}_{\Pi}(z_{f}) - 4 \frac{F(i\omega_{n};\beta,2 \Delta_{G})}{F(i\omega_{n};\beta,\Delta_{\Pi})} \mathcal{C}_{G}^{2}(z_{f}) \Bigg) \Bigg( \mathcal{C}_{G}(z_{f}) \nn && - \frac{1}{2} \frac{\frac{F(i\omega_{n};\beta,\Delta_{\Sigma}) F(i\omega_{n};\beta,\Delta_{G})}{ F(i\omega_{n};\beta,\Delta_{\Pi}) F(i\omega_{n};\beta,2 \Delta_{G})}}{\mathcal{C}_{\Pi}(z_{f}) - 4 \frac{F(i\omega_{n};\beta,2 \Delta_{G})}{F(i\omega_{n};\beta,\Delta_{\Pi})} \mathcal{C}_{G}^{2}(z_{f})} \mathcal{C}_{\Sigma}(z_{f}) \Bigg) \Bigg(\partial_{z_{f}} \mathcal{C}_{G}(z_{f}) - \frac{1}{2} \frac{\frac{F(i\omega_{n};\beta,\Delta_{\Sigma}) F(i\omega_{n};\beta,\Delta_{G})}{ F(i\omega_{n};\beta,\Delta_{\Pi}) F(i\omega_{n};\beta,2 \Delta_{G})}}{\mathcal{C}_{\Pi}(z_{f}) - 4 \frac{F(i\omega_{n};\beta,2 \Delta_{G})}{F(i\omega_{n};\beta,\Delta_{\Pi})} \mathcal{C}_{G}^{2}(z_{f})} \partial_{z_{f}} \mathcal{C}_{\Sigma}(z_{f}) \Bigg) - \frac{1}{2} \Bigg\} . \label{HDEFT_Free_Energy} \eqa
Here, we recall the field theoretic large$-N$ free energy for comparison, given by
\bqa F_{eff}^{(0)}(\beta) = \frac{N}{\beta} \sum_{i \omega_{n}} \Bigg\{ \!\!\! && - \frac{1}{2} \ln \Big( q_{d} \Gamma_{J} \mbox{sgn}(\omega_{n}) F(i\omega_{n};\beta,\Delta_{\Sigma}) \mathcal{C}_{\Sigma}(0) \Big) - q_{d} \Gamma_{J} F(i\omega_{n};\beta,\Delta_{\Sigma}) F(i\omega_{n};\beta,\Delta_{G}) \mathcal{C}_{\Sigma}(0) \mathcal{C}_{G}(0) \nn && + q_{d} \Gamma_{J} F(i\omega_{n};\beta,\Delta_{\Pi}) F(i\omega_{n};\beta,2 \Delta_{G}) \mathcal{C}_{\Pi}(0) \mathcal{C}_{G}^{2}(0) + \frac{ q_{d} \Gamma_{J} }{4} F^{2}(i\omega_{n};\beta,\Delta_{\Pi}) \mathcal{C}_{\Pi}^{2}(0) \Bigg\} . \label{Field_Theory_Large_N_Free_Energy} \eqa

Performing the Laplace transformation for the (grand-)canonical partition function with respect to the complexified inverse temperature $\beta = \beta_{r} + i \beta_{i}$ as
\bqa && \rho(\mathcal{E}) = \frac{1}{2 \pi i} \int_{\beta_{r} - i \infty}^{\beta_{r} + i \infty} d \beta Z(\beta) e^{\beta \mathcal{E}} , \eqa
we obtain the density of states in the microcanonical ensemble as follows
\bqa && \rho(\mathcal{E}) = \frac{1}{2 \pi i} \int_{\beta_{r} - i \infty}^{\beta_{r} + i \infty} d \beta \exp\Big[\beta \Big\{ \mathcal{E} - \mathcal{F}_{eff}(\beta) \Big\} \Big] . \label{DOS_LT} \eqa
Performing the saddle-point analysis, we obtain
\bqa && \beta_{s} \Big\{ \mathcal{E} - \mathcal{F}_{eff}(\beta_{s}) \Big\} - \mathcal{S}_{eff}(\beta_{s}) = 0 , \label{Laplace_Transformation} \eqa
where the inverse temperature is given by the energy, which transforms the canonical ensemble to the microcanonical one. Here, $\mathcal{S}_{eff}(\beta_{s}) = \beta_{s}^{2} [\partial_{\beta} \mathcal{F}_{eff}(\beta)]_{\beta_{s}}$ is entropy as a function of the energy, given by $\beta_{s} = \beta_{s}(\mathcal{E})$. Expanding Eq. (\ref{DOS_LT}) with respect to the inverse temperature up to the second order around this saddle-point, we perform the Gaussian integral to obtain
\bqa && \rho(\mathcal{E}) = \exp\Big\{ \mathcal{S}_{eff}[\beta_{s}(\mathcal{E})] - \frac{1}{2} \ln \Big( \partial_{\beta} \mathcal{S}_{eff}(\beta) \Big)_{\beta = \beta_{s}(\mathcal{E})} \Big\} . \eqa

First, we consider the field theoretic large$-N$ free energy of Eq. (\ref{Field_Theory_Large_N_Free_Energy}). Then, it can be rewritten as
\bqa && F_{eff}^{(0)}(\beta) = \frac{N}{\beta} \left\{  q_{d} \Gamma_{J} \left(  \mathit{z}_0  \mathit{y}_0^2+\frac{1}{4} \mathit{z}_0^4  \right) \sum_{i \omega_{n}} F_{\beta}^2(1) - q_{d} \Gamma_{J} \mathit{x}_0  \mathit{y}_0  \sum_{i \omega_{n}} F_{\beta}(1/2)  F_{\beta}(3/2)   - \frac{1}{2}  \sum_{i \omega_{n}}  \ln \Big( q_{d} \Gamma_{J}  \mathit{x}_0  \mbox{sgn}(\omega_{n}) F_{\beta}(3/2) \Big) \right\}. \nn \label{eq:85}
\eqa
Based on the regularization method for the frequency summation, detailed discussed in appendix section B, we obtain
	\bqa && - \beta F_{eff}^{(0)}(\beta) = N \left\{ 59.777  q_{d} \Gamma_{J} \left(  \mathit{z}_0  \mathit{y}_0^2+4 \mathit{y}_0^4  \right) +  0.232424 \right\}. \label{eq:89}
	\eqa 
%
%
Resorting to Eq. (\ref{eq:67}) and Eq. (\ref{eq:69}) with Eq. (\ref{qq:18}) and Eq. (\ref{qq:20}), Eq. (\ref{eq:89}) becomes
	\bqa && -\beta F_{eff}^{(0)}(\beta) =  0.232424 N , \label{rr:4}
	\eqa
where $ 59.777  q_{d} \Gamma_{J} \left(  \mathit{z}_0  \mathit{y}_0^2+4 \mathit{y}_0^4  \right) \approx 0.0028$ is negligible. More details on the above proof can be found in appendix section C. This free energy gives the following entropy
\bqa && \mathcal{S}_{eff} = 0.232424 N. \eqa
As a result, we find a constant density of states, $\rho(\mathcal{E}) = e^{\mathcal{S}_{eff}}$, where the numerical coefficient $0.232424$ coincides with that of Ref. \cite{SYK_Model_V}. This demonstration verifies our regularization scheme of appendix section B.

Finally, we investigate the onshell effective free energy functional of Eq. (\ref{HDEFT_Free_Energy}). Resorting to the IR boundary conditions of $\mathit{x}_f = \Lambda \mathfrak{X}_d$ and $z_f = 4 y_f^2$ [Eq. (\ref{eq:69})], we rewrite Eq. (\ref{HDEFT_Free_Energy}) as
\bqa  F_{eff}(\beta) 
%
%
&=& \frac{N}{\beta} \sum_{i \omega_{n}} \Big\{   - \frac{1}{2} \ln \Big( q_{d} \Gamma_{J} \mbox{sgn}(\omega_{n}) F_{\beta}(3/2)\Lambda^2 \frac{\mathfrak{X}_d^2}{ \mathit{x}_0} \Big) - q_{d} \Gamma_{J}F_{\beta}(1/2) F_{\beta}(3/2) \Lambda  \mathfrak{X}_d \mathit{y}_f  + 8 q_{d} \Gamma_{J} F^2_{\beta}(1)  y_f^4 \Big\} \nn &+& \frac{N}{\beta} \sum_{i \omega_{n}} \Big\{ 3 q_{d} \Gamma_{J}  y_f^2 (\mathit{z}_{\text{fd}}+4\mathit{y}_f \mathit{y}_{\text{fd}}) F^2_{\beta}(1) - q_{d} \Gamma_{J} \Lambda  \mathfrak{X}_d \mathit{y}_{\text{fd}} F_{\beta}(1/2) F_{\beta}(3/2) - \frac{1}{2} \Big\}  . \label{eq:92}
 \eqa   
Considering the leading contributions of Eqs. (\ref{yfd}) and (\ref{zfd}), and inserting $\mathit{y}_{\text{fd}} = - \Lambda \mathit{y}_f$ and $\mathit{z}_{\text{fd}} = - 2 \Lambda \mathit{y}_f^2$ into this free energy, we obtain
	\bqa   F_{eff}(\beta) &=& \frac{N}{\beta} \sum_{i \omega_{n}} \Big\{ - \frac{1}{2} \ln \Big( q_{d} \Gamma_{J} \mbox{sgn}(\omega_{n}) F_{\beta}(3/2)\Lambda^2 \frac{\mathfrak{X}_d^2}{ \mathit{x}_0} \Big) - q_{d} \Gamma_{J}F_{\beta}(1/2) F_{\beta}(3/2) \Lambda  \mathfrak{X}_d \mathit{y}_f  + 8 q_{d} \Gamma_{J} F^2_{\beta}(1)  y_f^4   \Big\} \nn &+& \frac{N}{\beta} \sum_{i \omega_{n}} \Big\{ -18 q_{d} \Gamma_{J}\Lambda y_f^4 F^2_{\beta}(1) + q_{d} \Gamma_{J} \Lambda^2 \mathit{y}_f \mathfrak{X}_d   F_{\beta}(1/2) F_{\beta}(3/2) - \frac{1}{2} \Big\} . \label{jj:1}
	\eqa
Based on the regularization method for the frequency summation, detailed discussed in appendix section B, we obtain
	\bqa   F_{eff}(\beta) &&= \frac{N}{\beta } \left(-18 \Lambda  (16 \log (\beta )-43.1576) q_d \mathit{y}_f^4 \Gamma _J+8 (16 \log (\beta )-43.1576) q_d \mathit{y}_f^4 \Gamma _J-0.4824\right) .  \label{jj:10}
	\eqa 
We point out that Eqs. (\ref{eq:69})--(\ref{eq:71}) give
	\bqa && \mathit{y}_f\approx \frac{\sqrt[4]{-a_0}}{2 \sqrt[4]{3} \sqrt[4]{\alpha }} , ~~~~~ \alpha=32 \pi  q_{d} \Gamma_{J} , ~~~~~ a_0=-\frac{96 \pi }{24 \log ^2(\beta )-148 \log (\beta )+333} . \label{eq:94}
	\eqa
%
%
%
As a result, we find 
 	\bqa - \beta F_{eff}(\beta) = \frac{N (\log (\beta ) (0.48 \log (\beta )+0.75 \Lambda )-(3.63 \log (\beta )+2.02 \Lambda )+8.47)}{(\log (\beta )-6.84) \log (\beta )+15.692} ,
 	\label{eq:96}
 	\eqa
where the $q_{d} \Gamma_{J}$ dependence disappears. We emphasize that the $\ln \beta$ dependence plays an essential role in the density of states.

\begin{figure}[!htb] 
   	\subfigure[]
   	{ \includegraphics[width=0.45\linewidth]{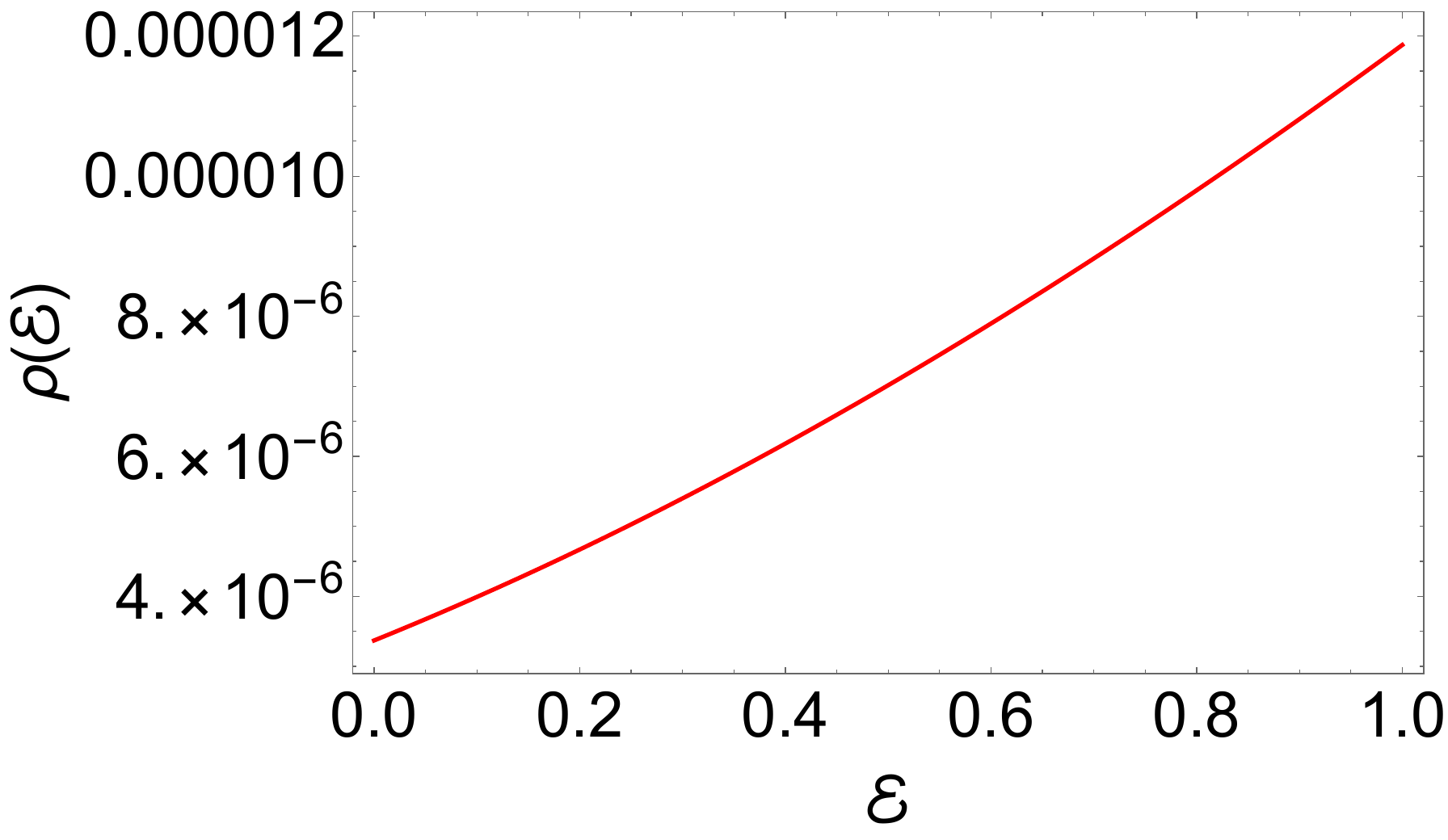}}  
   	\subfigure[]
   	{ \includegraphics[width=0.43\linewidth]{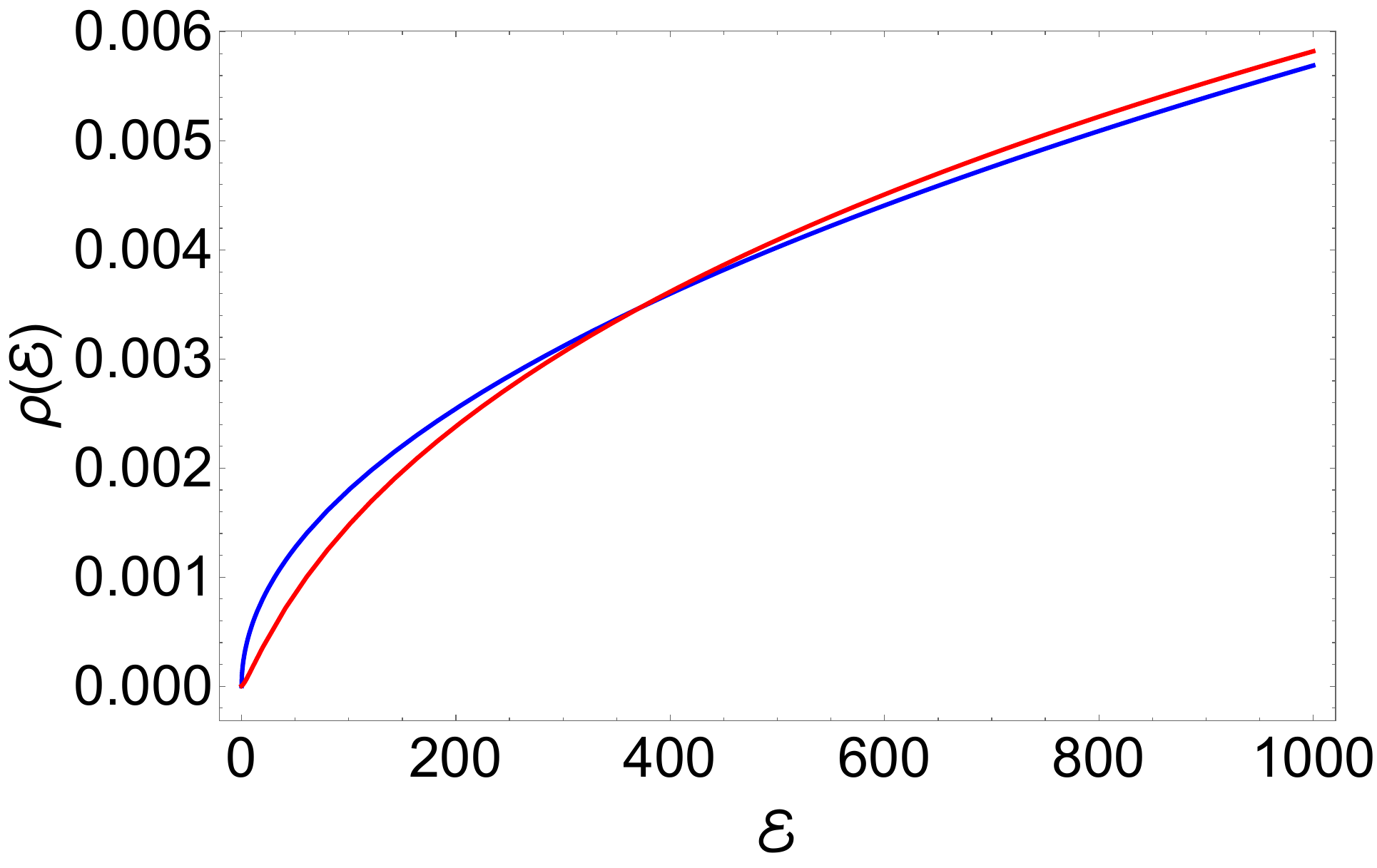}}  
   	\caption{ Vanishing density of states near $\mathcal{E} \approx 0$. Here, we consider $N=1$ and $\Lambda =100$.} 
   	\label{density}
   \end{figure}

The density of states is given by
\bqa \rho(\mathcal{E}) = e^{\mathcal{S}_{eff}[\beta_{s}(\mathcal{E})]} = \exp \left(\beta _s \mathcal{E} + \frac{N \left(\log \left(\beta _s\right) \left(0.75 \Lambda +0.48 \log \left(\beta _s\right)\right)-\left(2.02 \Lambda +3.63 \log \left(\beta _s\right)\right)+8.47\right)}{\left(\log \left(\beta _s\right)-6.84\right) \log \left(\beta _s\right)+15.692} \right) .
\label{eq:99}
\eqa 
Here, $\beta_{s}(\mathcal{E})$ is given by the Lapace transformation of Eq. (\ref{Laplace_Transformation}). This expression gives rise to the vanishing density of states as
\bqa 
\lim_{\mathcal{E} \rightarrow 0} \rho(\mathcal{E}) \approx \lim_{\Lambda \rightarrow \infty} \exp (-0.13 \Lambda  N) \longrightarrow 0 ,
\label{eq:98}
\eqa 
confirmed by Fig.~\ref{density}(a). 

One may point out that the Schwarzian path integral of the disk partition function gives $\rho(\mathcal{E}) \sim \sqrt{\mathcal{E}}$ near $\mathcal{E} = 0$ \cite{SYK_Schwarzian_I,SYK_Schwarzian_II,SYK_Schwarzian_III,SYK_Schwarzian_IV,SYK_Schwarzian_V,SYK_Schwarzian_VI,JT_GR_I,JT_GR_II,JT_GR_III,JT_GR_IV}. On the other hand, Fig.~\ref{density}(a) shows a convex behavior near $\mathcal{E} = 0$. To see a concave behavior of the density of states, we plot the density of states in a broad energy range as shown in Fig.~\ref{density}(b). Here, the red curve is the holographic density of states of Eq. (\ref{eq:99}) while the blue one is $\rho(\mathcal{E}) = 0.00018 \sqrt{\mathcal{E}}$ for comparison. Although we suspect that this convex-like behavior at extreme low energies might be related with possible non-perturbative effects of our holographic dual effective field theory, i.e., beyond $1/N$ quantum corrections, we cannot exclude that it would be a regularization artifact in the frequency summation. This point has to be more clarified near future.

%
%
  
\section{Summary and discussion}

%
%
%
%
%

Generalizing the mathematical structure of the $1/N$ expansion in an SYK model, we constructed a holographic dual effective field theory in terms of the three bi-local order-parameter fields, the self-energy collective field $\Sigma_{\tau}^{\tau'}(z) $, the Green's function collective field $G_{\tau}^{\tau'}(z)  $, and the polarization function collective field $\Pi_{\tau}^{\tau'}(z) $, respectively. To figure out the underlying structure of the holographic dual effective field theory, we took the UV limit and showed that it reproduces the leading $1/N$ expansion result. Beyond this UV limit, we discretized the emergent extra dimensional space and investigated how quantum corrections are introduced through the extra dimensional space. It turns out that field theoretic $1/N$, $1/N^{2}$, ... quantum corrections are resummed to form a holographic dual effective action in a `holographic' large $N$ limit. We point out that this non-perturbative nature has to be investigated more rigorously near future.

Taking the large $N$ limit in the holographic dual effective field theory, we obtained nonlinearly coupled second order bulk differential equations of motion for the three bi-local order-parameter fields, respectively. One may regard them as RG flows of the collective dual order parameter fields along the extra dimensional space. In addition, both UV and IR boundary conditions are derived self-consistently from the boundary effective action to complete the semiclassical analysis. Based on the conformal ansatz for the bi-local order-parameter fields, we obtain essentially the same solution as that of the field theoretic large $N$ limit at a given slice of the extra dimension. However, their overall coefficients $RG-flow$ along the extra dimensional space, respectively, reflecting effects of higher-order quantum corrections beyond the field theoretic large $N$ limit. 
%
%

It turns out that the resulting IR fixed point shows an insulating behavior, where the fermion self-energy diverges to cause pseudogap physics for the dynamics of Majorana fermions. To understand physical implications, we investigated thermodynamics. We derived an effective onshell action from the holographic dual field theory, identified with a renormalized free energy functional in terms of the bi-local order-parameter fields. We emphasize that this renormalized free energy functional is a solution of the Hamilton-Jacobi equation, not shown here. Based on this effective free energy functional, we obtained the thermodynamic entropy and found the density of states. The vanishing density of states indicates that the RG flows of the collective dual order-parameter fields give rise to deviation from the Bekenstein-Hawking entropy behavior of the field theoretic large $N$ limit.

An immediate question is whether our holographic dual effective field theory can describe quantum chaos \cite{Quantum_Chaos_Review} or not. Recently, the quantum chaos in the dual holography has been more deeply understood by the study of the spectral form factor \cite{Black_Hole_Random_Matrix}. It turns out that the Euclidean wormhole geometry was proposed to be responsible for the quantum chaos in the spectral form factor perspective, more precisely, the $t-$linear increase (`ramp' behavior) in the spectral form factor \cite{Euclidean_Wormhole}. Through the Euclidean wormhole, the spectra of both boundary conformal field theories are correlated, giving rise to so called the phenomena of level repulsion \cite{Sonner_QC_EFT}. To catch the quantum chaotic behavior from our holographic dual field theory, we suspect that similar quantum gravity corrections have to be introduced in the study of the spectral form factor. We leave it as a fascinating future study along this direction of the research.

\begin{acknowledgments}
K.-S. Kim was supported by the Ministry of Education, Science, and Technology (RS-2024-00337134) of the National Research Foundation of Korea (NRF) and by TJ Park Science Fellowship of the POSCO TJ Park Foundation. K.-S. Kim appreciates helpful discussions with A. Mitra, D. Mukherjee, and M. Nishida.
\end{acknowledgments}

\appendix

\section{Field theoretic analysis for an SYK model} \label{SYK_1_N_QCs}

\subsection{Large $N$ analysis in the frequency domain}

For completeness, we show the field theoretic anlysis for an SYK model, expected to clarify quantum corrections in the effective potential beyond the large $N$ analysis. We perform the Fourier transformtion in the partition function to obtain
\bqa && \mathcal{Z}_{SYK} = \int D \Sigma(i \omega_{n}) D \Pi(i \Omega_{n}) D G(i \omega_{n}) \int D \eta_{\alpha}(i \omega_{n}) \exp\Big[ - \Big\{ \sum_{i \omega_{n}} \eta_{\alpha}(-i \omega_{n}) \Big(- i \omega_{n} + q_{d} \Gamma_{J} \Sigma(i \omega_{n}) \Big) \eta_{\alpha} (i \omega_{n}) \nn && - N q_{d} \Gamma_{J} \sum_{i \omega_{n}} \Sigma(- i \omega_{n}) G(i \omega_{n}) + N q_{d} \Gamma_{J} \sum_{i \omega_{n}} \frac{1}{\beta} \sum_{i \Omega_{n}} \Pi(i \Omega_{n}) G(- i \omega_{n} - i \Omega_{n}) G(i \omega_{n}) + \frac{ N q_{d} \Gamma_{J} }{4} \sum_{i \Omega_{n}} \Pi(- i \Omega_{n}) \Pi(i \Omega_{n}) \Big\} \Big] . \nn \label{SYK_Frequency} \eqa
Taking the Gaussian integration for the Majorana fermion field, we obtain
\bqa && \mathcal{Z}_{SYK} = \int D \Sigma(i \omega_{n}) D \Pi(i \Omega_{n}) D G(i \omega_{n}) \exp\Big[ - N \Big\{ - \frac{1}{2} \sum_{i \omega_{n}} \ln \Big(- i \omega_{n} + q_{d} \Gamma_{J} \Sigma(i \omega_{n}) \Big) \nn &&- q_{d} \Gamma_{J} \sum_{i \omega_{n}} \Sigma(- i \omega_{n}) G(i \omega_{n})  + q_{d} \Gamma_{J} \sum_{i \omega_{n}} \frac{1}{\beta} \sum_{i \Omega_{n}} \Pi(i \Omega_{n}) G(- i \omega_{n} - i \Omega_{n}) G(i \omega_{n}) + \frac{ q_{d} \Gamma_{J} }{4} \sum_{i \Omega_{n}} \Pi(- i \Omega_{n}) \Pi(i \Omega_{n}) \Big\} \Big] . \eqa
In the large $N$ limit, we consider the saddle-point approximation and determine the fermion Green's function, self-energy, and polarization function as follows
\bqa && G(i \omega_{n}) = \frac{1}{2} \frac{1}{i \omega_{n} - q_{d} \Gamma_{J} \Sigma(i \omega_{n})} , \\ && \Sigma(i \omega_{n}) = 2 \frac{1}{\beta} \sum_{i \Omega_{n}} \Pi(i \Omega_{n}) G(- i \omega_{n} - i \Omega_{n}) , \\ && \Pi(i \Omega_{n}) = - 2 \frac{1}{\beta} \sum_{i \omega_{n}} G(i \omega_{n}) G(- i \omega_{n} - i \Omega_{n}) . \eqa
In the low frequency limit, one finds the conformal invariant solution, discussed in the main text.

\subsection{Beyond the large $N$ analysis in the frequency domain} 

To take effects of quantum corrections beyond the large $N$ limit, we introduce perturbations for collective dual fields as follows
\bqa && \Sigma(i \omega_{n}) \longrightarrow \Sigma(i \omega_{n}) + \delta \Sigma(i \omega_{n}) , \\ && \Pi(i \Omega_{n}) \longrightarrow \Pi(i \Omega_{n}) + \delta \Pi(i \Omega_{n}) , \\ && G(i \omega_{n}) \longrightarrow G(i \omega_{n}) + \delta G(i \omega_{n}) . \eqa 
Inserting these fluctuations into Eq. (\ref{SYK_Frequency}), we obtain
\bqa && \mathcal{Z}_{SYK}[\Sigma(i \omega_{n}), \Pi(i \Omega_{n}), G(i \omega_{n})] = \int D \delta \Sigma(i \omega_{n}) D \delta \Pi(i \Omega_{n}) D \delta G(i \omega_{n}) \int D \eta_{\alpha}(i \omega_{n}) \nn && \exp\Big[ - \Big\{ \sum_{i \omega_{n}} \eta_{\alpha}(-i \omega_{n}) \Big(- i \omega_{n} + q_{d} \Gamma_{J} \Sigma(i \omega_{n}) + q_{d} \Gamma_{J} \delta \Sigma(i \omega_{n})\Big) \eta_{\alpha} (i \omega_{n}) \nn && - N q_{d} \Gamma_{J} \sum_{i \omega_{n}} \Big(\Sigma(- i \omega_{n}) + \delta \Sigma(- i \omega_{n})\Big) \Big(G(i \omega_{n}) + \delta G(i \omega_{n})\Big) \nn && + N q_{d} \Gamma_{J} \sum_{i \omega_{n}} \frac{1}{\beta} \sum_{i \Omega_{n}} \Big(\Pi(i \Omega_{n}) + \delta \Pi(i \Omega_{n})\Big) \Big(G(- i \omega_{n} - i \Omega_{n}) + \delta G(- i \omega_{n} - i \Omega_{n})\Big) \Big(G(i \omega_{n}) + \delta G(i \omega_{n})\Big) \nn && + \frac{ N q_{d} \Gamma_{J} }{4} \sum_{i \Omega_{n}} \Big(\Pi(- i \Omega_{n}) + \delta \Pi(- i \Omega_{n})\Big) \Big(\Pi(i \Omega_{n}) + \delta \Pi(i \Omega_{n})\Big) \Big\} \Big] . \eqa 
Performing the Gaussian integration for the Majorana fermion field and expanding the resulting logarithmic term for $\delta \Sigma(i \omega_{n})$ up to the second order, we obtain
%
%
\bqa && \mathcal{Z}_{SYK}[\Sigma(i \omega_{n}), \Pi(i \Omega_{n}), G(i \omega_{n})] = \int D \delta \Sigma(i \omega_{n}) D \delta \Pi(i \Omega_{n}) D \delta G(i \omega_{n}) \exp\Big[ - N \Big\{ - \frac{1}{2} \sum_{i \omega_{n}} \ln \Big(- 2 [\mathcal{G}(i \omega_{n})]^{-1} \Big) \nn && + q_{d} \Gamma_{J} \sum_{i \omega_{n}} \mathcal{G}(i \omega_{n}) \delta \Sigma(- i \omega_{n}) + (q_{d} \Gamma_{J})^{2} \sum_{i \omega_{n}} \delta \Sigma(- i \omega_{n}) \mathcal{G}(- i \omega_{n}) \mathcal{G}(i \omega_{n}) \delta \Sigma(i \omega_{n}) - q_{d} \Gamma_{J} \sum_{i \omega_{n}} \Big(\Sigma(- i \omega_{n}) + \delta \Sigma(- i \omega_{n})\Big)\nn &&\times \Big(G(i \omega_{n}) + \delta G(i \omega_{n})\Big)  + q_{d} \Gamma_{J} \sum_{i \omega_{n}} \frac{1}{\beta} \sum_{i \Omega_{n}} \Big(\Pi(i \Omega_{n}) + \delta \Pi(i \Omega_{n})\Big) \Big(G(- i \omega_{n} - i \Omega_{n}) + \delta G(- i \omega_{n} - i \Omega_{n})\Big) \Big(G(i \omega_{n}) + \delta G(i \omega_{n})\Big) \nn && + \frac{ q_{d} \Gamma_{J} }{4} \sum_{i \Omega_{n}} \Big(\Pi(- i \Omega_{n}) + \delta \Pi(- i \Omega_{n})\Big) \Big(\Pi(i \Omega_{n}) + \delta \Pi(i \Omega_{n})\Big) \Big\} \Big] , \eqa
where $\mathcal{G}(i \omega_{n})$ is the Majorana fermion Green's function, given by
\bqa && \mathcal{G}(i \omega_{n}) = \frac{1}{2} \frac{1}{i \omega_{n} - q_{d} \Gamma_{J} \Sigma(i \omega_{n})} . \eqa 
Here, we regard $\Sigma(i \omega_{n})$, $\Pi(i \Omega_{n})$, and $G(i \omega_{n})$ as background fields, identified with collective order parameter fields dual to composites of original fermion fields and determined self-consistently after the Gaussian integrations of $\int D \delta \Sigma(i \omega_{n}) D \delta \Pi(i \Omega_{n}) D \delta G(i \omega_{n})$.

\subsection{Low energy effective action beyond the large $N$ limit in the frequency domain} 

Performing the Gaussian integral for $\int D \delta \Sigma(i \omega_{n})$, we obtain
\bqa && \mathcal{Z}_{SYK}[\Sigma(i \omega_{n}), \Pi(i \Omega_{n}), G(i \omega_{n})] = \int D \delta \Pi(i \Omega_{n}) D \delta G(i \omega_{n}) \exp\Big[ - N \Big\{ - \frac{1}{2} \sum_{i \omega_{n}} \ln \Big(- 2 [\mathcal{G}(i \omega_{n})]^{-1} \Big) \nn && + \frac{1}{2 N} \sum_{i \omega_{n}} \ln \Big( 2 (q_{d} \Gamma_{J})^{2} [\mathcal{G}(i \omega_{n})]^{2} \Big)  - \frac{1}{4} \sum_{i \omega_{n}} \Big(G(-i \omega_{n}) - \mathcal{G}(-i \omega_{n}) + \delta G(-i \omega_{n})\Big) [\mathcal{G}(i \omega_{n})]^{-2} \Big(G(i \omega_{n}) - \mathcal{G}(i \omega_{n}) + \delta G(i \omega_{n})\Big) \nn && - q_{d} \Gamma_{J} \sum_{i \omega_{n}} \Sigma(- i \omega_{n}) \Big(G(i \omega_{n}) + \delta G(i \omega_{n})\Big)  + q_{d} \Gamma_{J} \sum_{i \omega_{n}} \frac{1}{\beta} \sum_{i \Omega_{n}} \Big(\Pi(i \Omega_{n}) + \delta \Pi(i \Omega_{n})\Big) \Big(G(- i \omega_{n} - i \Omega_{n}) + \delta G(- i \omega_{n} - i \Omega_{n})\Big)\nn && \times \Big(G(i \omega_{n}) + \delta G(i \omega_{n})\Big)  + \frac{ q_{d} \Gamma_{J} }{4} \sum_{i \Omega_{n}} \Big(\Pi(- i \Omega_{n}) + \delta \Pi(- i \Omega_{n})\Big) \Big(\Pi(i \Omega_{n}) + \delta \Pi(i \Omega_{n})\Big) \Big\} \Big] . \eqa  
Here, $\frac{1}{2 N} \sum_{i \omega_{n}} \ln \Big( 2 (q_{d} \Gamma_{J})^{2} [\mathcal{G}(i \omega_{n})]^{2} \Big)$ corresponds to a $1/N$ correction, and $- \frac{1}{4} \sum_{i \omega_{n}} \Big(G(-i \omega_{n}) - \mathcal{G}(-i \omega_{n}) + \delta G(-i \omega_{n})\Big) [\mathcal{G}(i \omega_{n})]^{-2} \Big(G(i \omega_{n}) - \mathcal{G}(i \omega_{n}) + \delta G(i \omega_{n})\Big)$ gives a shift term due to the presence of the linear coupling term with $\delta \Sigma(i \omega_{n})$.

Performing the Gaussian integral for $\int D \delta G(i \omega_{n})$, we obtain
\bqa && \mathcal{Z}_{SYK}[\Sigma(i \omega_{n}), \Pi(i \Omega_{n}), G(i \omega_{n})] = \exp\Bigg[ - N \Bigg\{ - \frac{1}{2} \sum_{i \omega_{n}} \ln \Big(- 2 [\mathcal{G}(i \omega_{n})]^{-1} \Big) \nn && - q_{d} \Gamma_{J} \sum_{i \omega_{n}} \Sigma(- i \omega_{n}) G(i \omega_{n}) + q_{d} \Gamma_{J} \sum_{i \omega_{n}} \frac{1}{\beta} \sum_{i \Omega_{n}} \Pi(i \Omega_{n}) G(- i \omega_{n} - i \Omega_{n}) G(i \omega_{n}) + \frac{ q_{d} \Gamma_{J} }{4} \sum_{i \Omega_{n}} \Pi(- i \Omega_{n}) \Pi(i \Omega_{n}) \Bigg\} \Bigg] \nn && \int D \delta \Pi(i \Omega_{n}) \exp\Bigg[ - N \Bigg\{ \frac{1}{2 N} \sum_{i \omega_{n}} \ln \Big( 2 (q_{d} \Gamma_{J})^{2} [\mathcal{G}(i \omega_{n})]^{2} \Big) - \frac{1}{4} \sum_{i \omega_{n}} \Big(G(-i \omega_{n}) - \mathcal{G}(-i \omega_{n}) \Big) [\mathcal{G}(i \omega_{n})]^{-2} \Big(G(i \omega_{n}) - \mathcal{G}(i \omega_{n}) \Big) \nn && + \frac{1}{2 N} \sum_{i \omega_{n}} \frac{1}{\beta} \sum_{i \omega'_{n}} \ln \frac{1}{2} \Bigg( [\mathcal{G}(i \omega_{n})]^{-2} \delta(\omega_{n} - \omega'_{n}) - 4 q_{d} \Gamma_{J} \Big(\Pi(i \omega_{n} - i \omega'_{n}) + \delta \Pi(i \omega_{n} - i \omega'_{n})\Big) \Bigg) \nn && + \frac{1}{4} \sum_{i \omega_{n}} \frac{1}{\beta} \sum_{i \omega'_{n}} \frac{1}{\beta} \sum_{i \nu_{n}} \frac{1}{\beta} \sum_{i \nu'_{n}} \Bigg( \delta(\omega_{n} - \nu_{n}) \Big( [\mathcal{G}(i \omega_{n})]^{-2} \Big(G(i \omega_{n}) - \mathcal{G}(i \omega_{n}) \Big) + 2 q_{d} \Gamma_{J} \Sigma(i \omega_{n}) \Big) \nn && - 4 q_{d} \Gamma_{J} \Big(\Pi(i \omega_{n} - i \nu_{n}) + \delta \Pi(i \omega_{n} - i \nu_{n})\Big) \Bigg)  \Bigg( [\mathcal{G}(i \nu_{n})]^{-2} \delta(\nu_{n} - \nu'_{n}) - 4 q_{d} \Gamma_{J} \Big(\Pi(i \nu_{n} - i \nu'_{n}) + \delta \Pi(i \nu_{n} - i \nu'_{n})\Big) \Bigg)^{-1} \nn && \times \Bigg( \delta(\nu'_{n} - \omega'_{n}) \Big( [\mathcal{G}(i \omega'_{n})]^{-2} \Big(G(i \omega'_{n}) - \mathcal{G}(i \omega'_{n}) \Big) + 2 q_{d} \Gamma_{J} \Sigma(i \omega'_{n}) \Big) - 4 q_{d} \Gamma_{J} \Big(\Pi(i \nu'_{n} - i \omega'_{n}) + \delta \Pi(i \nu'_{n} - i \omega'_{n})\Big) \Bigg) \nn && + \frac{ q_{d} \Gamma_{J} }{4} \sum_{i \Omega_{n}} \delta \Pi(- i \Omega_{n}) \delta \Pi(i \Omega_{n}) + \frac{ q_{d} \Gamma_{J} }{2} \sum_{i \Omega_{n}} \delta \Pi(i \Omega_{n}) \Big( \Pi(- i \Omega_{n}) + 2 \frac{1}{\beta} \sum_{i \omega_{n}} G(- i \omega_{n} - i \Omega_{n}) G(i \omega_{n}) \Big) \Bigg\} \Bigg] , \eqa 
where the newly generated $\ln-$term corresponds to a $1/N$ correction, and the corresponding shift term of a lengthy expression follows.

To consider the Gaussian integral for $\int D \delta \Pi(i \Omega_{n})$ in the above expression, we expand both the $\ln-$term and the shift term up to the second order in $\delta \Pi(i \Omega_{n})$. Then, we obtain
\bqa && \mathcal{Z}_{SYK}[\Sigma(i \omega_{n}), \Pi(i \Omega_{n}), G(i \omega_{n})] \approx \exp\Big[ - N \Big\{ - \frac{1}{2} \sum_{i \omega_{n}} \ln \Big(- 2 [\mathcal{G}(i \omega_{n})]^{-1} \Big) \nn && - q_{d} \Gamma_{J} \sum_{i \omega_{n}} \Sigma(- i \omega_{n}) G(i \omega_{n}) + q_{d} \Gamma_{J} \sum_{i \omega_{n}} \frac{1}{\beta} \sum_{i \Omega_{n}} \Pi(i \Omega_{n}) G(- i \omega_{n} - i \Omega_{n}) G(i \omega_{n}) + \frac{ q_{d} \Gamma_{J} }{4} \sum_{i \Omega_{n}} \Pi(- i \Omega_{n}) \Pi(i \Omega_{n}) \Big\} \Big] \nn && \times \int D \delta \Pi(i \Omega_{n}) \exp\Bigg[ - N \Bigg\{ \frac{1}{2 N} \sum_{i \omega_{n}} \ln \Big( 2 (q_{d} \Gamma_{J})^{2} [\mathcal{G}(i \omega_{n})]^{2} \Big) + \frac{1}{2 N} \sum_{i \omega_{n}} \frac{1}{\beta} \sum_{i \omega'_{n}} \ln \frac{1}{2} \Bigg( [\mathcal{G}(i \omega_{n})]^{-2} \delta(\omega_{n} - \omega'_{n}) \nn &&- 4 q_{d} \Gamma_{J} \Pi(i \omega_{n} - i \omega'_{n}) \Bigg)  - \frac{1}{4} \sum_{i \omega_{n}} \Big(G(-i \omega_{n}) - \mathcal{G}(-i \omega_{n}) \Big) [\mathcal{G}(i \omega_{n})]^{-2} \Big(G(i \omega_{n}) - \mathcal{G}(i \omega_{n}) \Big) \nn && + \frac{1}{4} \sum_{i \omega_{n}} \frac{1}{\beta} \sum_{i \omega'_{n}} \frac{1}{\beta} \sum_{i \nu_{n}} \frac{1}{\beta} \sum_{i \nu'_{n}} \Bigg( \delta(\omega_{n} - \nu_{n}) \Big( [\mathcal{G}(i \omega_{n})]^{-2} \Big(G(i \omega_{n}) - \mathcal{G}(i \omega_{n}) \Big) + 2 q_{d} \Gamma_{J} \Sigma(i \omega_{n}) \Big) - 4 q_{d} \Gamma_{J} \Pi(i \omega_{n} - i \nu_{n}) \Bigg) \nn && \times \Bigg( [\mathcal{G}(i \nu_{n})]^{-2} \delta(\nu_{n} - \nu'_{n}) - 4 q_{d} \Gamma_{J} \Pi(i \nu_{n} - i \nu'_{n}) \Bigg)^{-1} \Bigg( \delta(\nu'_{n} - \omega'_{n}) \Big( [\mathcal{G}(i \omega'_{n})]^{-2} \Big(G(i \omega'_{n}) - \mathcal{G}(i \omega'_{n}) \Big) + 2 q_{d} \Gamma_{J} \Sigma(i \omega'_{n}) \Big) \nn &&- 4 q_{d} \Gamma_{J} \Pi(i \nu'_{n} - i \omega'_{n}) \Bigg) \Bigg\}  - N \Bigg\{ \frac{ q_{d} \Gamma_{J} }{4} \sum_{i \omega_{n}} \frac{1}{\beta} \sum_{i \omega'_{n}} \frac{1}{\beta} \sum_{i \nu_{n}} \frac{1}{\beta} \sum_{i \nu'_{n}} \delta \Pi(i \omega_{n} - i \nu_{n}) \Bigg( \delta(\omega_{n} - \nu_{n}) \delta(\omega'_{n} - \nu'_{n}) \nn &&+ 16 q_{d} \Gamma_{J}\Big( [\mathcal{G}(i \nu_{n})]^{-2} \delta(\nu_{n} - \nu'_{n}) - 4 q_{d} \Gamma_{J} \Pi(i \nu_{n} - i \nu'_{n}) \Big)^{-1} \Bigg)  \delta \Pi(i \nu'_{n} - i \omega'_{n}) \nn && + \frac{ q_{d} \Gamma_{J} }{2} \sum_{i \omega_{n}} \frac{1}{\beta} \sum_{i \omega'_{n}} \frac{1}{\beta} \sum_{i \nu_{n}} \frac{1}{\beta} \sum_{i \nu'_{n}} \delta \Pi(i \omega_{n} - i \nu_{n}) \Bigg( \delta(\nu'_{n} - \omega'_{n}) \Big( \Pi(i \nu_{n} - i \omega'_{n}) + 2 \frac{1}{\beta} \sum_{i \omega_{n}} G(- i \omega_{n} - i \nu_{n} + i \omega'_{n}) G(i \omega_{n}) \Big) \nn && - 4 \Big\{ [\mathcal{G}(i \nu_{n})]^{-2} \delta(\nu_{n} - \nu'_{n}) - 4 q_{d} \Gamma_{J} \Pi(i \nu_{n} - i \nu'_{n}) \Big\}^{-1} \Big\{ \delta(\nu'_{n} - \omega'_{n}) \Big( [\mathcal{G}(i \omega'_{n})]^{-2} \Big(G(i \omega'_{n}) - \mathcal{G}(i \omega'_{n}) \Big) + 2 q_{d} \Gamma_{J} \Sigma(i \omega'_{n}) \Big) \nn && - 4 q_{d} \Gamma_{J} \Pi(i \nu'_{n} - i \omega'_{n}) \Big\} \Bigg) \Bigg\} \Bigg] . \eqa
It is straightforward to take the Gaussian integral for $\int D \delta \Pi(i \Omega_{n})$, where the result is not shown here for the presentation perspective.

\section{Some regularizations and $\alpha$ $\&$ $a_{0}$}

\subsection{Fermionic case}

We consider the Fourier transformation of the fermionic conformal ansatz solution at finite temperature as 
  \bqa   F_{\beta}(\Delta) = \int_{-\beta}^{\beta} d \tau \; \mbox{sgn}(\tau) e^{i \omega \tau} \left|\frac{\frac{\pi}{\beta}}{\sin\left(\frac{\pi \tau}{\beta} \right)} \right|^{\Delta}. 
  \label{eq:4}\eqa 
With $x=\frac{\pi \tau}{\beta}$, Eq. (\ref{eq:4}) becomes   
  \bqa   F_{\beta}(\Delta) &=& \left(\frac{\pi}{\beta} \right)^{\Delta-1} \int_{-\pi}^{\pi} d x \; \frac{\mbox{sgn}(x) e^{i k x}}{\left|\sin\left(x\right) \right|^{\Delta} } =2i\left(\frac{\pi}{\beta} \right)^{\Delta-1} \int_{0}^{\pi} d x \; \frac{ \sin(kx)}{\sin\left(x\right)^{\Delta} } .   
  \label{eq:6}\eqa 
Here, $k=\frac{\omega \beta}{\pi}$.
%
%
For $x \to \pi$, we have $\frac{\sin(k x)}{\sin\left(x\right)^{\Delta}} \approx \frac{\sin(k \pi)}{(\pi -x)^{\Delta }}$. Then, we obtain $\int \frac{d x}{(\pi -x)^{\Delta }} = \frac{(\pi -x)^{1-\Delta }}{\Delta -1}  $. We  require $\Re(\Delta)<1$, otherwise, $\lim_{x \rightarrow \pi} \frac{(\pi -x)^{1-\Delta }}{\Delta -1} \rightarrow \infty$. We also have the following identity
  \bqa && \frac{1}{\sin(x)^{\Delta}}=\frac{1}{\Gamma(\Delta)} \int_{0}^{\infty}  d t\; t^{\Delta-1} \exp(-t \sin (x)  )
  \label{eq:8}\eqa 
with $\Re(\Delta)> 0$. Otherwise, the above integral does not converge at $t=0$.

  We substitute Eq. (\ref{eq:8}) into Eq. (\ref{eq:6}) and obtain
    \bqa F_{\beta}(\Delta) &=& \frac{2i}{\Gamma(\Delta)}\left(\frac{\pi}{\beta} \right)^{\Delta-1} \int_{0}^{\pi} d x \;  \sin(kx)  \int_{0}^{\infty}  d t\; t^{\Delta-1} \exp(-t \sin (x)  )\nonumber\\
   &=& \frac{2i}{\Gamma(\Delta)}\left(\frac{\pi}{\beta} \right)^{\Delta-1}  \int_{0}^{\infty}  d t\; t^{\Delta-1}  \int_{0}^{\pi} d x \;  \sin(kx)   \exp(-t \sin (x)) \nonumber\\
    &=& \frac{2i}{\Gamma(\Delta)}\left(\frac{\pi}{\beta} \right)^{\Delta-1}  \int_{0}^{\infty}  d t\; t^{\Delta-1}  \int_{0}^{\pi} d x \;  \sin(kx)  \sum_{n=0}^{\infty}  \frac{(-t)^n  \sin (x)^n}{n!} \nonumber\\
    &=& \frac{2i}{\Gamma(\Delta)}\left(\frac{\pi}{\beta} \right)^{\Delta-1}  \int_{0}^{\infty}  d t\; t^{\Delta-1} \sum_{n=0}^{\infty}  \frac{(-t)^n  }{n!}  \int_{0}^{\pi} d x \;  \sin(kx)  \sin (x)^n ,
  \label{eq:9}\eqa
where $0<\Re(\Delta)<1$.

If $n$ is even ($n=0,2,4,\cdots$), we obtain
 \bqa \int_{0}^{\pi} d x \;  \sin(kx)  \sin (x)^n = \frac{ (-1)^{\frac{n}{2}+1} n!   (\cos (\pi  k)-1) \Gamma \left(\frac{k}{2}-\frac{n}{2}\right) }{2^{n+1} \Gamma \left(\frac{k}{2}+\frac{n}{2}+1\right)}.
  \label{eq:10}\eqa
When $n$ is odd ($n=1,3,5,\cdots$), we have
  \bqa \int_{0}^{\pi} d x \;  \sin(kx)  \sin (x)^n =\frac{(-1)^{\frac{n+1}{2}} n! \sin (\pi  k) \Gamma \left(\frac{k}{2}-\frac{n}{2}\right)}{2^{n+1} \Gamma \left(\frac{k}{2}+\frac{n}{2}+1\right)}.
  \label{eq:11}\eqa

\noindent We substitute Eq. (\ref{eq:10}) and Eq. (\ref{eq:11}) into Eq. (\ref{eq:9}) and obtain
  \bqa F_{\beta}(\Delta) &=&   \frac{2i}{\Gamma(\Delta)}\left(\frac{\pi}{\beta} \right)^{\Delta-1}  \int_{0}^{\infty}  d t\; t^{\Delta-1} \sum _{n=0}^{\infty } \frac{ (-1)^{n+1} t^{2 n}  \left(\frac{(\cos (\pi  k)-1) \Gamma \left(\frac{k}{2}-n\right)}{\Gamma \left(\frac{k}{2}+n+1\right)}-\frac{t \sin (\pi  k) \Gamma \left(\frac{k-1}{2}-n\right)}{2 \Gamma \left(\frac{k+3}{2}+n\right)}\right)}{2^{2 n+1}}\nonumber\\
  &=&  \frac{i\pi}{\Gamma(\Delta)}\sin \left(\frac{\pi  k}{2}\right)\left(\frac{\pi}{\beta} \right)^{\Delta-1}  \int_{0}^{\infty}  d t\; t^{\Delta-1}     \left(\frac{2 \,\pFq[4]{1}{2}{1}{1-\frac{k}{2},\frac{k}{2}+1}{\frac{t^2}{4}}  }{\Gamma \left(1-\frac{k}{2}\right) \Gamma \left(\frac{k}{2}+1\right)}-\frac{t \, \pFq[4]{1}{2}{1}{\frac{3-k}{2},\frac{k+3}{2}}{\frac{t^2}{4}} }{\Gamma \left(\frac{3-k}{2}\right) \Gamma \left(\frac{k+3}{2}\right)}\right).
  \label{eq:12}\eqa

We recall the following identity \cite{Olve2010}
  \bqa \int \frac{z^m \, \pFq[4]{1}{2}{\mathit{a}}{\mathit{b},\mathit{c}}{z} }{\Gamma (\mathit{b}) \Gamma (\mathit{c})} \, dz= \frac{z^{m+1} \Gamma (m+1) \, \pFq[4]{2}{3}{\mathit{a}, m+1}{\mathit{b}, \mathit{c},m+2}{z} }{\Gamma (\mathit{b}) \Gamma (\mathit{c}) \Gamma (m+2)}.
  \label{eq:13}\eqa 
Utilizing this idenity, the indefinite integral in Eq. (\ref{eq:13}) reads
  \bqa 
  &&\int d t\; t^{\Delta-1} \left(\frac{2 \;  \pFq[4]{1}{2}{1}{1-\frac{k}{2}, \frac{k}{2}+1}{\frac{t^2}{4}}  }{\Gamma \left(1-\frac{k}{2}\right) \Gamma \left(\frac{k}{2}+1\right)}-\frac{t \;  \pFq[4]{1}{2}{1}{\frac{3-k}{2}, \frac{k+3}{2}}{\frac{t^2}{4}}  }{\Gamma \left(\frac{3-k}{2}\right) \Gamma \left(\frac{k+3}{2}\right)}\right) \nonumber\\
  &=& \frac{1}{2} t^{\Delta } \left(\frac{ 2 \Gamma \left(\frac{\Delta }{2}\right) \, \pFq[4]{2}{3}{1, \frac{\Delta }{2}}{1-\frac{k}{2}, \frac{k}{2}+1, \frac{\Delta }{2}+1}{\frac{t^2}{4}} }{\Gamma \left(\frac{\Delta }{2}+1\right) \Gamma \left(1-\frac{k}{2}\right) \Gamma \left(\frac{k}{2}+1\right)}-\frac{t\; \Gamma \left(\frac{\Delta +1}{2}\right) \,\pFq[4]{2}{3}{1, \frac{\Delta +1}{2}}{\frac{3-k}{2}, \frac{k+3}{2}, \frac{\Delta +3}{2}}{\frac{t^2}{4}} }{\Gamma \left(\frac{\Delta +3}{2}\right) \Gamma \left(\frac{3-k}{2}\right) \Gamma \left(\frac{k+3}{2}\right)}\right) .
  \label{eq:14}\eqa 

\noindent We recall the asymptotic formula for the $_2F_3$ hypergeometric function in the $|z|\rightarrow\infty$ limit as shown in Ref. \cite{Olve2010}:
  	\begin{eqnarray}
  		\pFq[4]{2}{3}{a_1, a_2}{b_1, b_2, b_3}{z} &=& \frac{\Gamma(b_1)\Gamma(b_2)\Gamma(b_3)}{2\sqrt{\pi}\Gamma(a_1)\Gamma(a_2)} (-z)^{\chi} \left(
  		\exp(-i(\pi \chi +2\sqrt{-z}))+\exp(i(\pi \chi +2\sqrt{-z}))+ \mathcal{O}\left(\frac{1}{\sqrt{-z}} \right)\right) \nonumber\\
  		&&+ \frac{\Gamma(b_1)\Gamma(b_2)\Gamma(b_3)\Gamma(a_2-a_1)}{\Gamma(b_1-a_1)\Gamma(b_2-a_1)\Gamma(b_3-a_1)\Gamma(a_2)}(-z)^{-a_1}\left(1+
  		\mathcal{O}\left(\frac{1}{z}  \right)   \right) \nonumber\\
  		&&+ \frac{\Gamma(b_1)\Gamma(b_2)\Gamma(b_3)\Gamma(a_1-a_2)}{\Gamma(b_1-a_2)\Gamma(b_2-a_2)\Gamma(b_3-a_2)\Gamma(a_1)}(-z)^{-a_2}\left(1+
  		\mathcal{O}\left(\frac{1}{z}  \right) \right)
  		\label{eq:15}
  	\end{eqnarray}
  at $\chi=\frac{1}{2}\left(a_1+a_2-b_1-b_2-b_3+\frac{1}{2}\right)$ and wherein the case of simple poles (i.e., $a_1- a_2 \not\in \mathbb{Z}$). 
   
Substituting Eq. (\ref{eq:15})  into Eq. (\ref{eq:14}), we have
   \bqa 
   &&\int_{0}^{\infty}\; t^{\Delta-1}     \left(\frac{2 \;  \pFq[4]{1}{2}{1}{1-\frac{k}{2}, \frac{k}{2}+1}{\frac{t^2}{4}}  }{\Gamma \left(1-\frac{k}{2}\right) \Gamma \left(\frac{k}{2}+1\right)}-\frac{t \;  \pFq[4]{1}{2}{1}{\frac{3-k}{2}, \frac{k+3}{2}}{\frac{t^2}{4}}  }{\Gamma \left(\frac{3-k}{2}\right) \Gamma \left(\frac{k+3}{2}\right)}\right) \nonumber\\
   &=& \lim_{t \to \infty} 2 \left(\frac{\pi  2^{\Delta } \csc (\pi  \Delta )}{\Gamma \left(\frac{1}{2} (k-\Delta +2)\right) \Gamma \left(-\frac{k}{2}-\frac{\Delta }{2}+1\right)}+t^{\Delta -2} \left(\frac{2 k \sin \left(\frac{\pi  k}{2}\right)}{\pi  (\Delta -2)}-\frac{2 t \cos \left(\frac{\pi  k}{2}\right)}{\pi -\pi  \Delta }+i \sqrt{\frac{2}{\pi }} e^{-t} \sqrt{t}\right)\right) \nonumber\\
   &=& \lim_{t \to \infty}  2 \left(\frac{\pi  2^{\Delta } \csc (\pi  \Delta )}{\Gamma \left(\frac{1}{2} (k-\Delta +2)\right) \Gamma \left(-\frac{k}{2}-\frac{\Delta }{2}+1\right)}+\frac{ 2 k \sin \left(\frac{\pi  k}{2}\right)  t^{\Delta -2}}{\pi  (\Delta -2)}-\frac{ 2 \cos \left(\frac{\pi  k}{2}\right)  t^{\Delta -1}}{\pi(1 - \Delta ) }\right).
   \label{eq:16}\eqa 
When $0<\Re(\Delta)<1$, Eq. (\ref{eq:16}) becomes
   \bqa 
    \int_{0}^{\infty}\; t^{\Delta-1}     \left(\frac{2 \;  \pFq[4]{1}{2}{1}{1-\frac{k}{2}, \frac{k}{2}+1}{\frac{t^2}{4}}  }{\Gamma \left(1-\frac{k}{2}\right) \Gamma \left(\frac{k}{2}+1\right)}-\frac{t \;  \pFq[4]{1}{2}{1}{\frac{3-k}{2}, \frac{k+3}{2}}{\frac{t^2}{4}}  }{\Gamma \left(\frac{3-k}{2}\right) \Gamma \left(\frac{k+3}{2}\right)}\right) = \frac{\pi  2^{\Delta +1} \csc (\pi  \Delta )}{\Gamma \left(\frac{1}{2} (k-\Delta +2)\right) \Gamma \left(-\frac{k}{2}-\frac{\Delta }{2}+1\right)}.
   \label{eq:17}\eqa 
If $1<\Re(\Delta)<2$, Eq. (\ref{eq:16}) becomes
    \bqa 
   &&\int_{0}^{\infty}\; t^{\Delta-1}     \left(\frac{2 \;  \pFq[4]{1}{2}{1}{1-\frac{k}{2}, \frac{k}{2}+1}{\frac{t^2}{4}}  }{\Gamma \left(1-\frac{k}{2}\right) \Gamma \left(\frac{k}{2}+1\right)}-\frac{t \;  \pFq[4]{1}{2}{1}{\frac{3-k}{2}, \frac{k+3}{2}}{\frac{t^2}{4}}  }{\Gamma \left(\frac{3-k}{2}\right) \Gamma \left(\frac{k+3}{2}\right)}\right) \nonumber\\
   &=& \frac{\pi  2^{\Delta +1} \csc (\pi  \Delta )}{\Gamma \left(\frac{1}{2} (k-\Delta +2)\right) \Gamma \left(-\frac{k}{2}-\frac{\Delta }{2}+1\right)}-\frac{ 4 \cos \left(\frac{\pi  k}{2}\right) }{\pi(1 - \Delta )}\lim_{t \to \infty} t^{\Delta -1}.
   \label{eq:18}\eqa 
   We substitute Eq. (\ref{eq:17}) and Eq. (\ref{eq:18}) into Eq. (\ref{eq:12}) and obtain 
  \bqa F_{\beta}(\Delta) &=&    \frac{i\pi}{\Gamma(\Delta)}\sin \left(\frac{\pi  k}{2}\right)\left(\frac{\pi}{\beta} \right)^{\Delta-1} \frac{\pi  2^{\Delta +1} \csc (\pi  \Delta )}{\Gamma \left(\frac{1}{2} (k-\Delta +2)\right) \Gamma \left(-\frac{k}{2}-\frac{\Delta }{2}+1\right)} 
   \label{eq:21}\eqa
for $0<\Re(\Delta)<1$ and
   \bqa F_{\beta}(\Delta) &=&    \frac{i\pi}{\Gamma(\Delta)}\sin \left(\frac{\pi  k}{2}\right)\left(\frac{\pi}{\beta} \right)^{\Delta-1}  \left( \frac{\pi  2^{\Delta +1} \csc (\pi  \Delta )}{\Gamma \left(\frac{1}{2} (k-\Delta +2)\right) \Gamma \left(-\frac{k}{2}-\frac{\Delta }{2}+1\right)}-\frac{ 4 \cos \left(\frac{\pi  k}{2}\right) }{\pi(1 - \Delta )}\lim_{t \to \infty} t^{\Delta -1} \right)
   \label{eq:22}\eqa
 for $1<\Re(\Delta)<2$. As a result, we obtain 
   \bqa  F_{\beta}\left(1/2 \right) = \frac{2 i \sqrt{2} \pi  \sin \left(\frac{\beta  \omega }{2}\right)}{\sqrt{\frac{1}{\beta }} \Gamma \left(\frac{3}{4}-\frac{\beta  \omega }{2 \pi }\right) \Gamma \left(\frac{\beta  \omega }{2 \pi }+\frac{3}{4}\right)},
   \label{eq:23}\eqa
   and 
   \bqa   F_{\beta}\left(3/2 \right)= -8 i \sqrt{\frac{1}{\beta }} \left(\frac{\sqrt{2} \pi ^2 \sin \left(\frac{\beta  \omega }{2}\right)}{\Gamma \left(\frac{\pi -2 \beta  \omega }{4 \pi }\right) \Gamma \left(\frac{2 \beta  \omega +\pi }{4 \pi }\right)}-  \sin (\beta  \omega )\lim_{t \to \infty} \sqrt{t} \right).
   \label{eq:24}\eqa


Putting $ \omega= \frac{\pi  (2 n+1)}{\beta }$ into  Eq. (\ref{eq:23}) and  Eq. (\ref{eq:24}) and taking the following infinite summation, we have
      \bqa   \lim_{N\to \infty }  \sum _{n=-N}^{N} F_{\beta}\left(1/2 \right)  F_{\beta}\left(3/2 \right) &&= - \lim_{N\to \infty }  \sum _{n=-N}^{N} \frac{16 \cos 2(\pi  n) \cos (2 \pi  n) \left(\pi ^2+ \lim_{t\to \infty }  \sqrt{2} \sqrt{t} \sin (\pi  n) \Gamma \left(-n-\frac{1}{4}\right) \Gamma \left(n+\frac{3}{4}\right)  \right)}{\pi } \nonumber\\
    &&  = -16 \pi  \lim_{N\to \infty } (2 N+1).
   \label{eq:27} \eqa   
The last expression will be regularized below.

\subsection{Bosonic case}

\subsubsection{A regularization for the $\log-$function near a singular point}

First, we consider
\bqa      
\log(x)=2\left(\xi +\frac{1}{3}\xi^3+ \frac{1}{5}\xi^5+\cdots \right) ,
\label{eq:32}\eqa
where $\xi=\frac{x-1}{x+1}$ and $\Re (x) >0$. Then, we have
\bqa      
\lim_{x \to \infty} \log(x)&=& 2 \sum_{n=0}^{\infty}  \frac{1}{2n+1} =2  \lim_{\epsilon \to 0}  \sum_{n=0}^{\infty}  \frac{\exp(-\epsilon n)}{2n+1}  =2  \lim_{\epsilon \to 0} \exp(\epsilon/2) \arctanh(\exp(-\epsilon/2))\nonumber\\
&=& 2\log(2) -  \lim_{\epsilon \to 0} \log(\epsilon) .
\label{eq:33}\eqa

\noindent To regularize $\lim_{\epsilon \to 0} \log(\epsilon)$, we consider
\bqa      
\log(x)=(x-1)-\frac{1}{2}(x-1)^2+ \frac{1}{3}(x-1)^3-\frac{1}{4}(x-1)^4+\cdots   ,
\label{eq:34}\eqa
where $0<x<2$. Then, we have
\bqa      
\lim_{\epsilon\to 0} \log(\epsilon)= -\lim_{N \to \infty}\sum_{n=1}^{N} \frac{1}{n}= - \zeta(1).
\label{eq:35}
\eqa

Finally, we regularize $\zeta(1)$. A functional relation of the Riemann zeta function is given by
\bqa      
\zeta(s)=2^s \pi^{s-1} \sin\left( \frac{\pi s}{2} \right) \Gamma(1-s) \zeta(1-s) ,
\label{eq:36_1}\eqa
where $0< \Re (s) <1$. Then, we have 
\bqa      
\lim_{s \to 1} \zeta(s)&= & 2 \zeta(0)  \lim_{s \to 1}  \Gamma(1-s)=-\lim_{s \to 1}  \Gamma(1-s)= \gamma - \lim_{s \to 1} \frac{1}{1-s}\nonumber\\
&=& \gamma +  \lim_{s \to 1} \sum_{n=0}^{\infty} s^n =\gamma + \zeta(0)=\gamma -\frac{1}{2} .
\label{eq:36_2}
\eqa 
Based on Eqs. (\ref{eq:35}) and  (\ref{eq:36_2}), we obtain
\bqa      \lim_{\epsilon\to 0} \log(\epsilon)= -\gamma +\frac{1}{2}. \label{eq:37}\eqa
As a result, Eq. (\ref{eq:33}) becomes  
\bqa      
\lim_{x \to \infty} \log(x) = 2\log(2)+\gamma -\frac{1}{2}.
\label{eq:38}\eqa

\subsubsection{Bosonic case}

We also consider the Fourier transformation of the bosonic conformal ansatz solution at finite temperature as 
\bqa   F_{\beta}(1) = \int_{-\beta}^{\beta} d \tau \;  e^{i \omega \tau} \left|\frac{\frac{\pi}{\beta}}{\sin\left(\frac{\pi \tau}{\beta} \right)} \right|= \frac{2 \pi    }{\beta }\int_0^{\beta } \frac{\cos (\tau  \omega )}{\sin \left(\frac{\pi  \tau }{\beta }\right)} \, d\tau. 
\label{eq:28}\eqa   
We should notice that $\frac{\cos (\tau  \omega )}{\sin \left(\frac{\pi  \tau }{\beta }\right)}$ does not converge at the end points of $\tau= 0$ and $\tau = \beta$. To regularize this expression, we take an indefinite integral of it, and consider the limit of $\tau =0$ and $\tau = \beta$. Then, we renormalize the resulting expression.

The indefinite integral of Eq. (\ref{eq:28}) is given by
	\bqa  && \frac{2 \pi    }{\beta }\int  \frac{\cos (\tau  \omega )}{\sin \left(\frac{\pi  \tau }{\beta }\right)} \, d\tau  
	=   -2 \pi  e^{\frac{i \tau  (\pi -\beta  \omega )}{\beta }} \left(\frac{e^{2 i \tau  \omega } \, _2F_1\left(1,\frac{\beta  \omega +\pi }{2 \pi };\frac{1}{2} \left(\frac{\beta  \omega }{\pi }+3\right);e^{\frac{2 i \pi  \tau }{\beta }}\right)}{\beta  \omega +\pi }+ \frac{\, _2F_1\left(1,\frac{\pi -\beta  \omega }{2 \pi };\frac{3}{2}-\frac{\beta  \omega }{2 \pi };e^{\frac{2 i \pi  \tau }{\beta }}\right)}{\pi -\beta  \omega }\right). \nn \label{eq:29}\eqa
As $\tau \rightarrow \beta$,   Eq. (\ref{eq:29}) becomes
\bqa   e^{-i \beta  \omega } \left(-\left(\left(1+e^{2 i \beta  \omega }\right) \left(   \lim_{\tau \to \beta} \log \left(1-\frac{\tau }{\beta }\right)+\gamma +\frac{i \pi }{2}+\log (2 \pi )\right)\right)-e^{2 i \beta  \omega } \psi  \left(\frac{\beta  \omega +\pi }{2 \pi }\right)-\psi  \left(\frac{\pi -\beta  \omega }{2 \pi }\right)\right).
\label{eq:30}\eqa
Here, $ \psi  (z)$ is the digamma function. As $\tau \rightarrow 0$, Eq. (\ref{eq:29}) becomes
\bqa      
2 \left(\log \left(-\frac{2 i \pi }{\beta }\right)+ \lim_{\tau \to 0}\log (\tau )+\gamma \right)+\psi  \left(\frac{\beta  \omega +\pi }{2 \pi }\right)+\psi  \left(\frac{\pi -\beta  \omega }{2 \pi }\right).
\label{eq:31}\eqa
We notice that $\lim_{\tau \to \beta} \log \left(1-\frac{\tau }{\beta }\right)$ and $\lim_{\tau \to 0}\log (\tau )$ do not converge. But, we can regularize them as discussed above.

Substituting Eq. (\ref{eq:37}) into Eq. (\ref{eq:30}) and Eq. (\ref{eq:31}), we obtain
\bqa  && e^{-i \beta  \omega } \left(-\left(\left(1+e^{2 i \beta  \omega }\right) \left(   \lim_{\tau \to \beta} \log \left(1-\frac{\tau }{\beta }\right)+\gamma +\frac{i \pi }{2}+\log (2 \pi )\right)\right)-e^{2 i \beta  \omega } \psi  \left(\frac{\beta  \omega +\pi }{2 \pi }\right)-\psi  \left(\frac{\pi -\beta  \omega }{2 \pi }\right)\right) \nonumber\\
&=& e^{-i \beta  \omega } \left(-\left(\left(\frac{1}{2}+\frac{i \pi }{2}+\log (2 \pi )\right) \left(1+e^{2 i \beta  \omega }\right)\right)-e^{2 i \beta  \omega } \psi \left(\frac{\beta  \omega +\pi }{2 \pi }\right)-\psi \left(\frac{\pi -\beta  \omega }{2 \pi }\right)\right),
\label{eq:39}\eqa
and
\bqa &&     
2 \left(\log \left(-\frac{2 i \pi }{\beta }\right)+ \lim_{\tau \to 0}\log (\tau )+\gamma \right)+\psi  \left(\frac{\beta  \omega +\pi }{2 \pi }\right)+\psi  \left(\frac{\pi -\beta  \omega }{2 \pi }\right) \nonumber\\
&=& \psi \left(\frac{\beta  \omega +\pi }{2 \pi }\right)+\psi \left(\frac{\pi -\beta  \omega }{2 \pi }\right)+2 \log \left(-\frac{2 i \pi }{\beta }\right)+1 ,
\label{eq:40}\eqa
respectively. Inserting Eq. (\ref{eq:39}) and Eq. (\ref{eq:40}) into Eq. (\ref{eq:28}), we obtain 
	\bqa   F_{\beta}(1) = && -(  1+\log (4)+2 \log (\pi ))  \cos (\beta  \omega ) -(\cos (\beta  \omega )+1) \left(\psi  \left(\frac{\beta  \omega +\pi }{2 \pi }\right)+\psi \left(\frac{\pi -\beta  \omega }{2 \pi }\right)\right) \nn &&+2 \log (\beta )-1-2 \log (2 \pi ).   
	\label{eq:41} \eqa  

Putting $ \omega= \frac{2 n \pi }{\beta }$ into Eq. (\ref{eq:41}) and taking the following infinite summation, we obtain 
\bqa && \lim_{N\to \infty }  \sum _{n=-N}^{N} F_{\beta}^2\left(1 \right) = \lim_{N \to \infty } \sum _{n=-N}^{N} \left(2 \log \left(\frac{\beta }{2 \pi }\right)-4 \gamma +1 -4 \psi \left(n+\frac{1}{2}\right)\right)^2 \nonumber\\
&&= \lim_{N \to \infty } \Bigg\{ k^2 (2 N+1)-8 k (N+1) \left(\psi \left(-N-\frac{1}{2}\right)-2\right)-8 k N \psi\left(N+\frac{3}{2}\right) \nonumber\\
&&+16 \psi \left(\frac{1}{2}\right)^2 +32 \sum _{n=1}^{N} \psi  \left(n+\frac{1}{2}\right)^2 \Bigg\} ,
\label{qq:6} \eqa
where $k=2 \log \left(\frac{\beta }{2 \pi }\right)-4 \gamma +1$. In  Eq. (\ref{qq:6}),  we consider the following identity 
\bqa  \lim_{N\to \infty } \psi  \left(-N-\frac{1}{2}\right)  = \lim_{N\to \infty }\psi  \left(N+\frac{3}{2}\right) = \log (N).
\label{qq:7} \eqa
We also have 
\bqa && 32 \sum _{n=1}^{N} \psi  \left(n+\frac{1}{2}\right)^2 \approx 32 \int_1^{N} \log ^2\left(n+\frac{1}{2}\right) \, dn \nonumber\\
&&=16 \left(2 \left(2+\log ^2(2)+\log (4)\right) (N-1)+3 \log (9) \left(1+\coth ^{-1}(7)\right)\right) \nonumber\\
&&+ 16 (2 N+1) \log (2 N+1) \left(\log \left(\frac{1}{4} (2 N+1)\right)-2\right).
\label{qq:8} \eqa

Substituting Eq. (\ref{qq:7}) and Eq. (\ref{qq:8}) into Eq. (\ref{qq:6}), we obtain
\bqa &&\lim_{N\to \infty }  \sum _{n=-N}^{N} F_{\beta}^2\left(1 \right) \nonumber\\
&& = \lim_{N \to \infty } \Bigg\{ k^2 (2 N+1)-8 k (N+1) \left(\log (N)-2\right)-8 k  N \log (N) \nonumber\\
&&+16 \psi \left(\frac{1}{2}\right)^2 +16 \left(2 \left(2+\log ^2(2)+\log (4)\right) (N-1)+3 \log (9) \left(1+\coth ^{-1}(7)\right)\right) \nonumber\\
&&+ 16 (2 N+1) \left(\log \left(N+\frac{1}{2}\right)+\log (2)\right) \left(\log \left(N+\frac{1}{2}\right)-2-\log (2)\right) \Bigg\} .
\label{qq:9} \eqa
Eq. (\ref{eq:38}) results in
\bqa      
\lim_{N \to \infty} \log(N) = \lim_{N \to \infty}\log \left(N+\frac{1}{2}\right)= 2\log(2)+\gamma -\frac{1}{2}.
\label{qq:10}\eqa 
In Eq. (\ref{qq:9}), we also consider the following regularization
\bqa  \lim_{N \to \infty} N = \lim_{N\to \infty } \sum _{n=1}^N 1 =\zeta(0) = -\frac{1}{2} . 
\label{qq:11}\eqa 
%
%
Substituting Eq. (\ref{qq:10}) and Eq. (\ref{qq:11}) into Eq. (\ref{qq:9}), we finally have
\bqa \lim_{N\to \infty }  \sum _{n=-N}^{N} F_{\beta}^2\left(1 \right) \approx 16 \log (\beta )-43.1576.
\label{qq:12} \eqa   

There is another way of regularization for Eq. (\ref{eq:28}),
\bqa   F_{\beta}(1) =  \frac{2 \pi    }{\beta }\int_0^{\beta } \frac{\cos (\tau  \omega )}{\sin \left(\frac{\pi  \tau }{\beta }\right)} \, d\tau = 2 \int_0^{\pi } \frac{\cos \left(\frac{\beta  x \omega }{\pi }\right)}{\sin (x)} \, dx, 
\label{qq:1}\eqa 
where $x=\frac{\pi  \tau }{\beta }$ has been introduced. The indefinite integral of Eq. (\ref{qq:1}) is
\bqa 
-2 \pi  e^{\frac{i x (\pi -\beta  \omega )}{\pi }} \left(\frac{e^{\frac{2 i \beta  x \omega }{\pi }} \, _2F_1\left(1,\frac{\beta  \omega +\pi }{2 \pi };  \frac{\beta  \omega }{2\pi }+\frac{3}{2}; e^{2 i x}\right)}{\beta  \omega +\pi }+\frac{\, _2F_1\left(1,\frac{\pi -\beta  \omega }{2 \pi };\frac{3}{2}-\frac{\beta  \omega }{2 \pi };e^{2 i x}\right)}{\pi -\beta  \omega }\right).
\label{qq:2}\eqa 
Then, we obtain
	\bqa && \mbox{Eq. (\ref{qq:2})} =_{\tau \rightarrow \pi} \frac{1}{2} e^{-i \beta  \omega } \left(-2 e^{2 i \beta  \omega } \psi \left(\frac{\beta  \omega +\pi }{2 \pi }\right)-2 \psi\left(\frac{\pi -\beta  \omega }{2 \pi }\right)-\left(1+e^{2 i \beta  \omega }\right) \left(2  \lim_{\tau \to \pi} \log \left(1-\frac{x}{\pi }\right)+2 \gamma +i \pi +\log \left(4 \pi ^2\right)\right)\right) ,\nn \label{qq:3} \\ 
 && \mbox{Eq. (\ref{qq:2})} =_{\tau \rightarrow 0}\psi\left(\frac{\pi -\beta  \omega }{2 \pi }\right)+\psi \left(\frac{\beta  \omega +\pi }{2 \pi }\right)+2\lim_{\tau \to 0} \log (x)+2 \gamma -i \pi +2 \log (2).
	\label{qq:4} \eqa
Following the same regularization scheme above, we have
\bqa F_{\beta}(1) = -(\cos (\beta  \omega )+1)(1+\log (4)) -2 \log (\pi ) \cos (\beta  \omega )-(\cos (\beta  \omega )+1) \left(\psi\left(\frac{\beta  \omega +\pi }{2 \pi }\right)+\psi\left(\frac{\pi -\beta  \omega }{2 \pi }\right)\right) .
\label{qq:5}\eqa

Putting $\omega= \frac{2 n \pi }{\beta }$ into Eq. (\ref{qq:5}) and taking the following infinite summation, we have 
\bqa &&\lim_{N\to \infty }  \sum _{n=-N}^{N} F_{\beta}^2\left(1 \right) = \lim_{N \to \infty } \sum _{n=-N}^{N} \left(4 \psi  \left(n+\frac{1}{2}\right)+2 (1+\log (4 \pi ))\right)^2 \nonumber\\
&&= \lim_{N \to \infty } \Bigg\{ t^2 (2 N+1)+8 t \left(-2 (N+1)+(N+1) \psi  \left(-N-\frac{1}{2}\right)+N \psi  \left(N+\frac{3}{2}\right)\right) \nonumber\\
&&\ \ \ \ \ \ \ \ \ \ \ \  +16 \psi \left(\frac{1}{2}\right)^2 +32 \sum _{n=1}^{N} \psi  \left(n+\frac{1}{2}\right)^2 \Bigg\} .
\label{qq:13} \eqa
Substituting Eq. (\ref{qq:7}) and Eq. (\ref{qq:8}) into Eq. (\ref{qq:13}), we obtain
\bqa &&\lim_{N\to \infty }  \sum _{n=-N}^{N} F_{\beta}^2\left(1 \right) \nonumber\\
&&= \lim_{N \to \infty } \Bigg\{ t^2 (2 N+1)+8 t (-2 (N+1)+(N+1) \log (N)+N \log (N))+16 \psi \left(\frac{1}{2}\right)^2 \nonumber\\
&&\ \ \ \ \ \ \ \ \ \ \ \ +16 \left(2 \left(2+\log ^2(2)+\log (4)\right) (N-1)+3 \log (9) \left(1+\coth ^{-1}(7)\right)\right) \nonumber\\
&&\ \ \ \ \ \ \ \ \ \ \ \ +16 (2 N+1) \left(\log \left(N+\frac{1}{2}\right)-2-\log (2)\right) \left(\log \left(N+\frac{1}{2}\right)+\log (2)\right) \Bigg\} .
\label{qq:14} \eqa
Taking Eq. (\ref{qq:10}) and Eq. (\ref{qq:11}) into Eq. (\ref{qq:14}), we have
\bqa \lim_{N\to \infty }  \sum _{n=-N}^{N} F_{\beta}^2\left(1 \right) \approx -59.777.
\label{qq:15} \eqa

\subsection{$\alpha$ $\&$ $a_{0}$} 

Finally, we introduce 
\bqa \alpha= - 2 q_{d} \Gamma_{J}\frac{\lim_{N\to \infty }  \sum _{n=-N}^{N} F_{\beta}\left(1/2 \right)  F_{\beta}\left(3/2 \right)}{\lim_{N\to \infty }  \sum _{n=-N}^{N}1},
\label{qq:16} \eqa
and
\bqa a_0= \frac{\lim_{N\to \infty }  \sum _{n=-N}^{N} F_{\beta}\left(1/2 \right)  F_{\beta}\left(3/2 \right)}{\lim_{N\to \infty }  \sum _{n=-N}^{N} F_{\beta}^2\left(1 \right)}.
\label{qq:17} \eqa
Inserting Eq. (\ref{eq:27}) into Eq. (\ref{qq:16}), we obtain
\bqa \alpha = 32 \pi  q_{d} \Gamma_{J}.
\label{qq:18} \eqa   
Let's take Eq. (\ref{eq:27}), Eq. (\ref{qq:11}), and Eq. (\ref{qq:12}) into Eq. (\ref{qq:17}) with $N \rightarrow \infty$. Then, we obtain
\bqa a_0= -\frac{16 \pi }{4 \log ^2(\beta )-27.355 \log (\beta )+62.767}.
\label{qq:19} \eqa
For the other regularization, we have 
\bqa a_0= -0.52636.
\label{qq:20} \eqa

\section{Regularizations for the mean-field free energy } \label{MFF}

By regularizing Eq. (\ref{eq:27}) through Eq. (\ref{qq:11}), we write the field theoretic large-$N$ free energy in Eq. (\ref{eq:85}) as follows
\bqa
	F_{eff}^{(0)}(\beta) = \frac{N}{\beta} \left\{  q_{d} \Gamma_{J} \left(  \mathit{z}_0  \mathit{y}_0^2+\frac{1}{4} \mathit{z}_0^4  \right) \sum_{i \omega_{n}} F_{\beta}^2(1)    - \frac{1}{2}  \sum_{i \omega_{n}}  \ln \Big( q_{d} \Gamma_{J}  \mathit{x}_0  \mbox{sgn}(\omega_{n}) F_{\beta}(3/2) \Big) \right\}. \label{tt:85}
\eqa
Substituting Eq. (\ref{qq:15}) into Eq. (\ref{tt:85}) with $z_0=4y_0^2$ in Eq. (\ref{eq:67}), we obtain 
	\bqa
		F_{eff}^{(0)}(\beta) &=& \frac{N}{\beta} \left\{ -59.777 \times 8  q_{d} \Gamma_{J}  \mathit{y}_0^4    - \frac{1}{2}  \sum_{i \omega_{n}}  \ln \Big( q_{d} \Gamma_{J}  \mathit{x}_0  \mbox{sgn}(\omega_{n}) F_{\beta}(3/2) \Big) \right\}.\label{tt:86}
  \eqa 

Inserting $a_0$ in Eq. (\ref{qq:18}) and $\alpha$ in  Eq. (\ref{qq:20}) into $\mathit{y}_0$ in Eq. (\ref{eq:67}), and taking this new $\mathit{y}_0$  into Eq.(\ref{tt:86}), we have
  \begin{eqnarray}
		F_{eff}^{(0)}(\beta)&=& \frac{N}{\beta} \left\{ - 0.0028 - \frac{1}{2}  \sum_{i \omega_{n}}  \ln \Big( q_{d} \Gamma_{J}  \mathit{x}_0  \mbox{sgn}(\omega_{n}) F_{\beta}(3/2) \Big) \right\}\nonumber \\
  &\approx&   -\frac{N}{2 \beta}   \sum_{i \omega_{n}}  \ln \Big( q_{d} \Gamma_{J}  \mathit{x}_0  \mbox{sgn}(\omega_{n}) F_{\beta}(3/2) \Big) \nonumber\\
		&= & -\frac{N}{2 \beta} \lim_{\mathbb{N} \to \infty}  \sum _{n=-\mathbb{N}}^\mathbb{N} \log \left(-\frac{  8 i \sqrt{2} \pi ^2 \sqrt{\frac{1}{\beta }} (-1)^n   \text{sgn}(2 n+1)}{\Gamma \left(-n-\frac{1}{4}\right) \Gamma \left(n+\frac{3}{4}\right)}\right)
		\nonumber\\
		&= & -\frac{N}{2 \beta} \left(\log \left(-\sqrt{\frac{1}{\beta }}\right) \lim_{\mathbb{N}\to \infty }  \sum _{n=-\mathbb{N}}^{\mathbb{N}} 1   +\lim_{\mathbb{N}\to \infty }   \sum _{n=-\mathbb{N}}^{\mathbb{N}} \log \left(\frac{\left(8 i \sqrt{2} \pi ^2 (-1)^n\right) \text{sgn}(2 n+1)}{\Gamma \left(-n-\frac{1}{4}\right) \Gamma \left(n+\frac{3}{4}\right)}\right) \right).  \label{eq:85_Appendix}
	\end{eqnarray}    
By applying the regularization 
$\lim_{\mathbb{N} \to \infty}\sum_{n=-\mathbb{N}}^{\mathbb{N}}1=0$, we obtain
	\begin{eqnarray} 
		F_{eff}^{(0)}(\beta) 	&= & -\frac{N}{2 \beta}  \lim_{\mathbb{N}\to \infty }   \sum _{n=-\mathbb{N}}^{\mathbb{N}} \log \left(\frac{\left(8 i \sqrt{2} \pi ^2 (-1)^n\right) \text{sgn}(2 n+1)}{\Gamma \left(-n-\frac{1}{4}\right) \Gamma \left(n+\frac{3}{4}\right)}\right) \nonumber\\
		&=&  -\frac{N}{2 \beta} \lim_{\mathbb{N}\to \infty }  \left(\sum _{n=1}^{\mathbb{N}} \log \left(\frac{8 i \sqrt{2} \pi ^2 (-1)^n}{\Gamma \left(-n-\frac{1}{4}\right) \Gamma \left(n+\frac{3}{4}\right)}\right)+\sum _{n=-\mathbb{N}}^{-1} \log \left(-\frac{8 i \sqrt{2} \pi ^2 (-1)^n}{\Gamma \left(-n-\frac{1}{4}\right) \Gamma \left(n+\frac{3}{4}\right)}\right)\right)\nonumber\\
		&&-\frac{N}{2 \beta}  \log \left(\frac{8 i \sqrt{2} \pi ^2}{\Gamma \left(-\frac{1}{4}\right) \Gamma \left(\frac{3}{4}\right)}\right)\nonumber\\
		&= &  -\frac{N}{2 \beta}  \lim_{\mathbb{N}\to \infty } \sum _{n=1}^{\mathbb{N}} \log \left(\frac{128 \pi ^4}{\Gamma \left(-n-\frac{1}{4}\right) \Gamma \left(\frac{3}{4}-n\right) \Gamma \left(n-\frac{1}{4}\right) \Gamma \left(n+\frac{3}{4}\right)}\right) -\frac{N}{2 \beta}\log \left(\frac{8 i \sqrt{2} \pi ^2}{\Gamma \left(-\frac{1}{4}\right) \Gamma \left(\frac{3}{4}\right)}\right).  	 
		 \label{rr:1}
	\end{eqnarray}    
Here, we considered
	\begin{eqnarray}
		&&  \lim_{\mathbb{N}\to \infty } \sum _{n=1}^{\mathbb{N}} \log \left(\frac{128 \pi ^4}{\Gamma \left(-n-\frac{1}{4}\right) \Gamma \left(\frac{3}{4}-n\right) \Gamma \left(n-\frac{1}{4}\right) \Gamma \left(n+\frac{3}{4}\right)}\right) \nonumber\\
		&=&  \lim_{\mathbb{N}\to \infty } \log \left(\frac{G\left(\mathbb{N}+\frac{5}{4}\right) G\left(\mathbb{N}+\frac{9}{4}\right)}{G\left(\mathbb{N}+\frac{3}{4}\right) G\left(\mathbb{N}+\frac{7}{4}\right)}\right)+\frac{C}{\pi }\nonumber\\
		&&+\lim_{\mathbb{N}\to \infty }\log \left(2^{11 \mathbb{N}+2} \pi ^{4 \mathbb{N}} \Gamma \left(\frac{1}{4}\right)^{-2 \mathbb{N}-\frac{3}{2}} \Gamma \left(\frac{3}{4}\right)^{\mathbb{N}+\frac{1}{2}} \left(\frac{5}{3} \Gamma \left(-\frac{5}{4}\right) \Gamma \left(-\frac{1}{4}\right) \Gamma \left(\frac{7}{4}\right)\right)^{-\mathbb{N}}\right)\nonumber\\
		&=&  \frac{C}{\pi }+\lim_{\mathbb{N}\to \infty } \left((11 \mathbb{N}+2) \log (2)+  \mathbb{N} \log \left(\frac{3 \pi ^4}{5 \Gamma \left(-\frac{5}{4}\right) \Gamma \left(-\frac{1}{4}\right) \Gamma \left(\frac{7}{4}\right)}\right) -\left(2 \mathbb{N}+\frac{3}{2}\right) \log \left(\Gamma \left(\frac{1}{4}\right)\right)+\left(\mathbb{N}+\frac{1}{2}\right) \log \left(\Gamma \left(\frac{3}{4}\right)\right)  \right)\nonumber\\
		&&+\lim_{\mathbb{N}\to \infty } \log \left(\frac{G\left(\mathbb{N}+\frac{5}{4}\right) G\left(\mathbb{N}+\frac{9}{4}\right)}{G\left(\mathbb{N}+\frac{3}{4}\right) G\left(\mathbb{N}+\frac{7}{4}\right)}\right).\label{eq:ccc}
	\end{eqnarray}  
$C= \sum _{k=0}^{\infty } \frac{(-1)^k}{(2 k+1)^2}\approx 0.915966$ is Catalan's constant, and $G\left(z\right)$ is the Barnes G-function \cite{Barnes_G_Function}. 

Following the same regularization scheme as shown in the previous appendix subsection, this expression is simplified as
	\begin{eqnarray}
		&&  \lim_{\mathbb{N}\to \infty } \sum _{n=1}^{\mathbb{N}} \log \left(\frac{128 \pi ^4}{\Gamma \left(-n-\frac{1}{4}\right) \Gamma \left(\frac{3}{4}-n\right) \Gamma \left(n-\frac{1}{4}\right) \Gamma \left(n+\frac{3}{4}\right)}\right) \nonumber\\ 
		&=&  \frac{1}{2} \left(\log (2 \pi )+ \frac{2 }{\pi }+i \pi +\log \left(-\frac{5 \Gamma \left(-\frac{5}{4}\right) \Gamma \left(-\frac{1}{4}\right) \Gamma \left(\frac{7}{4}\right)}{384 \pi ^4 \Gamma \left(\frac{1}{4}\right)}\right)\right) , \label{eq:c}
	\end{eqnarray}  
where we used
	\begin{eqnarray}
		\lim_{\mathbb{N}\to \infty } \log \left(\frac{G\left(\mathbb{N}+\frac{5}{4}\right) G\left(\mathbb{N}+\frac{9}{4}\right)}{G\left(\mathbb{N}+\frac{3}{4}\right) G\left(\mathbb{N}+\frac{7}{4}\right)}\right) &\approx& \lim_{\mathbb{N}\to \infty }  \log \left(\exp \left(-\frac{23}{96 \mathbb{N}}+\mathbb{N} (\log (\mathbb{N})-1)+\frac{1}{2} (\log (\mathbb{N})+\log (2 \pi ))\right)\right) \nonumber\\
		&\approx& \lim_{\mathbb{N}\to \infty } \log \left(e^{\left(\mathbb{N}+\frac{1}{2}\right) \log (\mathbb{N})+\frac{1}{2} \log (2 \pi )}\right)=  \lim_{\mathbb{N}\to \infty } \left(\mathbb{N}+\frac{1}{2}\right) \log (\mathbb{N})+\frac{1}{2} \log (2 \pi ) \nonumber\\
		&=& \frac{1}{2} \log (2 \pi ).\label{eq:d}
	\end{eqnarray}  

%
%

Substituting Eq. (\ref{eq:c}) into Eq. (\ref{rr:1}), we have
	\begin{eqnarray}
		F_{eff}^{(0)}(\beta) &=& -\frac{N \left(\frac{4 C}{\pi }+\log (2)\right)}{8 \beta }=-\frac{0.232424 N}{\beta }, \label{eq:ccc3}
	\end{eqnarray}    
where the numerical coefficient $0.232424$ coincides with that of Ref. \cite{SYK_Model_V}, precisely.

\end{document}